%% file: CompTools.tex
\documentclass[
    a4paper,
    preprintnumbers,
    floatfix,
    superscriptaddress,
    pra,
    twocolumn,
    showpacs,
    longbibliography,
    nofootinbib,        
]{revtex4-1}

\usepackage[utf8]{inputenc}
\usepackage[english]{babel}

\usepackage{
    microtype,      
    dcolumn,        
}

\usepackage{
    etoolbox,       
    xparse,         
    xspace,         
    mathtools,      
}

\usepackage{
    amsmath,
    amsthm,         
    amsfonts,
    amssymb,        
    bbm,            
    braket,         
    breqn,          
}
\usepackage[
    load-configurations=abbreviations,
    separate-uncertainty,
    per-mode=symbol-or-fraction,
    table-number-alignment=center,
    binary-units=true,
]{siunitx}

\usepackage{
    graphicx,       
    booktabs,       
    listingsutf8,   
    algpseudocode,  
    algorithm,      
    float,          
    multirow,       
    rotating,       
    subcaption,     
}
\usepackage[export]{adjustbox}  

\usepackage{
    color,          
    xcolor,         
    attachfile,     
}

\usepackage{
    caption,        
    ragged2e,       
}

\usepackage{hyperref}

\usepackage{titlesec}

\titleformat{\part}
[display]
{\centering\normalfont\large\bfseries}
{\partname~\thepart}
{0pt}
{\huge}
\makeatletter

\let\c@table\c@figure
\makeatother

\DeclareCaptionJustification{justified}{\justifying}
\newfloat{algorithm}{t}{lop}

\algtext*{EndWhile}
\algtext*{EndIf}
\algtext*{EndFor}
\algtext*{EndFunction}

\let\orgautoref\autoref
\renewcommand{\autoref}
        {\def\equationautorefname{Equation}%
         \def\figureautorefname{Figure}%
         \def\subfigureautorefname{Figure}%
         \def\Itemautorefname{Item}%
         \def\tableautorefname{Table}%
         \def\sectionautorefname{Section}%
         \def\subsectionautorefname{Section}%
         \def\subsubsectionautorefname{Section}%
         \def\chapterautorefname{Chapter}%
         \def\partautorefname{Part}%
         \orgautoref}

\newcommand{\algrhs}[1]{\hfill \parbox[t]{0.7\linewidth}{#1}}

\makeatletter
\AtEndEnvironment{algorithm}{\kern2pt\hrule\relax\vskip3pt\@algcomment}
\let\@algcomment\relax
\newcommand\algcomment[1]{\def\@algcomment{\footnotesize#1}}
\renewcommand\fs@ruled{\def\@fs@cfont{\bfseries}\let\@fs@capt\floatc@ruled
  \def\@fs@pre{\hrule height.8pt depth0pt \kern2pt}%
  \def\@fs@post{}%
  \def\@fs@mid{\kern2pt\hrule\kern2pt}%
  \let\@fs@iftopcapt\iftrue}
\makeatother

\newcommand\alignl[2]{\mathrlap{#2}\phantom{#1}}
\newcommand\alignc[2]{\ooalign{$\phantom{#1}$\cr\hidewidth$#2$\hidewidth}}

\newcommand{\mm}[1]{\ensuremath{#1}\xspace}

\DeclarePairedDelimiterX{\norm}[1]{\lVert}{\rVert}{#1}
\DeclarePairedDelimiter\abs{\lvert}{\rvert}%
\DeclarePairedDelimiter\floor{\lfloor}{\rfloor}

\DeclarePairedDelimiter\mean{\langle}{\rangle}

\DeclareMathOperator\Pa{Pa}     
\DeclareMathOperator\argmin{arg\,min}

\newcommand\spanof[1]{\mm{\langle #1 \rangle}}
\newcommand{\ketbra}[2]{\ket{#1}\hspace{-0.2em}\bra{#2}}

\renewcommand{\vec}[1]{\ensuremath{\boldsymbol{#1}}}    
\newcommand{\tran}{^\intercal}                          
\newcommand{\dvec}[1]{{\vec #1}\tran}                   
\newcommand{\dotp}[2]{{\vec #1}\tran{\vec #2}}          

\newcommand\Mid{\alignc{{},{}}{|}}

\newcommand\R{\mathbbm{R}}
\newcommand\CC{\mathbbm{C}}

\newcommand{\BigO}{\mathcal{O}}
\newcommand\Cn[1]{\mm{\mathcal C_{#1}}}

\newcommand\A{A}
\newcommand\B{B}
\newcommand\C{C}
\newcommand\X{X}

\newcommand{\ProcName}[1]{\mm{\textsc{#1}}}
\newcommand{\Rotate}{\ProcName{rotate}}
\newcommand{\ToFacet}{\ProcName{tofacet}}
\newcommand{\ToFacets}{\ProcName{tofacets}}
\newcommand{\InitSimplex}{\ProcName{basis\,simplex}}
\newcommand{\FindVertex}{\ProcName{find\,vertex}}
\newcommand{\PtToFacet}{\ProcName{point\,2\,facet}}

\newcommand{\CHSHE}{\mm{\operatorname{CHSH_E}}}
\newcommand{\AFI}[1]{\mm{\operatorname{AFI_{#1}}}}
\newcommand{\RFD}[1]{\mm{\operatorname{RFD_{#1}}}}

\DeclareDocumentCommand{\poly}{ O{P} }{\mm{\mathcal{#1}}}
\newcommand{\marg}{\mm{\mathcal M}}         
\newcommand{\ents}{\mm{\mathcal R_n}}       
\newcommand{\subs}{\mm{\mathbb R^d}}        
\newcommand{\Pinp}{\mm{\poly[P]}}           
\newcommand{\Pfin}{\mm{\pi_\df(\Pinp)}}     
\newcommand{\Qi}{\mm{\poly[Q]_0}}           
\newcommand{\Qf}{\mm{\poly[Q]}}             
\newcommand{\di}{\mm{d+e}}                  
\newcommand{\df}{\mm{d}}                    

\newcommand{\con}[2]{\mm{{#1}_{#2}}}
\newcommand{\conV}[2]{\con{\vec{#1}}{#2}}
\newcommand{\conM}[2]{\con{#1}{\vec{#2}}}
\newcommand{\La}{\conM La}
\newcommand{\fb}{\conV fb}
\newcommand{\Lxga}{\mm{L\vec x \ge \vec a}}
\newcommand\Qunits[1]{\mm{\CC_{#1}^{\otimes 3}}}

\newcommand\minimize[1]{\argmin\, \vec{#1} \mapsto}
\newcommand\st{\text{subject to }}

\newcommand\SecRef[1]{Sec.~\ref{#1}}

\makeindex

\makeatletter
\let\cat@comma@active\@empty
\makeatother

\begin{document}
\title{Computational tools for solving a marginal problem with applications in Bell non-locality and causal modeling}
\date{\today}

\author{T. Gläßle}
\affiliation{69221 Heidelberg, Germany}
\author{D. Gross}
\affiliation{Institute for Theoretical Physics, University of Cologne, 50937 Cologne, Germany}
\author{R. Chaves}
\affiliation{International Institute of Physics, Federal University of Rio Grande do Norte, 59070-405 Natal, Brazil}

\begin{abstract}

    Marginal problems naturally arise in a variety of different fields: basically, the question is whether some marginal/partial information is compatible with a joint probability distribution. To this aim, the characterization of marginal sets via quantifier elimination and polyhedral projection algorithms is of primal importance.
    In this work, before considering specific problems, we review polyhedral
    projection algorithms with focus on applications in information theory,
    and, alongside known algorithms, we also present a newly developed
    geometric algorithm which walks along the face lattice of the polyhedron
    in the projection space.
    One important application of this is in the field of quantum non-locality,
    where marginal problems arise in the computation of Bell inequalities.
    We apply the discussed algorithms to discover many tight entropic Bell
    inequalities of the tripartite Bell scenario as well as more complex networks arising in the field of causal inference. Finally, we analyze the usefulness of these inequalities as nonlocality witnesses by searching for violating quantum states.

\end{abstract}

\maketitle

\section{Introduction}
Starting point of this paper is the marginal problem: given joint distributions of certain subsets of random variables $X_1,\dots, X_n$, are they compatible with the existence of any joint distribution for all these variables? In other words, is it possible to find a joint distribution for all these variables, such that this distribution marginalizes to the given ones? Such a problem naturally arises in several different fields. From the classical perspective, applications of the marginal problem range, just to cite a few examples, from knowledge integration in artificial intelligence and database theory \cite{studeny1994marginal,lauritzen1988} to causal discovery \cite{Pearlbook,Spirtesbook} and network coding protocols \cite{Yeung2008}. Within quantum information perhaps the most famous example --the one that will be the main focus in this paper-- is the phenomenon of nonlocality \cite{Bell1964}, showing that quantum predictions for experiments performed by distant parties are at odds with the assumption of local realism.

As shown by Bell in his seminal paper \cite{Bell1964}, the assumption of local realism imposes strict constraints on the possible probability distributions that are compatible with it. These are the famous Bell inequalities that play a fundamental role in the understanding of nonlocality since it is via their violation that we can unambiguously probe the nonlocal character of quantum correlations. Given its importance, very general frameworks have been developed for the derivation of Bell inequalities \cite{Pitowsky1989,Brunner2014}. Unfortunately, however, finding all Bell inequalities is a very hard problem given that its computational complexity increases very fast as the scenario of interest becomes less simple \cite{Pitowsky1989,Pitowsky1991}. The situation is even worse for the study of nonlocality in complex quantum networks, where on the top of local realism one also imposes additional constraints \cite{Branciard2010,Branciard2012,Fritz2012,Tavakoli2014,
Chaves2015b,Chaves2016,Tavakoli2016,Lee2015,Wolfe2016,Carvacho2017experimental,Andreoli2017}. In this case, the derivation of Bell inequalities involves the characterization of complicated non-convex sets for which even more computationally demanding tools from algebraic geometry \cite{Geiger1999,Garcia2005,Chaves2016,Lee2015} seem to be the only viable alternative.

In order to circumvent some these issues, an alternative route that has been attracting growing attention lately is the one given by entropic Bell inequalities \cite{Braunstein1988,Cerf1997,Chaves2012,Fritz2012,Fritz2013,Chaves2013,Chaves2014,
Henson2014,Kurzy2014,Poh2015,chaves2016nonsignalling,
weilenmann2016entropy,weilenmann2016non,Budroni2016indis,weilenmann2017analysing,miklin2017entropic,Pienaar2017}. In this case, rather than asking if a given probability distribution is compatible or not with local realism, we ask the same question but for the Shannon entropies of such distributions. The novelty of the entropic approach have both conceptual and technical reasons. Entropy is a key concept in our understanding of information theory, thus having a framework that focuses on it naturally leads to new insights and applications \cite{Pawlowski2009,Short2010,Dahlsten2012,Barnum2010,Fritz2012,Henson2014,Chaves2015,Chaves2015c}. In turn, entropies allow for a much simpler and compact characterization of Bell inequalities --at the cost of not being a complete description-- in a variety of scenarios, most notably in the aforementioned quantum networks \cite{Chaves2014,Henson2014,Chaves2015,chaves2016nonsignalling,
weilenmann2017analysing,miklin2017entropic,Pienaar2017}. Not surprisingly, however, this approach is also hampered by computational complexity issues. As will be discussed in details through this paper, current methods for the derivation of entropic Bell inequalities (see for instance \cite{Budroni2012,Chaves2014,Henson2014,chaves2016nonsignalling,weilenmann2016non,weilenmann2017analysing,miklin2017entropic} mostly rely on quantifier elimination algorithms \cite{Williams1986} that in practice is limited to a few simple cases of interest.

Within this context the aim of this paper is three-fold. First, to review existing methods for solving the marginal problem, particularly those relevant for the derivation of Bell inequalities. Given the importance of it, not surprisingly there is a rich literature and methods aiming at its solution \cite{duffin1974,Williams1986,huynh1990,davenport1988,lassez1990qc,lassez1990quantifier,huynh1992,imbert1993choose,simon2005sparsity,jones04equalityset}. Most of these results, however, appear in quite diverse contexts and most prominently in the field of convex optimization and computer science. Thus, we hope that researchers on quantum information and in particular working on Bell nonlocality will benefit from the concise and unified exposition of the computational tools that we present here. Our second aim is to give our own contribution by developing new computational tools that complement and in some case improve existing algorithms. That is of particular relevance to our third and final aim with this paper: to apply our improved method to derive new tests for witnessing quantum nonlocality. Here we limit ourselves to the derivation of new entropic Bell inequalities. Notwithstanding, the tools we review and introduce are quite general and can in principle also lead to new results and insights in the derivation of usual Bell inequalities. As we show, employing our new techniques we manage to derive new entropic inequalities for the tripartite Bell scenario and in some cases provide a complete characterization. In particular, we derive inequalities that involve marginal information only, that is, they only contain the entropies of at most two out of the three parties involved in the Bell test. We then show that is possible to witness quantum nonlocality in a tripartite scenario from local bipartite marginal distributions, thus partially extending to the entropic regime the results obtained in Ref. \cite{wurflinger2012}.

The first part of this paper is concerned with the computational tools themselves. In Secs.~\ref{sec:definitions}-\ref{sec:chm} we provide a review of known algorithms for the projection of convex polyhedra and their application in the derivation of Bell inequalities. In \SecRef{sec:afi} we introduce a new algorithm that we call adjacent facet iteration, study some of its properties and make a comparison with previous algorithms. These sections will be quite technical and can be skipped for the reader only interested in the applications of the new methods for the derivation of Bell inequalities.

In the second part, we deal with applications in the context of Bell nonlocality.  In \SecRef{sec:Bell_marginal} we introduce more formally the marginal problem and cast the study of nonlocality and the derivation of Bell inequalities as a particular case of it. In \SecRef{sec:entropic} we also introduce the entropic approach and give the basic elements in information theory and convex polyhedra required to understand it. We then proceed to apply the methods discussed in the first part to find Bell inequalities of the triparte Bell scenario \SecRef{sec:tripartite_ineqs}. In \SecRef{sec:nonlocality_marginals} we discuss the aforementioned application for detecting nonlocality from separable and local marginals. We also extend our analysis in the Appendix \autoref{sec:ancestor} to compute the full marginal characterizations of causal models beyond Bell scenarios and that had remained uncharacterized until now. We end the paper in \SecRef{sec:discussion} with an discussion of our findings and an outlook for promising future directions.

We highlight that all our results were obtained with a suite of programs that were developed during the course of this work by one of authors.\footnote{Questions and comments should be addressed to Thomas Gläßle, reachable at \href{mailto:thomas@coldfix.de}{thomas@coldfix.de}.} In the hope that it might be useful to others, we have released these programs as a free software package under the GNU General Public License (GPLv3) that can be found online on a public repository \cite{pystif}. For the convenience of the reader, all inequalities found in this work are listed in Appendix \ref{sec:ancestor}, and \ref{chap:facetlisting} and are also available online \cite{pystif}.

\begin{widetext}
    \part{Polyhedral Projection}
\end{widetext}

As we will see in \SecRef{sec:entropic}, Bell scenarios in the entropy
formalism are described by systems of linear inequalities and the marginal
problem boils down to variable elimination. From a geometric perspective, this
can be understood as the orthogonal projection of convex polyhedra to
subspaces. In this part, we first establish more formal definitions
(\autoref{sec:definitions}), proceed to discuss known projection algorithms:
Fourier-Motzkin elimination (\autoref{sec:fme}), Extreme Point Method
(\autoref{sec:epm}), Convex Hull method (\autoref{sec:chm}), Equality Set
Projection (\autoref{sec:esp}), and finally present a new algorithm in
\autoref{sec:afi} that acts similar to Equality Set Projection
~\cite{jones04equalityset}. We give a short summary in
\autoref{sec:algsummary}.

\vspace{3ex}

\section{Definitions and notation}
\label{sec:definitions}

\subsection{Convex polyhedra}

A \emph{convex polyhedron} can be written in the form
\begin{align}
    \label{eq:input}
    \Pinp &= \Big\{ \vec x \in \R^d : \Lxga \Big\},
\end{align}
using a constraint matrix $L$ and a vector $\vec a$ of inhomogeneities.
Equation \ref{eq:input} is the so-called \emph{half-space} representation of
convex polyhedra. By the Minkowski-Weyl theorem every polyhedron can
alternatively be represented in \emph{vertex representation} as the convex
hull of a set of vertices and extreme rays.
An algorithm that computes the half-space representation from the vertex
representation is called \emph{convex hull algorithm}. The converse problem is
known as \emph{vertex enumeration}.

The homogeneous case $L\vec x \ge 0$, i.e.~$\vec a = \vec 0$, corresponds to so-called
\emph{convex polyhedral cones}. We frequently encounter this form in the
context of entropy inequalities where we also know that $\vec x \ge 0$ for entropic
vectors $\vec x \in \Pinp$, i.e.~entropy cones reside in the positive orthant.

A bounded convex polyhedron is called \emph{polytope}. Algorithms
for cones and polytopes can easily be obtained from each other by standard
linear programming techniques. In particular, we note that entropy cones are
fully characterized by the polytope that is their intersection with the unit
simplex. Another possibility to transform an entropy cone to an equivalent
polytope is to simply limit all variables below a certain threshold. This can
be more convenient at times. On the other hand, a polytope can be transformed
into a cone by absorbing the inhomogeneity $\vec a$ into coefficients for a
new variable with the constant value $x_0 = 1$. With this in mind, we will
from now on consider projection of cones and polytopes an equivalent problem
and will not be concerned with the distinction.

Notationwise, we will often write \fb to denote a single affine constraint
$\dotp fx \ge b$. Addition and scaling shall be understood
on the level of coefficients, i.e.~
\begin{align*}
\beta\conV fb + \gamma\conV gc \equiv \con{(\beta\vec f + \gamma\vec g)}{\beta b +  \gamma c}.
\end{align*}
Likewise, \La shall denote a system of affine constraints. We also assume that
both matrix and set operations are understood, e.g.~for adding constraints to
the system as $\La \cup \{ \fb \}$.

\subsection{Face lattice}

We say that a linear constraint $\dotp fx\ge b$ is valid (implied) if it is
true for all $\vec x\in \poly$. A \emph{face} of \poly is its intersection
with a valid constraint
\begin{align}
    \label{eq:def-face}
    \poly[F] &= \Big\{ \vec x \in \poly : \dotp fx = b \Big\},
\end{align}
including the empty set $\emptyset$ and \poly itself. Faces of faces of \poly
are faces of \poly. According to \eqref{eq:def-face}, a face is defined by the
conjunction of linear inequalities and thus a polyhedron itself. In
particular, faces of cones are cones and faces of
polytopes are polytopes. For a polyhedron $\poly$ that is not contained
in any hyperplane we refer to a $k$-dimensional face briefly as a $k$-face.
The empty set is defined as $(-1)$-face.
The faces of dimension $(d{-}1)$, $(d{-}2)$, $1$, $0$ are called \emph{facets},
\emph{ridges}, \emph{edges} and \emph{vertices}, respectively. The vertices
are the same as the \emph{extreme points} of \poly, i.e.~those points
which can not be represented as convex combinations of other points in \poly.
Unbounded edges starting from an extreme point are also called \emph{extreme
rays}.

The faces of \poly, partially ordered by set inclusion, form the \emph{face
lattice}. The corresponding graph is sometimes called \emph{skeleton}. The
\emph{$k$-skeleton} is defined by all faces up to dimension $k$.

\subsection{Polyhedral projection}

Given a polyhedron $\Pinp = \{ \vec x \in \R^\di : \Lxga \}$, its orthogonal
projection $\Pfin$ to $\R^\df$ is defined by the pointwise projection
\begin{align*}
    \Pfin &= \Big\{ \pi_\df(\vec x) : \vec x \in \Pinp \Big\}.
\end{align*}
We have $\vec x \in \Pfin \Leftrightarrow \exists \vec y\in\R^e : (\vec x,\vec
y)\in \Pinp$. For this reason, the operation is also called
\emph{quantifier elimination}.


In vertex representation the
orthogonal projection is trivially computed by the pointwise projection of
vertices. This means we could theoretically compute the projection of a convex
polytope by enumerating vertices and subsequently computing the convex hull of
their projection. In practice, however, going from half-space to vertex
representation can yield exponentially many vertices and vice-versa
\cite{mcmullen1970}, which makes this approach not generally applicable.
\subsection{Linear programming}

\label{sec:lp}

\begin{figure}
    \includegraphics{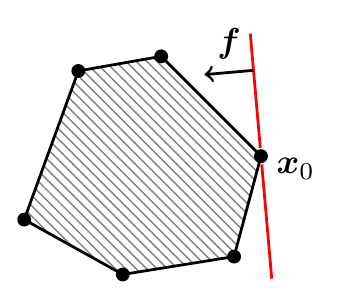}
\parbox{0.4\textwidth}{
 \caption{Linear program with 6 constraints that minimizes $b = \dotp fx$. All points on the red line have the same value $b$ for the objective function. The vertex $\vec x_0$ has the optimal value among all points in the feasible region.}
 \label{fig:lp}
}
\end{figure}

Linear programming (LP) is the problem of optimizing a linear objective
function within the boundaries specified by a set of linear inequalities. The
problem can be formulated in the following standard form
\begin{align}
    \label{eq:lp}
    \begin{split}
    &\text{minimize } \dotp fx \\
    &\text{subject to } \Lxga,
    \end{split}
\end{align}
where $\vec f \in \R^d$ is given and $\vec x \in \R^d$ is
sought. We will often write
\begin{align}
    \begin{split}
    &\minimize{x} \dotp fx\\
    &\st \Lxga
    \end{split}
\end{align}
to indicate that we are interested in the vector $\vec x_0$ that minimizes the
objective.
Noting that the constraints correspond exactly to the half-space
representation of a polyhedron $\poly = \{ \vec x\in\R^d: \Lxga \}$
the problem receives a geometric interpretation. Write $\dotp fx = b$, then
varying $\vec x$ amounts to shifting a hyperplane along its normal vector
$\vec f$. The optimization problem is then to find a point on the boundary of
\poly such that the hyperplane is shifted as far as possible in the direction
of its negative normal vector, see \autoref{fig:lp}.

The problem is called feasible if the polyehdron \poly is non-empty. Depending
on the shape of the polyhedron $\poly$ and optimization direction $\vec f$,
the problem may either be bounded or unbounded. If \poly is bounded, the
optimization problem is always bounded.

Although LP is known to have polynomial
complexity~\cite{khachiyan1980,karmarkar1984}, the most widespread algorithm
in use is \emph{simplex}~\cite{dantzig1990}, which has exponential worst case
scaling but performs quite well in practice~\cite{spielman2001}.

\subsection{Machine-proving constraints}

We frequently encounter the need to check whether a given constraint $\dotp fx
\ge b$ follows from a given set of inequalities $\Lxga$. This can be decided
by the LP \eqref{eq:lp}. In this notation it is a trivial observation that the
constraint $\fb$ is valid if and only if the final objective function $\dotp
fx$ agrees with $\fb$, i.e.~$\dotp fx \ge b$. The same condition applies for
infinite objective values.


We now give a review of known elimination methods.

\section{Fourier-Motzkin elimination (FME)}
\label{sec:fme}

Fourier-Motzkin elimination (FME) is perhaps the most straightforward
approach for solving systems of linear inequalities. The basic procedure is to
algebraically eliminate one variable after the other, as in \autoref{alg:fme}.
It is comparable with
Gaussian elimination known for systems of linear equalities. What is
different from that case is that linear inequalities are not scalable by
negative factors. Hence, after having selected one variable for elimination,
the elimination step is carried out by partitioning the inequalities according
to the sign of the coefficient for the active variable. Those inequalities
with zero coefficient can be added unmodified to the result of the current step.
Then, each pairing between inequalities with positive and negative sign
is scaled and added up to eliminate the selected variable; after
which the resulting inequality is appended to the result of the current step. For a
detailed explanation including examples and additional mentions for
applications see for example~\cite{Williams1986} or~\cite{huynh1990}.

\begin{algorithm}[H]
    \caption{Fourier-Motzkin elimination to compute the projection $\Pfin$ of an arbitrary polyhedron $\Pinp = \{ \vec x \in \R^\di : \Lxga \}$.}
\label{alg:fme}
\begin{algorithmic}[1]
\Function{FME}{\La, $d$, $e$}
    \If {$e = 0$}
        \State \Return \La
    \EndIf
    \State $\alignl{L^{+}}{j} \gets d+e$
    \State $\alignl{L^{+}}{L^{0}} \gets \{ (\conM La)_{i} : L_{ij} = 0 \}$
    \State $\alignl{L^{+}}{L^{\pm}} \gets \{ (\conM La)_{i} : L_{ij} \gtrless 0 \}$
    \State $\alignl{L^{+}}{L'}    \gets \{
        p_j\,{\conV qb} + q_j\,{\conV pa} : {\conV pa} \in L^{+}, \conV qb \in L^{-} \}$
    \State \Return $\ProcName{FME}(L^{0} \cup L',\ d,\ e{-}1)$
\EndFunction
\end{algorithmic}
\algcomment{Basic form of FME without redundancy removal nor other improvements.}
\end{algorithm}

\subsection{Redundancy elimination and improvements}

The main problem which FME suffers from is the matter of redundant
intermediate representations and output: Some of the considered combinations
of positive and negative input inequalities can produce constraints that are redundant with
respect to the resulting system. This leads to intermediate systems being
larger than necessary to fully characterize the projection --- which in turn
can lead to even more redundant constraints in the
following steps. Without strategies for redundancy removal the problem can
quickly grow out of control in terms of both time and memory requirements.
These redundancies can occur independently of the inherent complexity of the
intermediate systems, i.e.~even if the minimally required number of inequalities is
small. Several methods have been suggested to accommodate for redundancies. A
good overview is given by Imbert in~\cite{imbert1993choose}.

The choice of redundancy detection is usually a trade-off between cost and
effectiveness and can make the difference between finishing the computation in
a matter of seconds, hours, years, or not at all in feasible time. A rigorous
elimination of all redundancies can be achieved by using linear programing
(LP) to check one by one all inequalities of the resulting set. This is a
rather expensive operation, but can be well worth the cost in our experience.


There are various additional ways to improve the performance of FME.
Techniques for exploiting sparsity of the underlying linear system have been
suggested by Simon et al.~in~\cite{simon2005sparsity}. This paper also
mentions the matter of elimination order and suggests to use the standard rule
from~\cite{duffin1974} to select at each step the variable which presumably
delays the growth of intermediate systems most effectively: Assuming that for
variable $i$ the number of rows with positive and negative coefficients is
$E_i^+$ and $E_i^-$ respectively, the rule says to eliminate the variable which
minimizes $E_i^+ E_i^- - (E_i^+ + E_i^-)$, i.e.~worst-case size of the next
intermediate system.

In fact, we found this heuristic in combination with a full LP-based redundancy
elimination at every step to be highly effective. It was with this strategy,
that we could finally compute the full marginal characterizations of the \Cn4
and \Cn5 pairwise hidden ancestor models that had previously remained uncracked
in earlier work. For more details, see \autoref{sec:ancestor}.

\subsection{Approximate solutions}
\label{sec:fme-partial}

Despite the improvements mentioned above many problems are intractable using
FME. In this case FME can still be used to compute outer approximations of the
projection polyhedron, if no exact solution is required.  One way to do this
is by dropping inequalities from the input problem. The choice which
inequalities to remove can be guided by other insights, e.g.~always keeping a
certain known set of non-interior points outside the corresponding polyhedron,
but could also be arbitrary. After removing sufficiently many constraints, FME
can be performed efficiently on the smaller system.

There is a related, more sophisticated strategy, described in
\cite{simon2005sparsity}, that keeps the number of constraints below a certain
threshold after each elimination step.

The result of such methods may in general be a rather crude outer
approximation that can be improved upon by performing the procedure multiple
times and combining the results. In any case, the resulting inequalities need
not be facets of the actual solution. This fact can be mitigated by computing
from each output inequality a set of facets such that the inequality becomes
redundant.  We describe a method to perform this calculation in
\autoref{sec:roa}.

\subsection{Complexity}

Consider eliminating a single variable from a system of $N$ inequalities. In
the worst-case half of the inequalities have a positive coefficient for the
elimination variable and the other half has a negative coefficient,
i.e.~$E_i^+=E_i^- = N/2$.  In that situation the first elimination step
results in a new system with $N' = (N/2)^2$ inequalities.  Hence, performing
$e$ successive elimination steps can result in up to $4 (N/4)^{2^e}$
inequalities – which can easily be seen to be true by inserting this
expression in the above recursive formula. Although performing redundancy
removal can mitigate this problem in many practical cases, there are problems
for which the output size –and hence computation time– is inherently doubly
exponential in the problem dimension~\cite{davenport1988,monniaux2010}.

In information theoretic applications, the size of the input system is
exponential in the number $k$ of random variables, i.e.~$\di=(2^k-1)$ and we have
at least the elemental inequalities, which are $N = k + \binom{k}{2} 2^{k-2}
\sim k^2 2^k$ in number. If projecting to a much lower-dimensional subspace,
i.e.~$d \ll e$, then on the order of $e \approx 2^k$ elimination steps are required.
Thus, naively applying
FME the worst-case time and space requirements roughly build up to a
triple-exponential tower,
\begin{align*}
    (k^2 2^k) ^ {2^{2^k}} \sim 2^{2^{2^k}}.
\end{align*}

\section{Extreme Point Method (EPM)}
\label{sec:epm}

The Extreme Point Method (EPM) has been formulated in~\cite{lassez1990qc} and
is presented more accessibly in~\cite{huynh1992}. It is based on a geometric
perspective on the algebraic structure of the problem – viewing the set of
possible \emph{combinations} of the original constraints as a polytope of its own.

\subsection{The base algorithm}

Recall that FME constructs new constraints as non-negative linear combinations
of the original constraints such that the coefficients for the eliminated
variables vanish.  This can be formulated in terms of another problem. Let
$\poly = \{ \vec x \in \R^\di : \Lxga \}$ the original
polyhedron where the constraint matrix has $r$ rows.  The set of non-negative
linear combinations of rows of $L$ that eliminates all variables with index
$i > d$ is the pointed convex cone defined by
\begin{align*}
    \Qf &= \Big\{ \vec q\in \R^r : \vec q \ge \vec 0,\ (\vec q\tran L)_i = 0\ \forall i > d,\ \textstyle{\sum_j} q_j = 1 \Big\}
\end{align*}
It is sufficient to consider normalized $\vec q$ here since constraints can be
scaled using non-negative factors.
Every $\vec q\in \Qi$ corresponds to a face $\dotp fx \ge b$ of the projection
$\Pfin \subset \subs$ with $\vec f = \dvec qL$ and $b =
\dotp qa$. Furthermore, like any polyhedron, $\Qi$ is the convex combination of its
extreme rays. Hence, any face with $\vec f = \dvec qL$ for non-extremal $\vec
q = \sum \lambda_i \vec q_i$ is a non-negative sum $\vec f = \sum \lambda_i
\vec f_i$ of faces $\vec f_i = \dvec{q_i}L$ and therefore necessarily
redundant.  This means that the set of facets of \Pfin can be obtained from
the extreme points of $\Qf$.  More precisely, the facets of \Pfin are in
one-to-one correspondence to extreme points of the \emph{image} $\Qf\tran L$.
In~\cite{lassez1990qc} this problem is called a \emph{generalized linear
program}.

Observe that this perspective transforms the problem to a polytope and allows any
vertex enumeration method to be applied to solve the projection problem. The
transformed domain is bounded, independently of whether the original
polyhedron is bounded or not. This is especially interesting in the light of
some projection methods that do not work well for general unbounded polyhedra
such as the convex hull method to be presented in \ref{sec:chm}.

Note that the map $\vec q \rightarrow \dvec qL$ is in general not
injective and hence many extreme points of $\Qf$ may correspond to the
same face $\dotp fx\ge b$ and need not even be facets of \Pfin.
In other words, there may be many possible combinations of the original
constraints to obtain any particular facet of the projection body. This can
lead to extreme degrees of degeneracy and means that the amount of extreme
points of $\Qf$ may become impractically large to iterate over.

\subsection{Partial solutions}
\label{sec:epm-partial}

If a complete enumeration of all vertices of $\Qf$ is infeasible the
structure of EPM can be exploited to obtain approximate solutions. In fact,
since every vertex or boundary point of $\Qf$ corresponds to a face of the projection
polyhedron \Pfin, any subset of such points corresponds to an outer
approximation of \Pfin. Points on the hull of $\Qf$ can be
directly obtained by solving the LP
\begin{align}
    \label{eq:epm-random}
    \begin{split}
        &\minimize{q} \dotp pq \\
        &\st \vec q \in \Qf
    \end{split}
\end{align}
while imposing arbitrary but fixed vectors $\vec p \in \R^r$.

The objective vectors $\vec p$ can be sampled at random or guided by more
physical insight. One particularly useful approach is to use known points
$\vec x \in \R^\df \subset \R^\di$ on the boundary or exterior of the
projection \Pfin in the output space and letting $\vec p = L \vec x$. Then the
LP \eqref{eq:epm-random} finds a face $\fb = \dvec q\La$ of \Pfin which proves
that $\vec x$ is not in the interior of \Pfin, i.e.~$\dotp fx \le b$.

Whenever the output is
only a regular face and not a facet of \Pfin, a corresponding set of facets can
be derived using the strategy that will be discussed in \autoref{sec:roa}.  In
other words, this method allows to convert an outer approximation specified as
the convex hull of a set of points to an outer approximation in terms of
facets. See \autoref{alg:p2f} for further reference.

While this last property can be particularly useful, we note that in our
experiments, the partial EPM yielded only a small fraction of the facets that
we could discover using other methods.
This can be understood given the expected
degeneracies that was already mentioned in the preceding subsection.
Some facets $\vec f = \dvec qL$ are in
correspondence with \emph{many} vertices $\vec q$ of $\Qf$ and are therefore
very likely to be encountered over and over. For other facets only
one or a few combinations $\vec q\in\Qf$ may exist. Considering a large
possible total number of vertices of $\Qf$ this makes them unlikely
to ever be produced by the randomized EPM as presented here.

\subsection{Complexity}
\label{sec:epm-complexity}

A complete solution using EPM depends on the problem of vertex enumeration
which is known to be hard~\cite{khachiyan2008,bremner1998}. Furthermore, as
briefly mentioned above, the number of extreme points of the combination
polytope $\Qf$
can become impractically large as to prevent basic iteration even if no
additional computation cost for construction and redundancy removal were
needed. In fact, for general $d$-dimensional polyhedra with $v$ vertices
McMullen's Upper Bound Theorem~\cite{mcmullen1970} together with the
Dehn-Sommerville equations~\cite{sommerville1927} gives a tight upper bound on
the number of its facets $f$. By duality
\begin{align*}
    v &\le \textstyle{\binom{f-d-s}{s} + \binom{f-s-1}{d-s-1}} = \BigO(f^s),
\end{align*}
where $s=\floor{d/2}$. However, problems in information theory exhibit more
structure than general polyhedra and hence can be subjected to tighter bounds.
Depending on the exact structure of the problem, one of the bounds
in~\cite{elbassioni2006} is applicable. The best of these bounds that applies
to $\Qf$ arising from unconstrained Shannon cones $\poly[C] = \{\vec x :
L\vec x \ge \vec 0\}$ limits the number of vertices of $\Qf$ to
\begin{align*}
    v &\le 8^d,
\end{align*}
where $d+1 = r$ is the number of rows of $L$. In a system with $n$ random
variables, we have $r = n + \binom n2 2^{n-2}$ elemental inequalities and thus
the number of vertices of $\poly[Q']$ is bounded by an expression that is
doubly exponential in $n$.

\section{Convex Hull Method (CHM)}
\label{sec:chm}

The Convex Hull Method (CHM) uses a geometric approach to perform the subspace
projection without going through the descriptions of any intermediate systems.
The method was shortly mentioned in~\cite{taylor1988} and more thoroughly
treated in~\cite{lassez1990quantifier} and~\cite{huynh1992}. Since we came up
with this algorithm independently and without knowledge of their work, our
specification of the algorithm is slightly different from theirs and is listed
in more mathematical –less imperative– notation. This may be useful as to provide
an alternative reference on the problem.

The algorithm in the form discussed here is applicable when the output is a
\emph{polytope}. Pointed convex cones, such as Shannon cones, can be
considered polytopes by adding limiting constraints in the unbounded direction. Detailed
considerations about the application of this method to convex cones and general unbounded polyhedra can be found in \cite{lassez1990quantifier}.

\subsection{Description of the algorithm}

The algorithm works in two phases. The first phase finds an initial set of
vertices of the projection \Pfin that spans its full subspace. The convex hull
of a subset of vertices is an inner approximation. The goal of the
second phase is to incrementally improve the current inner approximation to
finally arrive at the full facetal description of \Pfin.

\begin{algorithm}[H]
    \caption{Compute the facets of a projection polytope $\Pfin$
    where $\Pinp = \{ \vec x \in \R^\di : \Lxga \}$.}
\label{alg:chm}
\begin{algorithmic}[1]
\Function{chm}{\La, $d$}
    \State $\mathcal V \gets \InitSimplex(\La,\ d)$
    \State \Return $\textsc{expand}(\La,\ d,\ \mathcal V)$
\EndFunction
\end{algorithmic}
\end{algorithm}

The initial step of the CHM algorithm is to find a simplex that serves as a
fully dimensional inner approximation to the projection $\Pfin \subset \R^\df$.
It works by repeatedly solving the LP
\begin{align}
    \label{eq:chm-init}
    \begin{split}
    &\minimize{x} \dotp gx \\
    &\st \Lxga,
    \end{split}
\end{align}
for every basis direction $\vec g \in \subs$ of the projection subspace
in order to find points on the boundary of $\Pfin$. The exact procedure is
described by \autoref{alg:chm-init}. The recommended implementation replaces the LP \eqref{eq:chm-init} with a dedicated function \FindVertex to make sure that every point in the result is a vertex of \Pfin.


\begin{algorithm}[H]
    \caption{Compute a fully dimensional simplex of vertices of the projection polytope $\Pfin$ where $\Pinp = \{ \vec x \in \R^\di : \Lxga \}$.}
\label{alg:chm-init}
\begin{algorithmic}[1]
    \Function{basis\,simplex}{\La, $d$}
    \State $\vec g \gets $ choose $\vec g \in \subs$
    \State $\vec x_0 \gets \FindVertex(\La,\ d,\ \vec g)$
    \State $\mathcal B \gets \{ \vec x_0 \}$
    \State $\mathcal N \gets \emptyset$
    \For {$i$ in $1 \ldots d$\,}
        \State $\vec g \gets $ choose $\vec g \in \mathcal N^\perp \subset \subs$
        \State $\vec x_i \gets \FindVertex(\La,\ d,\ \vec g)$
        \If {$\dvec g(\vec x_i - \vec x_0) = 0$}
            \State $\vec x_i \gets \FindVertex(\La,\ d,\ -\vec g)$
        \EndIf
        \If {$\dvec g(\vec x_i - \vec x_0) = 0$}
            \State $\mathcal N \gets \mathcal N \cup \{\vec g \}$
        \Else
            \State $\mathcal N \gets \mathcal N \cup \{\vec x_i - \vec x_0\}$
            \State $\mathcal B \gets B \cup \{ \vec x_i \}$
        \EndIf
    \EndFor
    \State \Return $\mathcal B$
\EndFunction
\end{algorithmic}
\algcomment{}
\end{algorithm}

\begin{algorithm}[H]
    \caption{Find a vertex of $\Pfin$ that minimizes the objective $\dotp qx$}
\label{alg:findvertex}
\begin{algorithmic}[1]
\Function{find\,vertex}{\La, $d$, $\vec q$}
    \State $\vec q_1 \ldots \vec q_d \gets \text{ ONB of } \subs \text{ with } \vec q_1 = \vec q$
    \For{$i$ in $1 \ldots d$}
        \State $\vec x_i \gets \minimize{x} \dvec q_i\vec x\ $ s.t.\ $\Lxga$
        \State \phantom{$\vec x_i \gets \minimize{x} \dvec q_i\vec x\ $ s.t.}
            $\dvec q_j\vec x = \dvec q_j\vec x_j\, \forall j < i$
    \EndFor
    \State \Return $\pi_{d}(\vec x_d)$
\EndFunction
\end{algorithmic}
    \algcomment{Use this function in place of a simple minimization of $\dotp
    qx$ subject to $\Lxga$, where a true vertex is desired as the result.
    This function is not strictly required for the correctness of CHM (in
    neither of the two phases), but constitutes an important optimization by
    keeping the number of points on which the convex hull operation in the
    second phase is performed down to a minimum.}
\end{algorithm}

As far as dimensionality is concerned, any potential equality constraints are
automatically detected during the initialization phase (null space is
described by the contents $\mathcal N$). It is then sufficient to perform the
second phase of the convex hull operation in the subspace in which \Pfin is
fully dimensional.

This algorithm has a corollary use since it effectively computes the
(affine) subspace in which the projection \Pfin is contained. In particular,
this can be used to compute the rank of a face $\dotp fx \ge b$ by performing
the initialization step after augmenting the constraint list with $\dotp fx =
b$, see \autoref{alg:facerank}.

\begin{algorithm}[H]
    \caption{Determine the rank of a face $\dotp fx\ge b$ of \Pfin. This is a
    corollary use of the initialization step of CHM.}
\label{alg:facerank}
\begin{algorithmic}[1]
\Function{face\,rank}{\La, $d$, $\vec f$}
    \State $\mathcal S \gets \InitSimplex(\La \cup \{ -\fb \},\ d)$
    \State \Return $\abs{\mathcal S} - 1$
\EndFunction
\end{algorithmic}
\end{algorithm}

\subsection{Incremental refinement}

\definecolor{invalidconstraint}{rgb}{0.9,0.1,0.1}
\definecolor{validconstraint}{rgb}{0.1,0.1,0.8}
\begin{figure}[b]
    \includegraphics{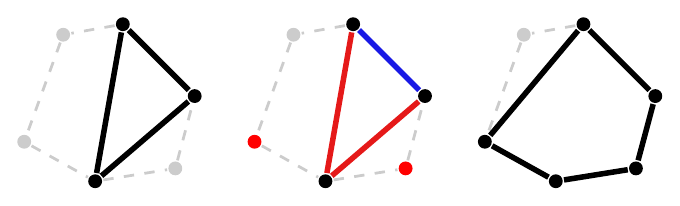}
\parbox{0.4\textwidth}{
    \caption{The facet construction phase of the Convex Hull Method. Start with a $\df$-simplex of vertices and compute their convex hull. Check each element of the hull if it corresponds to a \textcolor{validconstraint}{\textbf{valid}} constraint of the system using a linear program. For each \textcolor{invalidconstraint}{\textbf{invalid}} constraint, the LP outputs an additional point that violates this constraint. Add all vertices and recompute the convex hull. Continue until all faces are valid constraints.}
    \label{fig:chm}
}
\end{figure}

The second phase of CHM iteratively improves the current inner approximation.
This is based on the fact that each facet of the convex hull of the currently
known vertices is either a facet of \Pfin – or is violated by a vertex of
\Pfin. In the second case, the vertex is added to the list of known vertices
and the step can be repeated until no facet of the convex hull is violated.
The process is depicted in \autoref{fig:chm}. A rule of thumb is to add as
many vertices as possible before starting the computation of the next hull.
Known symmetries of the output polytope can be exploited by immediately taking
into account all points in the orbit of a newly discovered point.

\begin{algorithm}[H]
    \caption{Second phase of the CHM algorithm to compute the facets of the
    projection polytope $\Pfin$. Given a set of vertices $\mathcal V$, finds
    missing vertices of $\Pfin$ until all are known.}
\label{alg:chm-phase2}
\begin{algorithmic}[1]
\Function{expand}{\La, $d$, $\mathcal V$}
    \State $\mathcal U \gets \emptyset$
    \State $\mathcal F \gets \textsc{convex\,hull}(\mathcal V)$
    \For {$\fb \in \mathcal F$}
        \State $\vec x \gets \minimize{x} \dotp fx\ $ s.t $\ \Lxga$
        \If {$\dotp fx < b$}
            \State $\mathcal U \gets \mathcal U \cup \{ \FindVertex(\La,\ d,\ \vec f) \}$
        \EndIf
    \EndFor
    \If {$\mathcal U = \emptyset$}
        \State \Return $\mathcal F$
    \EndIf
    \State \Return $\textsc{expand}(\La,\ d,\ \mathcal V \cup \mathcal U)$
\EndFunction
\end{algorithmic}
\end{algorithm}

\subsection{Complexity}

Contrary to FME, the runtime of CHM does not depend on the size of
intermediate projections. Further, while it is hard to cast useful runtime
predictions in terms of the input parameters (dimension, size) of the
problem,\footnote{The runtime \emph{can} be upper-bounded in terms of the input
size, but the runtime can vary dramatically among problems of the same input
size.} it can be specified relatively well in terms of the output size,
i.e.~the dimension and number of vertices of the output polytope. This
characterizes CHM as \emph{output sensitive} and makes it a promising algorithm
for many projection problems that appear intractable with FME. In practice CHM only
performs well if the dimension of the output space is sufficiently low and the
number of vertices is small.

To make this more precise, assume that \Pfin is a $\df$
dimensional polytope with $v$ vertices and $f$ facets. For $\df>3$ the
computation of the convex hull then takes time $\BigO(v^{\floor{\df/2}})$.
Let's imagine we have access to an online convex hull algorithm that spits out
exactly one new facet of the convex hull $\operatorname{CH}(\mathcal V)$ each time that we ask it to. Every one of these
facet candidates needs to be checked using a single LP instance. It can either
be a valid facet of \Pfin – or the result of the
LP is a \emph{new} point on the boundary, which can be converted to a vertex.
This means that, in addition to the convex hull operation itself, only $(v+f)$
LPs need to be solved before we arrive at the final solution. The total work
required is thus~\cite{chazelle1993convexhull}
\begin{align*}
    \BigO(v^{\floor{\df/2}}) + (v+f) \cdot \mathrm{LP}.
\end{align*}
Perceivably, the bottle neck of this algorithm is the computation of the
convex hull in the output space.

The actual implementation used to obtain our results is based
on a non-incremental convex hull solver – i.e.~the hull is computed multiple
times without incorporating the knowledge of previous computations. At every
step each facet of the current inner approximation is checked using an LP. An
invalid constraint will lead to the discovery of a new vertex. Therefore, at
least one new vertex is added at each step as long as the description is
incomplete. If only a single vertex is added at each step, the total work
spent on convex hulls is:
\begin{align*}
    \sum_{k=0}^{v} \BigO(k^{\floor{\df/2}})
        &\lesssim \BigO(v \cdot v^{\floor{\df/2}}) \\
        &= \BigO(v^{\floor{\df/2}+1}).
\end{align*}
It can easily be seen that this worst-case corresponds to stopping when the
first invalid facet candidate vector is encountered at each step. In this
case, the total number of LP instances is $(v+f)$, again. If the strategy is
changed to allow testing more than one invalid facet candidate, the number
of LP instances can be limited to be any desired value above $(v+f)$.
Typically, testing all new candidates is the best strategy as it maximizes
the probability to find as many new vertices as possible in each iteration
and therefore minimizes the number of iterations, i.e.~convex hull operations.



\section{Equality Set Projection (ESP)}
\label{sec:esp}

In a 2004 paper~\cite{jones04equalityset}, Jones et al.~described an algorithm
called Equality Set Projection (ESP). Similar to CHM, this output-sensitive
method computes the projection of polytopes directly in the output space
without going through intermediate representations. It is based on the
underlying principle that the faces of a polytope form a connected graph by
the subset relation. Two $n$-dimensional faces are called adjacent if they
share an $(n{-}1)$-dimensional subface. The principle of ESP is to first find
an arbitrary facet, and then compute its adjacencies. This computation can be
understood as a rotation of the facet around the ridge, as displayed in
\autoref{fig:afi}.

Mathematically, ESP relies on the insight that every face of a polytype can be
uniquely identified with a so-called \emph{equality set}, which is the set of
all input inequalities that are satisfied with equality on all points of the
given face. The ESP rotation operation to find adjacent facets and the
discovery of an initial facet can be performed using linear algebra on
submatrices corresponding to the rows in the equality set and require the
solution of only very few linear programs.

An in-depth description of the mathematics required to implement ESP is beyond
the scope of this paper. The interested reader is well advised to read the
article~\cite{jones04equalityset}. Instead, we will present in the next
section a related method that we came up with before we knew about the
existence of ESP. It has the same geometric interpretation and is easier to
implement, but offers less potential optimizations. One interesting property
of ESP compared to our method is that it can skip the recursion to compute
lower dimensional faces of the projection polytope in the non-degenerate case.

Another noteworthy side-effect of the identification with equality sets is
that faces of the projection polytope can be labeled with a tuple of numbers.
This allows a constant time lookup operation for already computed faces (using
a hash-table). This is important for avoiding recomputation of subfaces with
multiple parents. Our AFI implementation in contrast depends on a linear
number of matrix multiplications to achieve the same.

\section{A new method: Adjacent facet iteration}
\label{sec:afi}

We now present our own method, which is very similar to Equality Set
Projection, but uses a different set of primitives that appears easier to
implement. Contrary to ESP, we always require a recursive solution in the
lower dimensional subspace.

\subsection{The base algorithm}

\begin{figure}[b]
    \includegraphics{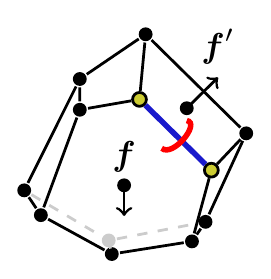}
\parbox{0.4\textwidth}{
    \caption{\emph{Adjacent facet iteration}: First compute the subfacets of a given facet $\vec f$ and then use every subfacet to obtain an adjacent facet $\vec f'$.}
 \label{fig:afi}
}
\end{figure}

The core idea behind AFI is to traverse the face graph of the output polytope
by moving along adjacencies (see Figure \ref{fig:afi}). More precisely, two facets of a $\df$-dimensional
polytope are said to be adjacent if their intersection is a
$(\df{-}2)$-dimensional face (ridge). With this notion of adjacency the facets
of a polytope form a connected graph by duality of Balinski's
theorem~\cite{balinski1961}.  Knowing a facet and one of its ridges the
adjacent facet can be obtained using an LP-based rotation operation. Hence,
presuming the knowledge of an arbitrary initial facet of a $\df$-dimensional
polytope all further facets can be iteratively determined by computing the
ridges of every encountered facet, i.e.~solving the projection problem for
several $(\df{-}1)$-dimensional polytopes. The lower dimensional projection
algorithm can be chosen at will. For example, all AFI-based computations
mentioned in this article were carried out by a $k$-level AFI recursion on top
of CHM, i.e.
\begin{align}
    \begin{split}
        \AFI{k} &:= \textsc{afi}\left[\Pi=\AFI{k-1}\right], \\
        \AFI{0} &:= \textsc{chm}.
    \end{split}
\end{align}
Choosing $k$ is not only a matter of performance but also allows to control
the amount of information that is recovered about the polytope. For example,
\AFI{1} lists all vertices, ridges and facets; \AFI{d} outputs the entire face
skeleton.

\begin{algorithm}[H]
    \caption{Compute the facets of \Pfin using AFI.}
\label{alg:afi}
\begin{algorithmic}[1]
    \Require $\Pi$ computes $\pi_{d-1}(\poly')$
    \Function{afi$\left[\Pi\right]$}{\La, $d$}
    \State $\mathcal R \gets \emptyset$
    \State $\mathcal Q \gets \{ \textsc{getfacet}(\La,\ d) \}$
        \Comment{uninspected facets}\label{alg:afi:initQ}
    \While{$\mathcal Q \ne \emptyset$}
        \State $\fb \gets $ pop from $\mathcal Q$       \label{alg:afi:selectF}
        \If{$\fb \not\in \mathcal R$}
            \State $\mathcal R \gets \mathcal R \cup \{ \fb \}$ \label{alg:afi:updateR}
            \For{$\vec s_c \in \Pi(\La \cup \{-\fb\},\ d)$}
                \State $\vec f'_{b'}\ \vec s'_{c'} \gets \textsc{rotate}(\La,\ {-}\fb,\ \vec s_c)$
                \State $\mathcal Q \gets \mathcal Q \cup \{ \vec f'_{b'} \}$
            \EndFor
        \EndIf
    \EndWhile
    \State \Return $\mathcal R$
\EndFunction
\end{algorithmic}
\algcomment{The algorithm as listed here is kept simple for clarity and has no
protection from calculating the projection of the same face multiple times. It
is advisable to add a cache to AFI implementations that prevents this from
happening.}
\end{algorithm}

\paragraph{Initial facet}

\begin{figure}[b]
    \includegraphics{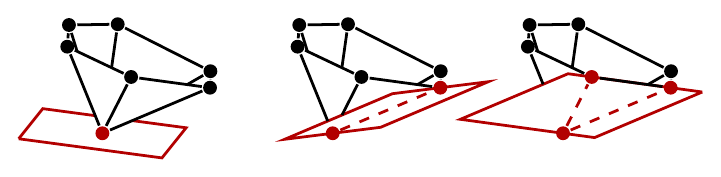}
\parbox{0.4\textwidth}{
    \caption{Obtain an initial facet by successively fitting an arbitrary face onto the polyhedron -- at each step increasing the face rank by one.\label{fig:initfacet}}
}
\end{figure}

AFI requires a facet of the output polyhedron as initializer. In cases where
no facet is known a priori, one can be obtained by successively increasing the
rank of an arbitrary face. Geometrically, this can be pictured as rotating the
face onto the polyhedron as shown in \autoref{fig:initfacet}. The algorithm is
described in \autoref{alg:getfacet}.

The outcome of this procedure depends on a several arbitrary choices and can
for appropriate values have \emph{any} facet as a result. Thus, it is
suggestive to employ this method repeatedly with the expectation that after
enough iterations every facet will eventually be discovered. In practice, it
turns out that this strategy typically recovers only a small number of facets
even after long searches – similar to the randomized EPM discussed in
\autoref{sec:epm-partial}.


\begin{algorithm}[H]
    \caption{Compute an (arbitrary) facet of \Pfin.}
    \label{alg:getfacet}
\begin{algorithmic}[1]
\Function{getfacet}{\La, $d$}
    \State $\vec p \gets $ choose $\vec p \in \R^m$
    \State $\vec q \gets \minimize{q} \dotp pq\ $ s.t.\ $
        \vec q \ge \vec 0,
        \ \textstyle{\sum_j} q_j = 1$,
    \State \phantom{$\vec q \gets \minimize{q} \dotp pq\ $ s.t.}
        $(\dvec q L)_i = 0\ \forall i > d$
    \State \Return $\ToFacet(\La,\ d,\ \dvec q\La)$
\EndFunction
\end{algorithmic}
    \algcomment{For production use you should additionally take care to handle
    the case $\dvec qL = 0$.}
\end{algorithm}

\begin{algorithm}[H]
    \caption{Turn a valid constraint on \Pfin into a facet.}
    \label{alg:tofacet}
    \begin{algorithmic}[1]
\Function{tofacet}{\La, $d$, $\vec f_b$}
    \State $\mathcal P \gets \InitSimplex(\La,\ d)$
    \State $\mathcal F \gets \InitSimplex(\La \cup \{ -\fb \},\ d)$
    \If {${\abs{\mathcal F}} = \abs{\mathcal P} - 1$}
        \State \Return $\vec f_b$
    \EndIf
        \State $\vec s \gets$ choose $\vec s \in
            \spanof{\mathcal P -\mathcal P_0} \cap
            \spanof{\mathcal F - \mathcal F_0, \vec f}^\perp,\ \vec s\ne 0$
    \State $\vec x \gets \minimize{x} \dotp sx\ $ s.t.\ \Lxga
    \State $c \gets \dvec s{\mathcal F_0}$
    \If {$\dotp sx\ge c$}
        \State $\vec s_c \gets -\vec s_c$
    \EndIf
    \State $\vec f_b,\ \vec s_c \gets \Rotate(\La,\ {-}\vec f_b,\ {-}\vec s_c)$
    \State \Return $\ToFacet(\La,\ d,\ \vec f_b)$
\EndFunction
    \end{algorithmic}
\end{algorithm}


\paragraph{Computing adjacencies}

\begin{figure}[b!]
    \includegraphics{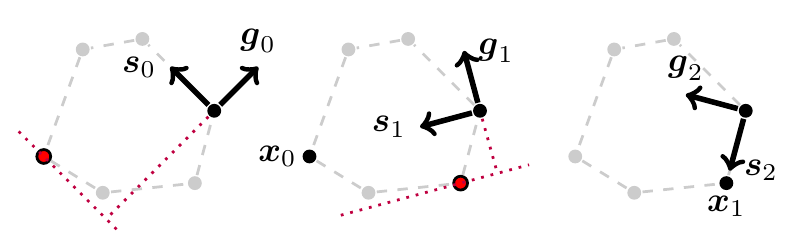}
\parbox{0.4\textwidth}{
    \caption{Computation of the adjacent facet $\vec f'$ from the facet $\vec f$ and a subface $\vec s$. Start with $\vec g_0 {=} {-}\vec f$, $\vec s_0 {=} \vec s$. Iteratively obtain $\vec x_i$ from the minimization of $\dotp {g_i}x$ and then proceed by constructing new $\vec s_{i+1} \perp \vec g_{i+1}$ such that $\vec g_{i+1}$ is orthogonal to both $\vec x_i$ and \poly[K].\label{fig:rotate}}
}
\end{figure}

At the heart of AFI is an LP-based rotation operation that computes the
adjacent facet $\vec f'_{b'}$ given an initial facet $\vec f_b$ and one of its
subfacets $\vec s_c$. The procedure is based on the fact that any
$(\df{-}2)$-face of an $\df$-dimensional convex polyhedron \poly is the
intersection of precisely two adjacent facets of
\poly~\cite{gruenbaum1967convex}. The normal vector $\vec f'$ must be
orthogonal to the $(\df{-}2)$-dimensional (affine) subspace in which the
subface denoted by $\vec s_c$ is fully dimensional. This means $\vec f'$ must
lie in the 2D plane spanned by the vectors $\vec s$ and $\vec f$. Hence, the
problem is effectively to find the correct rotation angle of $\vec f'$.  Our
strategy is to successively obtain candidates for $\vec f'_{b'}$ that increase
the rotation angle in only one sense and converge toward the real value. This
is achieved by searching for $\vec x_i$ that violates the current candidate
and then constructing a new candidate that contains $\vec x_i$.
\autoref{fig:rotate} shows this process in the $\spanof{\vec f, \vec s}$
plane. The method is like a partial CHM in 2D where we only care about the
outermost face instead of computing the whole convex hull.
With the precise specification of rotation operation as listed in
\autoref{alg:rotate} the adjacent facet vector is obtained as $\vec f'_{b'} =
\Rotate(\La,\ {-}\fb,\ \vec s_c)$.
In a sense \Rotate{} represents a very local operation in the output space.
This is indicated by the fact that it depends only on the input constraints as
well as the face and subface vectors, but e.g.~not on the global output space
$\R^d$ directly.

\begin{algorithm}[H]
    \caption{Rotate a face of \Pfin around one of its subfaces.}
\label{alg:rotate}
\begin{algorithmic}[1]
\Require candidates $\vec g_b$ for the rotated face and a subface $\vec s_c$
    where $\vec s\perp\vec g$.
\Function{rotate}{\La, $\vec g_b$, $\vec s_c$}
    \State $\vec x \gets \minimize{x} \dotp gx\ $ s.t.\ \Lxga
    \If {$\dotp gx \ge b$}\label{alg:rotate:break}
        \State \Return $\vec g_b$, $\vec s_c$
    \EndIf

    \State $\vec g_b \gets \vec g_b/\norm{\vec g}$,  \algrhs{$\gamma \gets \dotp gx-b$}
    \State $\vec s_c \gets \vec s_b/\norm{\vec s}$,  \algrhs{$\sigma \gets \dotp sx-c$}

    \State \Return $ \Rotate(\La,
        \ \sigma\vec g_b {-} \gamma\vec s_c,
        \ \gamma\vec g_b {+} \sigma\vec s_c)$
\EndFunction
\end{algorithmic}
\algcomment{Given an \emph{invalid} constraint $\vec g_b$ the result is a face
    that touches the polytope in the $\spanof{\vec g,\vec s}$ plane. This
    property is exploited by the \ToFacet{} routine to successively increase
    the face rank.}
\end{algorithm}

The update rules for $\vec g_b$ and $\vec s_c$ in \Rotate{} can be derived
using the following simple argument. Assume normalized constraints $\vec g_b$
and $\vec s_c$ with $\vec g\perp\vec s$. At each step, we search for $\vec g'$
in the 2D plane spanned by $\vec g$ and $\vec s$, i.e.\ $\vec g' = \alpha \vec
g + \beta \vec s$. Let now $\vec y = b\vec g + c\vec s$ at the intersection of
$\vec g_b$ and $\vec s_c$.  Then, since $\vec g'_{b'}$ should contain $\vec x$
and $\vec y$, we must have for $\vec z = \vec x-\vec y$
\begin{align*}
    0 &= \dotp z{g'} = \alpha \dotp zg + \beta \dotp zs
    = \alpha\gamma + \beta\sigma,
\end{align*}
with $\gamma = \dotp gx-b$ and $\sigma = \dotp sx-c$.
To enforce a rotation \emph{toward} $\vec s$, demand
\begin{align*}
    0 &\stackrel{!}{<} \dotp s{g'}
      = \alpha \underbrace{\dotp sg}_{=0} + \beta \underbrace{\dotp
      ss\vphantom{\vec g}}_{=1}
      = \beta.
\end{align*}
Furthermore, we know $\gamma = \dotp gx-b < 0$ due to the break condition
in line~\ref{alg:rotate:break} of \autoref{alg:rotate}. Therefore, the
canonical choice for $\vec g'$ is
\begin{align*}
    \vec g' &= \sigma\vec g - \gamma\vec s.
\end{align*}
The invariant $\vec s' \perp \vec g'$ is retained by using the projection of
$\vec z$ as the new $\vec s'$:
\begin{align*}
    \vec s' &= \pi_{\spanof{\vec s, \vec g}}(\vec z)
            = \left({\vec s\dvec s} + {\vec g\dvec g}\right) \vec z
            = \sigma \vec s + \gamma \vec g.
\end{align*}
This also ensures a monotonous rotation sense.
The inhomogeneities are obtained using corresponding linear combinations
\begin{align*}
    b' &= \dotp yg' = \sigma b - \gamma c, \\
    c' &= \dotp ys' = \gamma b + \sigma c.
\end{align*}

\subsection{Exploiting symmetries}

If the symmetry group $G$ of the output polytope is known it is sufficient to
compute the adjacencies of only one representative $\vec f$ of every orbit
$G\vec f$ since
\begin{align*}
    \vec f'(g\vec f, g\vec s) = g\vec f'(\vec f, \vec s)
\end{align*}
for every $g\in G$.
This can be implemented by changing line~\ref{alg:afi:updateR} in \autoref{alg:afi} to
\begin{align*}
    \mathcal R \leftarrow \mathcal R \cup G\vec f.
\end{align*}
This modification can speed up AFI by the average orbit size $\abs{G\vec f}$.

\subsection{Randomized facet discovery (RFD)}
\label{sec:afi-randomized}

When increasing the output dimension of the projection problem an AFI-based
complete computation quickly becomes infeasible. However, a randomized variant
of AFI can be highly effective for computing partial descriptions. The
randomized facet discovery $\RFD{k,n}$ recursively computes partial
projections. It is defined similar to $\AFI{k}$:
\begin{align*}
    \RFD{k,n}(\ldots) &:= \textsc{afi}(\ldots, \RFD{k-1,n}),
\intertext{with the difference that CHM is carried out only $n$ times in total, i.e.}
    \RFD{0,n}(\ldots) &:= \textsc{chm}(\ldots) \quad\text{and decrease}\quad n \rightarrow n-1,\\
    \RFD{0,0}(\ldots) &:= \emptyset.
\end{align*}
Alternatively, carry out CHM $n$ times within \emph{every} \RFD{1,n} call.


To improve the exhaustiveness of the RFD output, the routine can be invoked
multiple times while preserving the knowledge about recovered polyhedral
substructure. This means populating the queue $\mathcal Q$ on
line~\ref{alg:afi:initQ} of \autoref{alg:afi} with the known facets of the
current polytope and selecting in line~\ref{alg:afi:selectF} the facet with
the least known substructure.

The effectiveness of RFD is based on the observation that AFI usually recovers
the full projection after only a few steps but takes longer to finish in order
to make sure that no further facets are missing. In other words, to obtain the
full solution it is sufficient to compute the adjacencies of a small subset of
the facets. This can be understood as the result of a high connectivity among
the facets: every facet is adjacent to at least $d$ other facets.


\subsection{Refining outer approximations (ROA)}
\label{sec:roa}

We have seen that randomized variants of FME and EPM can provide outer
approximations that can in general contain non-facetal elements, see
\autoref{sec:fme-partial} and \autoref{sec:epm-partial}, respectively. With
the \ToFacet{} routine (\autoref{alg:tofacet}) any given face can be
transformed into a facet which contains the face as a subset. However, the
result –being a single constraint– can not imply the input
constraint (unless the input was a facet and is therefore returned
unchanged) and even after mapping all faces of the outer approximation
individually to facets this method can provide no guarantee that the result is
a strictly tighter approximation. To remedy this issue we can construct a
modified procedure $\ToFacets$ that returns a \emph{set} of facets that
provide a sufficient replacement for the input constraint,
see~\autoref{alg:tofacets}.

This procedure has another remarkable use case: the \PtToFacet routine
returns facets which document that a given point is not in the interior of
$\Pfin$. This allows to turn an outer approximation in vertex representation
into a (generally non-equivalent) outer approximation in half-space
representation, see \autoref{alg:p2f}, facilitated using the LP
\eqref{eq:epm-random} from \autoref{sec:epm-partial}.

\begin{algorithm}[H]
    \caption{Convert an arbitrary face of \Pfin to a set of implying facets.}
\label{alg:tofacets}
\begin{algorithmic}[1]
\Require $R_{\vec c}$ set of known facets (can be initially empty)
\Function{tofacets}{$\La$, $d$, $\fb$, $R_{\vec c}$}
    \State $\vec x \gets \minimize{x} \dotp fx\ $ s.t.\ $R\vec x\ge \vec c$
    \If {$\dotp fx \ge b$}
        \State \Return $R_{\vec c}$
    \EndIf
    \State $\vec g_c \gets \ToFacet_2(\La,\ d,\ \fb,\ \vec x)$
    \State \Return $ \ToFacets(\La,\ d,\ \fb,\ R_{\vec c} \cup \{ \vec g_c \})$
\EndFunction
\end{algorithmic}
\end{algorithm}

\begin{algorithm}[H]
    \caption{Convert a valid constraint on \Pfin into a facet using a
    non-interior control point.}
    \label{alg:tofacet2}
\begin{algorithmic}[1]
\Function{tofacet$_2$}{$L$, $d$, $\fb$, $\vec y$}

    \State $\mathcal P \gets \InitSimplex(\La,\ d)$
    \State $\mathcal F \gets \InitSimplex(\La \cup \{ -\fb \},\ d)$
    \If {${\abs{\mathcal F}} = \abs{\mathcal P} - 1$}
        \State \Return $\vec f_b$
    \EndIf
        \State $\vec s \gets$ choose $\vec s \in
            \spanof{\mathcal P -\mathcal P_0} \cap
            \spanof{\mathcal F - \mathcal F_0, \vec f}^\perp,\ \vec s\ne 0$
    \State $\vec x \gets \minimize{x} \dotp sx\ $ s.t.\ \Lxga
    \State $c \gets \dvec s{\mathcal F_0}$
    \If {$\dotp sx\ge c$}
        \State $\vec s_c \gets -\vec s_c$
    \EndIf
    \State $\vec f_b,\ \vec s_c \gets \Rotate(\La,\ -\vec f_b,\ -\vec s_c)$

    \If {$\dotp gy > c$}\label{alg:tofacet2:diffstart}
        \State $\vec f_b,\ \vec s_c \gets \Rotate(\La,\ -\vec f_b,\ \vec s_c)$
    \EndIf
    \State \Return $\ToFacet_2(\La,\ d,\ \vec f_b,\ \vec y)$\label{alg:tofacet2:diffstop}
\EndFunction
\end{algorithmic}
\algcomment{The resulting facet $\vec f_c$ proves that $\vec y$ is not an
    interior point of the projection polytope, i.e.~$\dotp fy \le c$. The
    only difference from the regular \ToFacet{} procedure are the
    lines~\mbox{\ref{alg:tofacet2:diffstart}-\ref{alg:tofacet2:diffstop}}.}
\end{algorithm}

\begin{algorithm}[H]
    \caption{Turn an non-interior point into a set of facet, each of which
    proves that the point is non-interior.}
\label{alg:p2f}
\begin{algorithmic}[1]
\Function{\PtToFacet}{$\La$, $d$, $\vec y$}
    \State $\vec q \gets \minimize{q} \dvec qL\vec y $\ s.t.\ $
        \vec q \ge \vec 0,
        \ \textstyle{\sum_j} q_j = 1,$
    \State \phantom{$\vec q \gets \minimize{q} \dvec qL\vec y $\ s.t.}
        $(\dvec q L)_i = 0\ \forall i > d$
    \State \Return $\ToFacets(\La,\ d,\ \dvec q\La,\ \emptyset)$
\EndFunction
\end{algorithmic}
\algcomment{The routine can alternatively be implemented using directly the
$\ToFacet_2$ primitive.}
\end{algorithm}

\subsection{Relation to other algorithms}

The fully recursive \AFI{d} method uses the same geometric traversal strategy
that is also used for the vertex enumeration method described by McRae and
Davidson~\cite{mcrae1973algorithm}. By duality, their method can be viewed as
a convex hull algorithm. However, the algorithm differs from AFI in that it
was not conceived as a projection algorithm and is therefore based on the
assumption that the full half-space representation (or vertex representation
in the case of convex hull) of the investigated polytope is already available
to begin with. Hence, their rotation primitive is based on an algebraic
approach which is not available for the projection problem.

In general, one class of convex hull algorithms is based on a pivoting
operation that can be understood as a rotation around a ridge just as the one
used in AFI, see e.g.~\cite{chand1970,avis1997}. In this sense, AFI can be
comprehended as an \emph{online} variant of convex hull that operates without
knowing the full list of vertices in advance, but rather determines them as
needed using LPs. Note that this represents an inversion of control when
compared to online convex hull algorithms in the usual sense. These algorithms
are operated in push-mode, i.e.~vertices are fed to the algorithm from an
outside operator (e.g.~CHM) when they become available and the algorithm is
expected to update its half-space representation after each transaction. In
contrast, AFI operates in pull-mode, i.e.~fetches vertices with specified
properties as they are needed to compute a new facet.

\section{Comparison of algorithms}
\label{sec:algsummary}

This section contains a short summary of the discussed algorithms and their
strengths and weaknesses.

\subsection{Fourier-Motzkin elimination}

FME is an algebraic variable elimination by appropriately combining input
inequalities. It works arbitrary polyhedra without further preparations,
and is easy to understand and implement. FME can outperform other methods
dramatically, in particular when the number of input inequalities is small and
only a small number of variables have to be eliminated from the input problem.
Despite its untracktable worst-case scaling, FME with redundancy elimination
can therefore be a valueable tool.  For many practical problems, however, FME
falls victim to the combinatorial explosion. In such cases, consider applying
a randomized FME as a first step to obtain outer approximations which can then
be improved upon with other methods.

\subsection{Extreme Point Method}

This method works by searching for combinations of input inequalities that sum
up to eliminate all but the output coordinates. While this is an insightful
take on the problem, we found it impractical for our use-cases due to the
degeneracy of the combination polytope with respect to the projection
polyhedron. One interesting aspect of the algorithm is its use to search for
random faces of the projection.

\subsection{Convex Hull Method}

CHM is a geometric method to compute the projection of \emph{polytopes}. It
works directly in the output space without going through intermediate systems
like FME. Also contrary to FME, is output-sensitive and particularly efficient
when the dimension of the output space is low. While performing CHM, you also
acquire all vertices of the projection polytope.

\subsection{Equality Set Projection}

Like CHM, the ESP algorithm is an output-sensitive method suited to compute
the projection of \emph{polytopes}. In non-degenerate cases, ESP can be
efficiently performed without having to solve the projection problem
recursively for lower-dimensional faces. For such cases, ESP can be used even
if there is a large number of vertices that would make CHM unpractical.

\subsection{Adjacent Facet Iteration}

AFI is a geometric approach that walks along the face lattice similar to ESP.
Compared to ESP, it takes a cruder approach to compute adjacencies, relying
entirely on LP based primitives. By our measure, this also makes the method
easier to understand and implement. However, it also falls short on potential
optimizations. It always requires recursion and can output the entire face
skeleton of a polytope. Advantages of the AFI primitives are their corollary
uses, e.g.~to compute facets from known faces or non-interior points of the
projection polytope, as well as the potential for application in randomized
contexts.

\begin{widetext}
\part{\hfill Quantum Nonlocality \hfill}
\end{widetext}

\section{Bell inequalities and marginal problems}
\label{sec:Bell_marginal}
We start describing the simplest non-trivial marginal problem. Consider you have three dichotomic random variables $X_1,X_2,X_3$ with $X_i\in \{\pm 1\}$ but can only sample two of them at a given time. Further, suppose that you observe perfect correlations between $X_1$ and $X_2$ and between $X_1$ and $X_3$, that is, $\mean{X_1X_2}=\mean{X_1X_3}=1$. Intuitively, since correlations are transitive, we would expect that in such a case the variables $X_2$ and $X_3$ should also be perfectly correlated, that is $\mean{X_2X_3}=1$. However, suppose we observe perfect anti-correlations $\mean{X_2X_3}=-1$. What does our intuition tell about such correlations?

Underlying the intuitive description of this simple experiment is the idea that there is a well defined (normalized and positive) joint probability distribution $p(x_1,x_2,x_3)$ --- even if empirically we can only sample two variables at a time, that is, we only have access to $p(x_i,x_j)$. As it turns out, the existence of a joint distribution $p(x_1,x_2,x_3)$ implies strict constraints on the possible marginal distributions that can be obtained from it. In particular, it follows that
\begin{equation}
\mean{X_1X_2}+\mean{X_1X_3} \leq 1 +\mean{X_2X_3},
\end{equation}
an inequality that is violated if
\begin{equation}
\label{eq.cac}
\mean{X_1X_2}=\mean{X_1X_3}=-\mean{X_2X_3}=1,
\end{equation}
thus showing that such correlations cannot arise from an underlying joint distribution among the three variables.

An alternative description for the existence of a joint distribution among all the variables is given by
\begin{equation}
p(x_i,x_j)=\sum_{\lambda} p(\lambda) \, p(x_i \Mid \lambda) \, p(x_j \Mid \lambda),
\end{equation}
that is, there is an underlying process, described by the unobserved (hidden) variable $\lambda$ that specifies the values of all variables $x_i$ independently of which variables we decide to sample at a given round of the experiment. What this shows is that correlations \eqref{eq.cac} cannot arise from such a process and can only be generate if $\lambda$ is correlated with our choices of which variables to sample at given run.

\emph{Bell's theorem} concerns a scenario very similar to this one. A Bell experiment involves two distant (ideally space-like separated) parties, Alice and Bob, that upon receiving physical systems can measure them using different observables. We denote the measurement choices and measurement outcomes of Alice and Bob by random variables $X$ and $Y$ and $A$ and $B$, respectively. A classical description of such experiment, imposes that the observable distribution $p(a,b\Mid x,y) $ can be decomposed as
\begin{eqnarray}
\label{eq.LHV}
\nonumber
p(a,b\Mid x,y)= & & \sum_{\lambda}p(a,b,\lambda \Mid x,y) \\ \nonumber
= & & \sum_{\lambda} p(a,b \Mid x,y,\lambda)\,p(\lambda \Mid x,y) \\
= & & \sum_{\lambda} p(a \Mid x,\lambda)\,p(b \Mid y,\lambda)\,p(\lambda),
\end{eqnarray}
where the variable $\lambda$ represents the source of particles and any other local mechanisms in Alice and Bob laboratories that might affect their measurement outcomes. In \eqref{eq.LHV} we have imposed the conditions $p(\lambda \Mid x,y)=p(\lambda)$, $p(a \Mid x,y,b,\lambda)= p(a \Mid x,\lambda)$ and $p(b \Mid x,y,a,\lambda)=p(b \Mid y,\lambda)$. The first condition refers to measurement independence (or ``free-will'') stating that the choice of which property of the physical system the observers decide to measure is made independently of how the system has been prepared. The second conditions related to local causality, that is, the probability of the measurement outcome $A$ is fully determined by events in its causal past, in that case the variables $X$ and $\lambda$ (and similarly to the measurement outcome $B$).

This classical description in terms of local hidden variable (LHV) model is equivalent to the existence of a joint distribution of measurement outcomes all possible measurements \cite{Fine1982}, that is, $p(a_1,\dots,a_{\abs x},b_1,\dots,b_{\abs y})$ where $\abs x$ and $\abs y$ stand for the total number of different observables to be measured by Alice and Bob, respectively. Since in a typical quantum mechanical experiment, the set of observables to be measured by Alice (or Bob) will not commute, at a given run of the experiment only one observable can be measured. That is, similarly to what we had in three variables example, the empirically accessible information is contained in the probability distribution $p(a_x,b_y)=p(a,b\Mid x,y)$.

In the simplest possible Bell scenario, each observer performs two possible dichotomic measurements $A,B\in\{0,1\}$. The LHV model \eqref{eq.LHV} is thus equivalent to the existence of the joint distribution $p(a_1,a_2,b_1,b_2)$, implying the famous Clauser-Horne-Shimony-Holz (CHSH) inequality \cite{chsh1969}
\begin{equation}
\label{eq.CHSH}
\mean{A_1B_1}+\mean{A_1B_2}+\mean{A_2B_1}-\mean{A_2B_2} \leq 2,
\end{equation}
an inequality that can be violated up to $2\sqrt{2}$ by appropriate local measurement on a maximally entangled state. This violation shows that quantum mechanics is incompatible with a LHV description, the phenomenon known as quantum nonlocality.

Inequalities like \eqref{eq.CHSH}, generally known as Bell inequalities, play a fundamental role in the study of nonlocality and its practical applications, since it is via their violation that we can witness the non-classical behaviour of experimental data. Given a generic Bell scenario, specified by the number of parties, measurement choices and measurement outcomes, there are two equivalent and systematic ways of deriving all the Bell inequalities characterizing it.

First, the LHV decomposition \eqref{eq.LHV} defines a convex set, more specifically a polytope that can be characterized by finitely many extremal points or equivalently in terms of finitely many linear inequalities, the non-trivial of which are exactly the Bell inequalities \cite{Brunner2014}. The extremal points of the Bell polytope can be easily listed, since they are nothing else than all the deterministic strategies assigning outputs to a given set of inputs. For instance, in the CHSH scenario there are 4 different functions $a=f_a(x)$ that take a binary input $x$ and compute a binary output $a$. Similarly, there are 4 functions $f_b(y)$. Thus, in the CHSH scenario we have a total of 16 extremal points defining the region of probabilities $p(a_x,b_y)$ compatible with \eqref{eq.LHV}. Given the description of the Bell polytope in terms of its extremal points, there are standard convex optimization algorithms to find its dual description in terms of linear (Bell) inequalities.

The second approach, equivalent to the first one but more clearly related to the marginal problem is the following. A well defined joint distribution $p(a_1,a_2,b_1,b_2)$ is characterized by two types of linear constraints: i) the normalization $\sum_{a_1,a_2,b_1,b_2}p(a_1,a_2,b_1,b_2)=1$ and ii) the positivity $p(a_1,a_2,b_1,b_2)\geq 0$. These constraints define a polytope, more precisely a unit simplex. However, since $p(a_1,a_2,b_1,b_2)$ is not directly observable, we are interested in the projection of this simplex to the subspace corresponding to the observable coordinates $p(a_x,b_y)$. That is, the problem at hand is equivalent to the projection of a convex polytope to a subspace of it, which can be achieved via quantifier elimination algorithms.

So far we have restricted our attention to Bell scenarios where the correlations between all parties is assumed to be mediated via a single common variable $\lambda$. There are, however, several scenarios of interest where the correlations can be mediated by several, typically independent, sources of states. This introduces further structure to our description and enormously complicates the problem. A typical example of such a scenario in quantum information is the so-called entanglement swapping experiment \cite{Zukowski1993}, where we have two independent pairs of entangled states shared between three parties. If the party in possession of one particle from each pair performs a joint measurement on both of them, it is possible to generate entanglement between the two other parties even though they have no common source of states. The independence of the sources implies that in a LHV description we should have two hidden variables $\lambda_1$ and $\lambda_2$ respecting the non-linear constraint $p(\lambda_1,\lambda_2)=p(\lambda_1)\,p(\lambda_2)$ \cite{Branciard2010,Branciard2012}. These non-linear constraints imply that the correlations compatible with them define non-convex sets, the characterization of which demand complicated and computationally unfeasible tools from algebraic geometry \cite{Geiger1999,Chaves2016,Lee2015}.

Through the remainder of the paper we will focus on entropies but the same
techniques can also be applied to characterize the Bell polytopes described in
this section.

\section{The entropic approach to marginal problems}
\label{sec:entropic}

As seen above, one of the issues in the derivation of Bell inequalities is their mathematical
complexity in probability space, most preeminently in scenario involving several sources of states. To circumvent that, as we will discuss in the following, it has been realized that a information theoretic description can simplify the mathematical structure of the problem, by moving from a probabilistic description to an entropic one. A more detailed account can be found in Refs.~\cite{Chaves2012,Fritz2013,Chaves2014,weilenmann2017analysing}.

The \emph{Shannon entropy} assigns to each probability distribution a real
number. Let $\Omega = \{1, \ldots, n\}$ the indices of all involved variables
$X_1, \dots, X_n$.
Then for non-empty $\alpha\subset\Omega$ we can compute the marginal entropies
$H(X_\alpha)$ from the marginal probability distributions $p(X_\alpha)$.
Therefore, every global probability distribution $p(X_1,\ldots,X_n)$ defines a
collection of $2^n-1$ real numbers in the entropic description. We write these
as the components of a $(2^n{-}1)$-dimensional vector $\vec h \in \ents$. A
vector $\vec h\in\ents$ is called \emph{entropic} if it is compatible with
some probability distribution. The set of all entropic vectors is denoted
$\Gamma_n^*\subset \ents$. A major topic in information theory is to find the
inequalities that describe the boundaries of $\Gamma_n^*$.

The so-called \emph{basic inequalities} are the non-negativities of the four
basic Shannon information measures (entropy, conditional entropy, mutual
information, conditional mutual information). The \emph{Shannon cone} is
defined as the region bounded by the basic inequalities
\begin{align}
    \Gamma_n &= \Big\{ \vec h\in\ents : B\vec h \ge \vec 0 \Big\} \supset \overline{\Gamma}_n^*,
\end{align}
where $B$ is a matrix defined by all the basic inequalities. This provides a
useful and finite outer approximation for the generally unknown $\Gamma_n^*$.

For computational tasks, it is usually inefficient to have more than needed
input constraints and so the question arises how to specify $\Gamma_n$ with as
few inequalities as possible. The basic inequalities contain lots of
redundancies\footnote{For example, $H(X,Y) \ge 0$ is an instance of
a basic inequality that also is trivially implied by two other basic
inequalities via $H(X,Y) = H(Y\mid X) + H(X) \ge 0$}
and it is therefore advisable to restrict attention to an equivalent set that
does not contain any redundancies. This subset is found in the \emph{elemental
inequalities} – the non-negativities of so-called \emph{elemental forms}
\begin{subequations}
    \label{eq:elementalineqs}
\begin{align}
    0 \le H(X_i \mid X_{\Omega - \{i\}})
\intertext{and}
    0 \le I(X_i : X_j \mid X_\omega),
\end{align}
\end{subequations}
where $\omega \subset \Omega - \{i, j\}$.
The total number of elemental inequalities is therefore
\begin{align*}
    m &= n +2^{n-2} \, \binom n2.
\end{align*}
By listing all elemental forms in a matrix $E$, the Shannon cone can be
specified as
\begin{align}
    \label{eq:shannon-cone}
    \Gamma_n &= \Big\{ \vec h\in\ents : E\vec h \ge \vec 0 \Big\},
\end{align}
and this specification is \emph{minimal}. Further constraints, as for example the independence constraint $p(x_1,x_2)=p(x_1)\,p(x_2)$ can be easily integrated in this framework. On the level of entropies, independencies amount to linear constraints, e.g., $H(X_1,X_2)=H(X_1)+H(X_2)$ that can be put together with the elemental inequalities, thus defining a new augmented constraint matrix.

In a Bell experiment, we can simultaneously observe only certain subsets of
variables. This is captured as \emph{marginal scenario} $\marg\subset\Omega$
of accessible terms.
We will have $\df = \abs\marg$ and denote the corresponding entropy subspace
$\subs\subset\ents$.
Since classical correlations correspond to the existence
of a global probability distribution, on the level of entropies this implies
that a marginal entropy vector $\vec h_\marg$ must be the projection $\vec
h_\marg = \pi_\df(\vec h)$ of some \emph{entropic} entropy vector $\vec h
\in \Gamma_n^*$, where often the Shannon cone $\Gamma_n$ is used as
approximation. Computationally, for single vectors $\vec h_\marg$ membership
can be tested using a linear feasibility check,
\begin{align*}
    &\minimize{h} 0\ \text{ over }\vec h\in\ents \\
    &\st E \vec h \ge 0,\ \vec h_i = (\vec h_\marg)_i\ \forall i\in\marg.
\end{align*}

To remove the need to solve an LP for each compatibility query, a marginalized
description of the causal model is required. This is exactly the problem of
computing the facets of a projection $\pi_\df(\poly)$ where \poly is the
geometric object described by the set of inequalities. The
projection is particularly important when specifically searching for
experimental violations of the model – or in order to compare two models in
the marginal space.

\section{Multipartite Bell scenarios}
\label{sec:tripartite_ineqs}

\begin{figure}
    \includegraphics{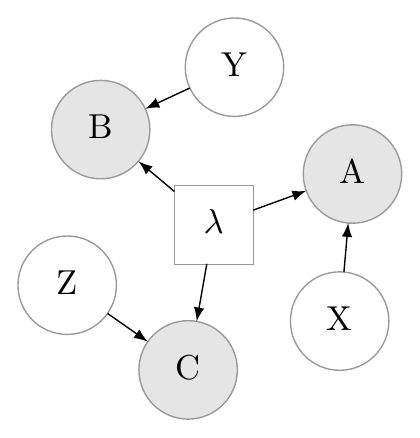}
\parbox{0.4\textwidth}{
 \caption{Tripartite Bell scenario.}
 \label{fig:tripartitebell}
}
\end{figure}

The Bell experiment can be generalized by allowing any number of parties $n$
and measurement choices $c$. We denote the $n\cdot c$ corresponding random
variables as $X_\Omega = (\A_1,\A_2,\ldots,\B_1,\B_2,\ldots)$.
As discussed in \SecRef{sec:entropic}, a LHV model is equivalent to the
existence of a probability distribution $p(X_\Omega)$ on all observable
variables. On the level of entropies, this means that we have a $(2^{nc}{-}1)$
dimensional space $\mathcal R_{nc}$ on which the elemental inequalities must
be satisfied. Entropic Bell inequalities arise as facets of the projection of
the Shannon cone $\Gamma_{nc}$ to subspaces $\mathcal R_\marg$ specified by a
marginal scenario $\marg = \{\omega_1,\ldots,\omega_l\}$ where each $\omega_i$
denotes a set of jointly measurable variables.
%
In general, the $c$ measurement operators of any single party do not
commute, so their precise value can not be measured simultaneously. This means
that we only have access to entropies in which the same party appears at most
once, i.e.~the marginal scenario $\marg$ can contain only combinations
$\omega\in\marg$ such that $\X_\omega \subset \{\A_i,\B_j,\C_k,\ldots\}$.
It's easy to see that for every $1 \le m \le n$ there are $\binom nm c^m$
different $m$-body terms $X_\omega$. The largest possible marginal scenario
that contains all theoretically observable combinations therefore has order
$(1{+}c)^n{-}1$, seen by applying the binomial formula $(x{+}y)^n =
\sum_{k=0}^n \binom{n}{k} x^k y^{n-k}$.

However, these multipartite correlations may be difficult to access
experimentally --- often only few-body or even two-body measurements are
available. Hence, this creates a natural interest in Bell inequalities that
require only few simultaneous measurements, see e.g.~\cite{tura2014}.
There is another practical reason to consider only few-body correlations.
Acquiring a sensible value for an entropy with $m$ arguments depends on the
knowledge of a corresponding probability distribution of $m$ random variables,
This in term requires a number of samples that grows exponentially with $m$.
This issue can be addressed in our entropic framework by limiting the marginal
scenario to only few-body combinations $\omega_i$, with $\abs{\omega_i} \le m$
for some small value of $m$, e.g.~$m=2$. Note that the case $m=1$, containing
only single-variable entropies, the corresponding projection reduces to the
positive orthant and is thus not too interesting.

Apart from experimental limitations, a third motivation to look at smaller
marginal scenarios is the \emph{computational} accessibility of the
corresponding local cone. As noted in the first part about polyhedral
projection, some of the projection algorithms are sensitive to the dimension
of the output space and will generally perform better when choosing a smaller
marginal scenario. This suggests to consider marginal scenarios such as the
set of all two-body terms while explicitly excluding one-body terms (even
though from the experimenters perspective, the one-body entropies can easily
be calculated from the measured two-body correlations).

\section{Computation of tripartite Bell inequalities}

In the following, we will be concerned with the tripartite Bell scenario with
two measurements per party. Specifically, we compute projections of the
$63$-dimensional Shannon cone $\Gamma_6\subset\mathcal R_6$ of the random
variables $X_\Omega = (\A_1,\A_2,\B_1,\B_2,\C_1,\C_2)$ to some experimentally
accessible subspaces.

The entropies corresponding to jointly observable combinations, are the $8$
three-body terms $H(\A_i,\B_j,\C_k)$, the 12 two-body terms $H(X_i,Y_k)$, and
the 6 one-body terms $H(X_i)$, corresponding to 8D, 12D, and 6D subspaces of
$\mathcal R_6$. The other ``symmetric'' subspaces by combining these sets are
14D, 18D, 20D, and 26D. To simplify notations we will refer to the individual
subspaces and the contained local cones using their dimensionality from here
on.

We have obtained full characterizations of the 8D, 12D, and 14D cones, as well
as several facets of the 18D, 20D and 26D cones. A summary of the results for
each examined local cone is given is given in Table \ref{tab:lcsummary}. A
comprehensive listing of the individual facets can be found in the Appendix
\ref{chap:facetlisting}. Furthermore, in the Appendix \ref{sec:facet-structure-constraint} we make a detailed analysis of the structure of these inequalities. In the following we give a few short notes on the techniques used during these computations:

\begin{figure}
    \centering
    \includegraphics{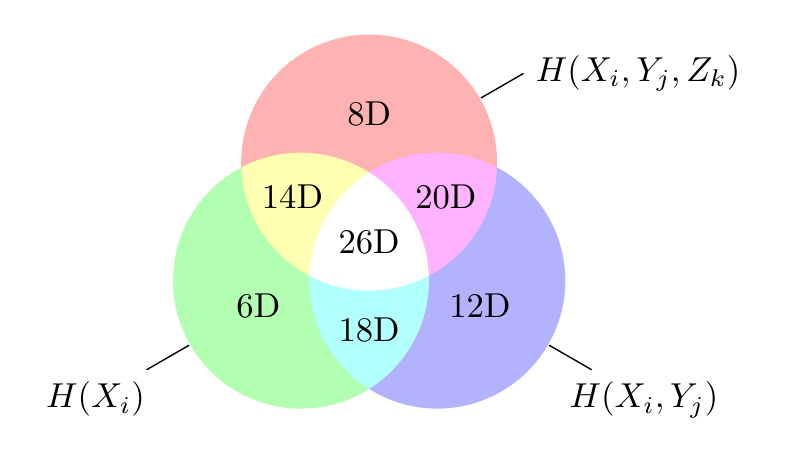}
\parbox{0.4\textwidth}{%
    \caption{Dimension of the local cones obtained by allowing all accessible $m$-body terms.}%
    \label{fig:localconedim}}
\end{figure}

\begin{table}
\begin{tabular}{rcccc}
    \toprule
    cone    & method    & facets    & classes   & \\
    \midrule
    8D      & CHM       & 104       & 7         & (complete) \\
    12D     & \AFI{1}   & 444       & 14        & (complete) \\
    14D     & \AFI{3}   & 566       & 22        & (complete) \\
    18D     & RFD       & 888       & 22        & (partial)  \\
    20D     & RFD       & 496       & 17        & (partial)  \\
    26D     & ROA       & 1360      & 37        & (partial)  \\
    \bottomrule
\end{tabular}
\parbox{0.4\textwidth}{\caption{\label{tab:lcsummary}Number of discovered facets. The second number doesn't count facets
    that can be obtained using a symmetry substitution.}}
\end{table}

\paragraph{Solving the 8D, 12D and 14D local cones}

The convex hull method (CHM) turns out to be sufficient to compute complete
descriptions of the 8D and 12D local cones in reasonable time. While the 8D
cone is solved in a matter of seconds, the 12D cone already takes about 40
minutes to finish on a \SI{2.4}{GHz} CPU and eats up several gigabytes of
RAM.
The 12D cone provides a prime example for the possible benefits of
supplementing CHM with a single-level adjacent facet iteration (AFI). Using
\AFI{1} cuts the calculation time down to about 20 seconds with a peak memory
usage of roughly \SI{180}{MiB}.
The 14D local cone was computed using a three-level AFI recursion in
\SI{37.5}{\hour} taking \SI{10}{\gibi\byte} of RAM.

\paragraph{Solving the 18D local cone}

The 18D subspace corresponding to the set of all one- and two-party
entropies proves to be too much to be fully solved by CHM/AFI. However,
this case admits to a treatment using a randomized facet discovery (RFD). We
have achieved the best results with a 5-level recursion strategy \RFD{5}.

\paragraph{Solving the 20D and 26D local cones}

The largest marginal scenarios are 20D and 26D and thus well out of
reach of CHM/AFI.  However, the projection problem is computationally
accessible using randomized strategies. In particular, we used the technique
discussed in \SecRef{sec:roa} (\autoref{alg:p2f}) to improve upon outer
approximations given by a list of known faces.
There are several ways to obtain valid input constraints. The most naive is
generate random ones by repeatedly applying the EPM based LP as discussed in
\SecRef{sec:epm-partial}. However, we could achieve a higher yield of facets
by generating the outer constraints constructively from valid
combinations of the input-space facets. This can be done systematically by
exploiting simple observations about how elemental inequalities must be combined
in order to obtain an expression in the output-space, see
\autoref{sec:facet-structure-constraint}.
A more physical approach is to directly improve upon a previously known
outer approximation. In our case, we could use the extreme rays of the
nonsignalling cone~\cite{chaves2016nonsignalling} as input constraints to find
even a few more new facets.

\section{Witnessing tripartite nonlocality}
\label{sec:nonlocality_marginals}

Having obtained a list of inequalities characterizing a given marginal Bell
scenario, our next interest is now which of these constraints can be violated
by quantum mechanical correlations obtained from appropriate local
measurements on a quantum state, and whether they can be used to indicate
nonlocal correlations that are not detected by known nonlocality tests.
Note that it's a priori not clear whether a given inequality can witness
quantum non-locality at all. Some of these inequalities are going to be
trivial, in the sense that they represent elemental inequalities of the form
\eqref{eq:elementalineqs} that are respected by any well defined probability
distribution. Other inequalities, though not of elemental form, might still be
trivial in the sense that they are respected by any non-signalling
correlations \cite{chaves2016nonsignalling,Budroni2016indis} (including quantum mechanical
correlations).

Secondly, an important question in the multipartite scenario is whether one
can witness non-locality if only few-body or even two-body measurements are
available. As discussed before, from the physical perspective the relevance of
these scenarios stems from the fact that there are typical setups where the
available measurements are very restricted and thus the empirical data is
limited to two-body correlators \cite{tura2014}. In this situation, we are
also interested to see if any of our found tripartite Bell inequalities
provides an advantage of the known bipartite nonlocality tests.

\subsection{Bipartite nonlocality tests}

There is a natural hierarchy to tackle this question. For setups in which
each party can perform only two possible measurements, each with two outcomes,
the canonical candidate for bipartite nonlocality tests is the CHSH inequality
\begin{align}
    2 &\ge \abs{E(\A_1,\B_1) + E(\A_1,\B_2) \nonumber \\
        &\quad + E(\A_2,\B_1) - E(\A_2,\B_2)}.
    \label{eq:chsh}
\intertext{
Its entropic counterpart, the \CHSHE inequality
}
    0 &\le H(\A_1,\B_1) + H(\A_1,\B_2) + H(\A_1,\B_1)\nonumber \\
        &\quad  - H(\A_2,\B_2) - H(\A_1) - H(\B_1)
    \label{eq:chshe}
\end{align}
is less tight but can also be applied in more general settings, e.g.~if the
number of outcomes is not fixed at two. In the case with more than 2 outcomes,
another option is given by the CGLMP family of Bell inequalities constructed
in~\cite{Collins2002}. These are constraints on the level of probabilities
(like CHSH) that are applicable for the bipartite scenario with two
measurements and $d$ outcomes per party (unlike CHSH). For example, the CGLMP
inequality for a $d=3$ is:
\begin{align}
    2 \ge \smash{\big(}
      &{} + P(A_1=B_1) + P(B_1=A_2+1) \nonumber \\
      &{} + P(A_2=B_2) + P(B_2=A_1)   \nonumber \\
      &{} - P(A_1=B_1-1) - P(B_1=A_2) \nonumber \\
      &{} - P(A_2=B_2-1) - P(B_2=A_1-1)
        \smash{\big)}.
    \label{eq:cglmp}
\end{align}

All of the above bipartite constraints –CHSH, \CHSHE and CGLMP– are tied to a
specified set of measurements. This means that even if the two-party locality
constraints are satisfied, it is in principle possible that the quantum state
could show bipartite non-local behaviour with a different set of measurements.
Since entanglement is a necessary precondition for non-locality, one way to
avoid this issue is to demand that all two-party subsystems are separable.
In general, deciding whether a quantum state is separable is a non-trivial
problem which has been shown to be
NP-hard~\cite{lewenstein2000,bruss2002,gurvits2003,gharibian2010}. A
sufficient condition for entanglement, however, is the Peres–Horodecki
criterion~\cite{peres1996,horodecki1996}. It is also called PPT, which stands
for \emph{positive partial transpose}. Given a density matrix
\begin{align*}
    \rho &= \sum_{ijkl} p_{ijkl} \ketbra ij \otimes \ketbra kl,
\end{align*}
it states that if its partial transpose,
\begin{align*}
    \rho^{T_B} &= \sum_{ijkl} p_{ijkl} \ketbra ij \otimes \ketbra kl,
\end{align*}
has any negative eigenvalues, $\rho$ is guaranteed to be entangled. The choice
of the subsystem B is arbitrary here. For $\CC_2 \otimes \CC_2$
and $\CC_2 \otimes \CC_3$ the criterion is both necessary and
sufficient,
\begin{align}
    \rho\ \text{ separable} &\Leftrightarrow \rho^{T_B} \ge 0,
\end{align}
which means that in a three-qubit system \Qunits{2}, by asserting PPT we can
limit the search for non-local states to only states that are
unentangled in any of the two-party subsystems.
In the three-qutrit system \Qunits{3} the criterion can be applied as well,
but provides only a necessary condition for separability.

\subsection{Search for non-locality witnesses}

To the aim of answering the above questions we have searched for violations of
each of the different inequality classes by means of numerical optimization.
More details on the numerical method can be found in the
Appendix~\ref{sec:search-violation}. We have considered projective
measurements on tripartite quantum states composed by either qubits or
qutrits, i.e.~states that live in one of the spaces \Qunits{2} or \Qunits{3}.

In both cases the search was first run unconstrained, i.e.~without imposing
that the violating quantum state should also fulfill further constraints on
the level of two parties. As can be seen in
\autoref{tab:nonlocality-witnesses-indices}, in all investigated marginal
scenarios, there are several facets (including non-trivial ones) for which we
could not find quantum mechanical violations.  Of course, this could be due to
the fact that we are limiting the dimension of the considered quantum states
and only looking to projective measurements.  This is clearly illustrated by
the fact that there are several inequalities for which we could not find
violations considering qubits but which were violated by qutrit states. For instance, we found a qutrit violation for the inequality $I_8$ in the $26D$ scenario but failed to find any qubit violations.

\newcolumntype{L}[1]{l}
\newcommand\Head[1]{\multicolumn{1}{c}{#1}}
\begin{table*}
\begin{tabular}{ll@{\quad}L{3.0cm}l@{\quad}L{3.0cm}l@{\quad}L{3.0cm}l@{\quad}L{3.0cm}l@{\quad}L{3.7cm}}
    \toprule
    system  & constraints   & \Head{12D}   && \Head{14D}   && \Head{18D} && \Head{20D}   && \Head{26D}   \\
    \midrule
    \input{qr/wittab.tex}
    \bottomrule
\end{tabular}
\parbox{0.8\textwidth}{\caption{\label{tab:nonlocality-witnesses-indices}Non-locality witnesses of the 12D/14D/18D/20D/26D cone that satisfy the given constraint on all two party subsystems. No non-locality witnesses have been identified for the 8D cone. The inequalities $I_k\vec h \ge 0$ are listed in \autoref{chap:facetlisting}. Recall that 14D/20D/26D do involve three-body correlators, while 12D/18D get along with two-body correlators.}}
\end{table*}

For each member of this established set of nonlocality witnesses we
proceeded by searching for nonlocal states while imposing all symmetries of
the known
bipartite locality constraints appropriate for the respective system.
Specifically, for the three-qu\emph{bit} system \CHSHE, CHSH and the PPT
criterion are applicable and provide increasingly tight bounds in this order.
The three-qu\emph{trit} system was subjected to CGLMP, \CHSHE and PPT. The
resulting sets of non-locality witnesses are listed in
\autoref{tab:nonlocality-witnesses-indices}.

We start by observing that the tripartite Bell inequalities seem to be more
advantageous compared to bipartite tests when going to higher dimensional
quantum systems, i.e.~systems with more outcomes. This is shown by the fact
that, by imposing CHSH or \CHSHE, we could find much fewer inequality violations with qubits. Of particular relevance is the usefulness of our inequalities as nonlocality
witnesses in the case that we have access to only two-body correlators (12D,
18D). As can be seen in \autoref{tab:nonlocality-witnesses-indices}, using qutrit states (and 3 measurement outcomes) we found examples of correlations such that the marginals violate neither \CHSHE nor
CGLMP, yet do violate some of the inequalities characterizing the 18D local
cone. That is, if we just trace out one of the parties (returning to a bipartite Bell scenario) the correlations are classical. However, if instead we look at all available bipartite information (considering the 3 pairs of bipartite distributions) we can witness its non-locality.

Finally, as pointed out by Wurflinger et al.~\cite{wurflinger2012}, multipartite entanglement can be inferred from marginal probabilities that are local in all two-party subsystems. We now ask a related question: Can our Bell inequalities detect the non-locality of tripartite states even if all two-party subsystems are separable? In such a case, no bipartite test can detect the nonlocality of
the state. Unfortunately, we could only achieve violations with separable/PPT marginals (for instance, a GHZ state of the inequalities that do involve 3-body terms. It thus remain an open question whether the results in~\cite{wurflinger2012} can be extended in its full generality for entropic Bell inequalities.

\section{Discussion and Outlook}
\label{sec:discussion}
The characterization of the set of correlations/probability distributions of a given marginal scenario is of central relevance in a variety of fields. Algebraic geometry \cite{huynh1990,Geiger1999,Garcia2005} and quantifier elimination methods \cite{lassez1990quantifier,davenport1988,monniaux2010} provide a very general tool for tackling the problem that in practice, unfortunately, is limited to very few cases of interest due to its double exponential computational complexity. In the particular case where such sets define convex regions such as the polytopes arising in the study of Bell non-locality \cite{Pitowsky1989} or the entropy cones arising in the study of information theory \cite{Yeung2008} or causal inference \cite{Chaves2014,Chaves2014b}, the complexity of the task is certainly reduced since the often efficient tools from convex optimization theory can be employed. Yet, even in the convex case, we also often encounter situations and marginal scenarios out of reach of current algorithms.

Within this context, we have provided a review of known algorithms for the projection of convex polyhedra and also proposed a new one, that we call adjacent facet iteration. To show its relevance and compare it with previous methods we have employed it for the derivation of entropic Bell inequalities in a tripartite scenario. As discussed, our method provided a significant time improvement over other usual methods and in some cases allowed for the characterization of marginal scenarios outside the reach of other algorithms.
With that we managed to derive several novel tripartite Bell inequalities that furthermore are facets of the associated entropy cone. Of particular relevance, are the inequalities involving at most bipartite information, that is, involving at most two observables. To our knowledge, these are first entropic Bell inequalities of this kind, thus extending the results of \cite{tura2014} in the context of Bell inequalities well suited for the analysis of many-body systems. Further, we have shown that such inequalities can be violated by probability distributions that appear to be local by other standard bipartite tests such as the CHSH and CGLMP inequalities \cite{Clauser1969,Collins2002}, an extension to the entropic regime of the results in \cite{wurflinger2012} and that clearly show the relevance of these new inequalities.

As for future research, we believe there are few promising directions. For instance, multipartite Bell inequalities involving at most two-body correlations have been proposed \cite{tura2014} to probe the non-classicality of many-body systems where the measurement of observables is very limited (see for instance an experimental realization of this idea in \cite{schmied2016bell}). Most of such inequalities, however, are derived for the particular of binary measurement outcomes. In contrast, since entropic Bell inequalities are valid for an arbitrary number of outcomes, they could provide a natural venue to extend such results for quantum systems and measurements of higher dimensions. In turn, we believe that the computational method we propose here could also find applications in the characterization of causal networks beyond the Bell network (see for instance \cite{Chaves2014,Henson2014,weilenmann2017analysing}). As an illustration of that, we provide in the Appendix the full characterization of two common ancestor causal structures that generalize the triangle causal structure \cite{Steudel2010b}  that has been the focus of much research in quantum foundations recently \cite{Fritz2012,Chaves2014,Henson2014,Wolfe2016}. We hope our results might trigger further research on this direction.

\bibliography{draft}

\begin{widetext}
    \part*{Appendix}
\end{widetext}

\section{Characterizing common-ancestors causal structures}
\label{sec:ancestor}

In addition to the Bell scenarios discussed in the previous sections, we have
also investigated applications for so-called pairwise hidden ancestor models.

\begin{figure}[b]
   \caption{Common ancestor models}
   \begin{subfigure}{0.23\textwidth}
       \centering
       \includegraphics[page=3]{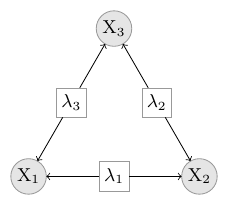}
       \caption{Triangle \Cn{3}}
       \label{fig:triangle}
   \end{subfigure}
   \begin{subfigure}{0.23\textwidth}
       \centering
       \includegraphics[page=2]{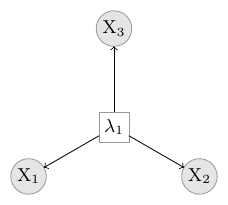}
       \caption{Single ancestor}
       \label{fig:single-ancestor}
   \end{subfigure}
   \begin{subfigure}{0.23\textwidth}
       \centering
       \includegraphics[page=4]{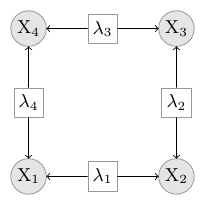}
       \caption{Square \Cn{4}}
       \label{fig:square}
   \end{subfigure}
   \begin{subfigure}{0.23\textwidth}
       \centering
       \includegraphics[page=5]{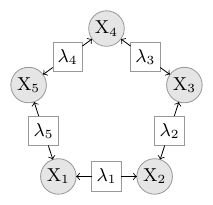}
       \caption{Pentagon \Cn{5}}
       \label{fig:pentagon}
   \end{subfigure}
\end{figure}

The \emph{triangle scenario} \Cn3 (\autoref{fig:triangle}) is a prominent
causal model for the relationship of three observable variables. It has been
considered both in classical context~\cite{steudel2010} as well as quantum
non-locality~\cite{Branciard2012,Fritz2012}. Contrary to the corresponding
single-common-ancestor model (\autoref{fig:single-ancestor}), it exhibits non-trivial constraints and is
therefore preferred by Occam's razor if both models are compatible with given
empirical data.

The triangle can be generalized as a regular polygone \Cn{n}, see the square
and pentagon in \autoref{fig:square} and \autoref{fig:pentagon}. The entropic
constraints associated with \Cn{n} are the mutual independence of the
ancestors $H(\lambda_\Omega) = \sum_{i\in\Omega} H(\lambda_i)$ and the local
Markov conditions
\begin{align*}
    0 = I(\X_i : X_{\Omega-\{i\}},\lambda_{\Omega-\Pa_i} \Mid \lambda_{\Pa_i}),
\end{align*}
for all observable variables $\X_i$, where $\Omega = [n]$ and $\Pa_i = \{
    i,\,(i\kern-2pt\mod n) + 1 \}$.

In~\cite{Fritz2013}, the authors employ a Fourier-Motzkin elimination using
the \texttt{PORTA} open source software package in an attempt to compute the
projection of \Cn{3} to the marginal scenario $\marg = 2^{\X_\Omega}$ that
includes all observable terms. However, their computations did not complete –
which shows that even such a simple model can prove difficult for FME if not
using proper protection redundancy elimination. This particular problem is
solved in~\cite{Chaves2014}, showing that there are only three
non-trivial classes of inequalities:
\begin{align*}
0 &\le 4 H_{12} + 4 H_{13} + 4 H_{23} - 2 H_{123} - 5 H_1 - 5 H_2 - 5 H_3, \\
0 &\le 3 H_{12} + 2 H_{13} + 2 H_{23} - \phantom{2} H_{123} - 3 H_1 - 3 H_2 - 3 H_3, \\
0 &\le  H_{12} + H_{23} - H_1 - H_2 - H_3.
\end{align*}

We now extend these results by computing the full projection of the
square~\Cn{4} and pentagon~\Cn{5}. The results are obtained using an FM
elimination with full LP-based redundancy removal. The implementation can be
found in~\cite{cfme}.
All classes of facets can be found online~\cite{pystif} and are additionally
listed in \autoref{chap:facetlisting}. In total, the marginal cone of \Cn4 has
12 distinct classes of facets, the \Cn5 marginal cone has 39 classes of facets.

Starting from a system with no redundancies, the computations take
\SI{7.4}{\second} and \SI{340}{\second} for \Cn{4} and \Cn{5}, respectively
(on a usual desktop PC). This underlines how LP-based redundancy removal can
be a major advantage in performing FM eliminations.

\section{List of inequalities}
\label{chap:facetlisting}

In the following we list the facets that were found using the techniques
discussed in the main text. Note that we list expressions $I_k$ which should
be read as inequalities $I_k \ge 0$. Furthermore, if multiple inequalities
are equivalent by a symmetry of the cone (i.e. relabeling of variables), we
list only one representative. The data can also be found online in the
repository of the \texttt{pystif} software package~\cite{pystif}.

\subsection{Machine readable data}

We include quick response (QR) codes to allow easy data acquisition. Due to the capacity limitations of QR codes the data is gzipped. The raw data files can be unpacked with a wide range of popular compression software such as \texttt{7z} on Windows or the prevalent \texttt{gunzip} tool on UNIX-like systems:
\begin{verbatim}
gunzip -c raw.dat > friendly.txt
\end{verbatim}
The unzipped data is also directly attached to the PDF and can be accessed by
clicking on the QR code images.  The uncompressed data is a self-documenting
text file that can e.g.~be directly used as input file for any of the
\texttt{pystif} utilities. However, the full \texttt{pystif} stack is not
required – the text format is suitable for machine reading and easy to parse.
For example, in python you can use
\begin{lstlisting}[language=python]
>>> numpy.loadtxt('friendly.txt')
\end{lstlisting}
to obtain a numpy array whose row vectors $\vec I$ correspond to linear
inequalities $\vec I\cdot \vec h \ge 0$ where $\vec h$ is an entropy vector.
The column order is generally based on number of arguments and party
affiliation but is also documented within the file.

\attachfilesetup{mimetype=text/plain}
\attachfilesetup{color=0 0 0}

\small
\subsection{8D cone}

Full characterization of the marginal cone with all one-body terms for the
tripartite Bell scenario.

\input{ineqs/bell-08D.tex}

    \begin{figure}[H]
        \centering
        \textattachfile{qr/bell-08D.txt}{\includegraphics[frame,width=0.4\textwidth]{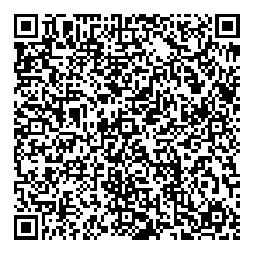}}
        \caption{8D cone}
    \end{figure}

\subsection{12D cone}

Full characterization of the marginal cone with all two-body terms for the
tripartite Bell scenario.

\input{ineqs/bell-12D.tex}

    \begin{figure}[H]
        \centering
        \textattachfile{qr/bell-12D.txt}{\includegraphics[frame,width=0.4\textwidth]{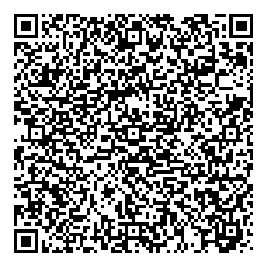}}
        \caption{12D cone}
    \end{figure}

\subsection{14D cone}

Full characterization of the marginal cone with all one- and three-body terms
for the tripartite Bell scenario.

\input{ineqs/bell-14D.tex}

    \begin{figure}[H]
        \centering
        \textattachfile{qr/bell-14D.txt}{\includegraphics[frame,width=0.4\textwidth]{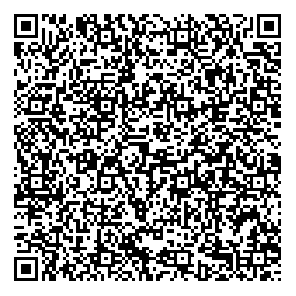}}
        \caption{14D cone}
    \end{figure}

\subsection{18D cone}

Listing of discovered facets of the marginal cone of the tripartite Bell
scenario where the marginal scenario contains all one- and two-body
terms. These facets do not provide a full characterization.

\input{ineqs/bell-18D.tex}

    \begin{figure}[H]
        \centering
        \textattachfile{qr/bell-18D.txt}{\includegraphics[frame,width=0.4\textwidth]{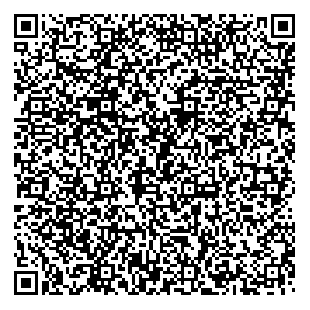}}
        \caption{18D cone}
    \end{figure}

\subsection{20D cone}

Listing of discovered facets of the marginal cone of the tripartite Bell
scenario where the marginal scenario contains all two- and three-body
terms. These facets do not provide a full characterization.

\input{ineqs/bell-20D.tex}

   \begin{figure}[H]
       \centering
       \textattachfile{qr/bell-20D.txt}{\includegraphics[frame,width=0.4\textwidth]{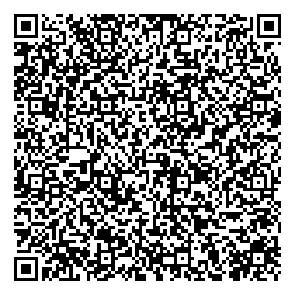}}
       \caption{20D cone}
   \end{figure}

\subsection{26D cone}

Listing of discovered facets of the marginal cone of the tripartite Bell
scenario where the marginal scenario contains all one-, two-, and three-body
terms. These facets do not provide a full characterization.

\input{ineqs/bell-26D.tex}

    \begin{figure}[H]
        \centering
        \textattachfile{qr/bell-26D.txt}{\includegraphics[frame,width=0.4\textwidth]{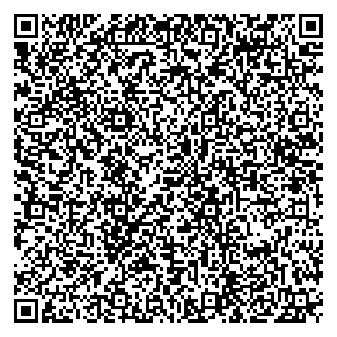}}
        \caption{26D cone}
    \end{figure}

\subsection{The square \texorpdfstring{$\mathcal{C}$4}{C4}}

Full characterization of the marginal cone of the pairwise hidden ancestor
model \Cn4 with 4 hidden ancestors and 4 observable variables.

\input{ineqs/cca-4.tex}

    \begin{figure}[H]
        \centering
        \textattachfile{qr/cca-4.txt}{\includegraphics[frame,width=0.4\textwidth]{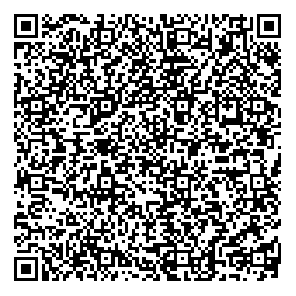}}
        \caption{The square \Cn4}
    \end{figure}

\subsection{The pentagon \texorpdfstring{$\mathcal{C}$5}{C5}}

Full characterization of the marginal cone of the pairwise hidden ancestor
model \Cn5 with 5 hidden ancestors and 5 observable variables.

\input{ineqs/cca-5.tex}

    \begin{figure}[H]
        \centering
        \textattachfile{qr/cca-5.txt}{\includegraphics[frame,width=0.4\textwidth]{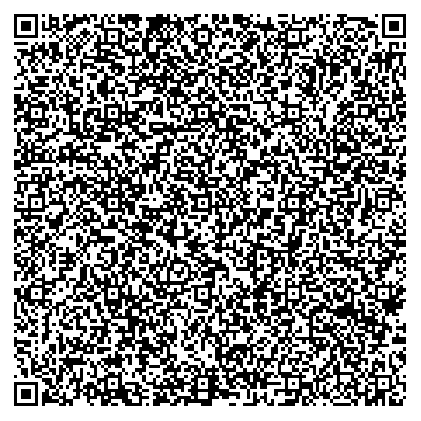}}
        \caption{The pentagon \Cn5}
    \end{figure}

\section{Details about the numerical search for the violation of entropic inequalities}
\label{sec:search-violation}

We search for quantum states that violate a particular Bell inequality $I_k
\ge 0$ via numeric optimization that minimizes the value of $I_k$ by varying
the state and measurement operators. In order to do this, we have to choose a
suitable parametrization for these objects.

Let the Hilbert space of the composite system be $\mathcal H = \Qunits{n}$.
First note that the subsystem dimensions give the maximum possible number of
outcomes of any measurement on the respective subsystem, in this case $n$.
Considering only non-degenerate measurements, the results can be labelled $\{1,
\ldots, n\}$. A single-party observable corresponds to a hermitian operator
and can thus be seen as a unitary rotation $UDU^\dagger$ of a diagonal
matrix $D$. Specifying a measurement hence amounts to
specifying a unitary matrix $U$ which can be done with $n^2$ real parameters
of which $n$ correspond to phase transformations on the components of the
state vector. When optimizing over both the state and measurements, only the
relative phase matters and so it not necessary to specify the
phases in both parameter sets. A convenient parametrization that allows to
effortlessly leave out the parameters corresponding to phase transformations
is given in~\cite{jarlskog2005}.

Pure states $\psi\in\mathcal H = \Qunits{n}$ can be represented as unit vectors with $n^3$
complex components. A straightforward approach to sample a random state is to either
normalize a vector created from $2n^3$ unconstrained (normally distributed)
real parameters; or to directly create a unit vector from $2n^3{-}1$ angles
using a spherical coordinate system. While the latter uses one less parameter,
we employ the former option as this involves less complexity and treats
every parameter equally.

We were unable to enforce the PPT constraint exactly using \emph{pure} states.
To avoid this issue, we removed the PPT constraint violations by replacing the
state vector with a density matrix and mixing with  ``white noise'' in a
second optimization pass:
\begin{align*}
    \psi \rightarrow \rho = \lambda \ketbra\psi\psi + (1-\lambda) \left(\tfrac 1{n^3} \mathbbm 1\right),
\end{align*}
where $\lambda \in [0, 1]$ is the mixing parameter. This adds one additional
optimization parameter to the set.

In the special case of a three-qubit system \Qunits{2} more
physically recognized parametrizations are available: Since we are only interested in
states that show non-local behaviour we only need to consider entangled
states. This can be achieved with a parametrization in terms of entanglement
parameters such as described in~\cite{acin2000}. The subsystems are
essentially spin one-half systems with spin measurements $\vec\sigma \cdot
\vec v$ along an arbitrary direction $\vec v$ where $\vec\sigma$ is the vector
of Pauli matrices. A measurement can thus be specified in terms of two angles.

\section{A structural constraint on facets}
\label{sec:facet-structure-constraint}

As can be seen in Appendix \ref{chap:facetlisting}, the listed inequalities fall into groups that share a distinct structural similarity. For instance, the $I_{26}$, $I_{27}$, $I_{28}$, and $I_{29}$ inequalities of the 26D local cone have the same number of positive and negative $N$-body terms.
Similar observations hold true for groups of facets in all of the computed
projections. This suggests to investigate the underlying structure.

We start noting that any valid Shannon-type inequality $I$ can be written as a
non-negative linear combination of elemental forms:
\begin{align}
    \label{eq:elemental-expansion}
    I = \sum_i c_i\,H(\X_i\Mid\X_{\Omega-\{i\}}) + \sum_{i<j} \sum_{\omega\subset\Omega} c_{ij\omega}\,I(\X_i:\X_j\Mid\X_\omega).
\end{align}
Such a combination can be retrieved as the dual solution of the LP that proves
the inequality, and hence we call such a combination a \emph{proof} for the
inequality.
For example, the facet $I_{17}$ of the 26D local cone,
\begin{align*}
    I_{17}^{\mathrm{26D}} = &H(\A_1\,\B_1\,\C_1) + H(\A_1\,\B_1\,\C_2) + H(\A_2\,\B_1\,\C_1) \\
      &- H(\A_2\,\B_1\,\C_2) - H(\A_1\,\B_1) - H(\B_1\,\C_1),
\end{align*}
can be proven by the following combination of
elemental forms:
\begin{alignat*}{2}
    I&_{17}^{\mathrm{26D}} = \\
    &+ H(\A_1\Mid\A_2,\B_1,\B_2,\C_1,\C_2)
     &&{}+ H(\B_1\Mid\A_1,\A_2,\B_2,\C_1,\C_2) \\
    &+ I(\A_1:\A_2\Mid\B_1,\B_2,\C_1,\C_2)
     &&{}+ I(\B_1:\C_2\Mid\A_1,\A_2,\B_2,\C_1) \\
    &+ I(\A_1:\B_1\Mid\A_2,\B_2,\C_1)
     &&{}+ I(\A_1:\C_2\Mid\B_1,\B_2,\C_1) \\
    &+ I(\B_1:\B_2\Mid\A_1,\C_1)
     &&{}+ I(\A_2:\B_2\Mid\B_1,\C_1).
\end{alignat*}
%
This expression is highly structured, and not by chance. In fact, every
entropic Bell inequality must have a similar looking proof: There must be the
same number $k$ of
elemental forms of every kind until reaching the output space. This
leads to the following result about the structure of facets of the
$(2n^2)$-dimensional marginal cone that contains all one- and two-body terms in the multipartite Bell scenario with two measurements per party.

\subsection{Definition of the constraint}

Consider linear information inequalities on $N$ random variables of the form
\begin{align}
    \label{eq:general2body}
    0 &\le I = \sum_{i} a_i H(\X_i) + \sum_{i < j} b_{ij} H(\X_i,\X_j).
\end{align}
Then, for $N\ge 2$ every minimal Shannon type inequality of this form falls into one of two possible categories up to a positive scale factor.  The
first one being the mutual informations among different random variables, i.e.:
\begin{align*}
    I = I(\X_i:\X_j) = H(\X_i) + H(\X_j) - H(\X_i,\X_j).
\end{align*}
In order to describe the second category, first rewrite the coefficients in
\eqref{eq:general2body} as the difference of two non-negative numbers $a_i =
a_i^+ - a_i^-$ and $b_{ij} = b_{ij}^+ - b_{ij}^-$ where only one of them is
allowed to be non-zero.  Then all remaining facets can be written with integer
coefficients fulfilling
\begin{align*}
               a_{i}^+ &= 0,   &   \sum_{i<j} b_{ij}^+ &= m + k,    \\
    \sum_{i}   a_{i}^- &= k,   &   \sum_{i<j} b_{ij}^- &= m,
\end{align*}
where $0 \le m \le k$ are integers.

For example, in
\begin{dgroup*}
\begin{dmath*}
    I_{9}^{\mathrm{18D}} = H(\A_1\,\B_2) + H(\A_2\,\B_2) + H(\A_1\,\C_2) + H(\A_2\,\C_1) + H(\B_1\,\C_1) + H(\B_1\,\C_2) - H(\A_1\,\C_1) - H(\A_2\,\C_2) - H(\A_1) - H(\B_1) - H(\B_2) - H(\C_1)
\end{dmath*}
\end{dgroup*}
we have:
\begin{align*}
               a_{i}^+ &= 0,   &   \sum_{i<j} b_{ij}^+ &= 6,    \\
    \sum_{i}   a_{i}^- &= 4,   &   \sum_{i<j} b_{ij}^- &= 2,
\end{align*}
consistent with this statement.

\subsection{Intuition}

Recall that the elemental inequalities are simply the non-negativities of
the so-called elemental forms
\begin{align*}
    0 \le H(\X_i \Mid \X_{\Omega - \{i\}})
       &= H(\X_\Omega) - H(\X_{\Omega-\{i\}}) \\
\intertext{and}
    0 \le I(\X_i : \X_j \Mid \X_\omega)
     &= - H(\X_{\omega \cup \{i,j\}}) - H(\X_\omega) \\&\quad+ H(\X_{\omega\cup\{i\}}) + H(\X_{\omega\cup\{j\}}),
\end{align*}
where $\Omega = [N]$ is the set of all indices and $\omega
\subset \Omega - \{i, j\}$.  The unconditional mutual informations
($\omega=\emptyset$) between two measurements of different parties lie
directly in the output space and therefore make up one category of its facets.

The true conditional mutual informations ($\omega\ne\emptyset$) have a
negative term containing $\abs{\omega}+2\ge 3$ random variables. Clearly, this
term is not part of the output space, and if such a term appears in the \emph{proof} of a facet, it has to be counteracted by
another elemental form. For $\abs{\omega}+2<N$ the only way to compensate the
negative high-order term is to add another conditional mutual information
(CMI) with $\abs{\omega'} = \abs{\omega}+1$.  For $\abs{\omega}+2=N$ this
requires a conditional entropy since there are no CMIs with $\abs{\omega'} >
\abs{\omega} = N-2$.  Hence, writing a facet as a sum of elemental forms leads
to a chain of equally many CMIs for each $\abs\omega$ and equally many
conditional entropies.

\subsection{Proof}

For $N=2$ the claim holds trivially since the output cone coincides with the
Shannon cone $\Gamma_2$ whose only facets are the elemental inequalities
$I(\X_1:\X_2) \ge 0$ and $H(\X_1,\X_2) - H(\X_i) \ge 0$. We consider
$N \ge 3$ in the following.

Let $I \ge 0$ be a valid Shannon-type inequality of the form
\eqref{eq:general2body}, then
%
\begin{align}
    \label{eq:elemental-expansion2}
    I = \sum_i c_i\,H(\X_i\Mid\X_{\Omega-\{i\}}) + \sum_{i<j} \sum_{\omega\subset\Omega} c_{ij\omega}\,I(\X_i:\X_j\Mid\X_\omega),
\end{align}
%
where all the coefficients are non-negative and $c_{ij\omega} = 0$ if $i\in
\omega$ or $j\in \omega$. Define
\begin{align}
    \Delta_p &= \sum_{i<j} \sum_{\abs\omega=p} c_{ij\omega}\,I(\X_i:\X_j\Mid\X_\omega) \\
    I_p &= I - \sum_{i<p} \Delta_i.
\end{align}
The first bit is to show that for $p\ge 2$ the following invariant holds:
\begin{align}
    \label{eq:constr-inductive-proof}
    I_p &= \sum_{\abs\alpha=p+1} c_\alpha^+ H(\X_\alpha) - \sum_{\abs\beta=p} c_\beta^ - H(\X_\beta)
\end{align}
for some non-negative coefficients with $\sum c_\alpha^+ = \sum c_\beta^- = k$.

Note that $I_{N-1}$ contains only the conditional entropy terms and hence
\eqref{eq:constr-inductive-proof} holds trivially for $p = N-1 \ge 2$. If
$N=3$, we are finished. Otherwise, assume that
\eqref{eq:constr-inductive-proof} is the case for $p\ge 3$ and consider
$I_{p-1}$. Clearly, the entropy terms in $I_p$ have at least $3$ arguments and
must therefore be cancelled in the final $I$. It is a straightforward
observation that this must be achieved by the terms in $\Delta_{p-1}$ and that
the term counts must match:
\begin{align}
    \sum_{i<j} \sum_{\abs\omega=p-1} c_{ij\omega} = \sum_{\abs\alpha=p+1} c^+_\alpha = k.
\end{align}
Conversely, this means that $k$ of the negative $(p{+}1)$-body terms and $k$ of
the positive $p$-body terms in $\Delta_{p+1}$ are sucked up by $I_p$ -- leaving
only $k$ positive $p$-body terms and $k$ negative $(p{-}1)$-body terms in
$I_{p-1}$. This proves the claim \eqref{eq:constr-inductive-proof} for
$p'=p-1\ge 2$.

The situation changes slightly when eliminating terms from $I_2$ which
consists only of two- and three-body terms. Just as before, the three-body
terms must be compensated by terms in $\Delta_1$ but the two-body entropies
are in the output space and can therefore be left untouched. Assuming that $m$
of the two-body terms are nevertheless cancelled, $I_1$ is of the form
%
\begin{alignat*}{3}
    I_1 = \sum_{i < j} b_{ij}^+ &H(\X_i,\X_j) &{}-{} \sum_{i < j} b_{ij}^- &H(\X_i,\X_j) &{}-{}  \sum_{i} a_i^+ &H(\X_i) \\
\intertext{with non-negative coefficients fulfilling}
    \sum_{i<j} b_{ij}^+ &= 2k - m, &
    \sum_{i<j} b_{ij}^- &= k - m, &
    \sum_{i}   a_{i}^- &= k.
\end{alignat*}
%

Finally, $\Delta_0$ consists of unconditional mutual informations. This means
that both $I_1$ and $\Delta_0$ are valid Shannon type inequalities  in the
output space and therefore an inequality $0 \le I=I_1+\Delta_0$ can only be
minimal if $\Delta_0=0$ or $I_1 = 0$.  Furthermore, the mutual informations
(having only positive one-body terms) can not be obtained as non-negative
combinations of facets of the other category and must therefore be facets on
their own.

This concludes the proof for the output space that includes \emph{all} one-
and two-body entropy terms.
Furthermore, this result extends naturally to the local cone of
\emph{simultaneously accessible} quantities in a multipartite Bell scenario.
These inequalities make up a subset of the general case by grouping the $\X_i$
into parties $P_j = (\X_{2j},\X_{2j+1})$ and forcing that no two-body terms of a
single party may be present. In order to achieve this, two issues must be
considered:

First, $\Delta_1$ must cancel all negative two-body terms in $I_2$ if they
contain variables of the same party. This puts an additional constraint on
$\Delta_1$ but doesn't affect the argument otherwise.

Second, we must take care for inaccessible two-body terms added by $\Delta_1$
because they would need to be cancelled by a non-zero $\Delta_0$ term. Without
loss of generality this means that the expansion
\eqref{eq:elemental-expansion2} for $I$ contains a subexpression of the form:
\begin{align*}
    I&(\X_2:\X_3\Mid\X_1) + I(\X_1:\X_2) \\
    &= - H(\X_1,\X_2,\X_3) + H(\X_1,\X_3) + H(\X_2).
\end{align*}

However, observe that the resulting inequality can not be minimal since an
alternative expansion that eliminates the same three-body terms but doesn't
require a non-zero $\Delta_0$ is
%
\begin{align*}
    I&(\X_1:\X_2\Mid\X_3) \\
    &= - H(\X_1,\X_2,\X_3) + H(\X_1,\X_3) + H(\X_2,\X_3) - H(\X_3) \\
     &\le - H(\X_1,\X_2,\X_3) + H(\X_1,\X_3) + H(\X_2) \\
     &= \phantom{-{}} I(\X_2:\X_3\Mid\X_1) + I(\X_1:\X_2).
\end{align*}
%
The inequality sign in the second line is a consequence of the independence
bound $H(\X_2,\X_3) \le H(\X_2)+H(\X_3)$.


\subsection{Bounding \texorpdfstring{$k$}{k}}

Looking at the obtained inequalities, we can see that that for all discovered facets, expansions in the
sense of \eqref{eq:elemental-expansion} were available with relatively low
coefficients $c \lesssim 3$. For most quantities even $c=1$. That is, when eliminating
$v$ variables from a system of linear inequalities, every minimal inequality
in the resulting system can be obtained as a linear combination of at most
$v+1$ of the original facets. This observation suggests upper bounds on the maximum value of $k$.

\subsection{Enumerating constraints}
Based on the insights, one can implement an algorithm that enumerates all possible linear combinations leading to constraints in the desired marginal space. This is achieved in a top-down manner where at each step we only need to consider those combinations of CMIs that eliminate the according terms of the current step. This enumeration can be done exhaustively only for low $k$ (number of elemental forms at each layer).

To show the practical relevance of these observations, we use this method to ensure that the listed descriptions of the 18D and 26D cones contain every facet with $k\le 4$. For the 26D cone this allowed
to find 14 of the 37 facets discovered in total. Note that this works more efficiently for the 26D cone than for the 18D, since this requires one less recursion.

\end{document}

%% file: qr/wittab.tex
\multirow{4}{*}{$\Qunits{2}$} & none & \multicolumn{1}{l}{\begin{adjustbox}{valign=t}$\kern-\nulldelimiterspace\begin{alignedat}[l]{4}&I_{9},\  & &I_{10},\  & &I_{11},\  & &I_{12}\end{alignedat}$\end{adjustbox}} && \multicolumn{1}{l}{\begin{adjustbox}{valign=t}$\kern-\nulldelimiterspace\begin{alignedat}[l]{4}&I_{5},\  & &I_{6},\  & &I_{7},\  & &I_{8},\\ &I_{9},\  & &I_{10},\  & &I_{15},\  & &I_{17},\\ &I_{18},\  & &I_{20},\  & &I_{21}\end{alignedat}$\end{adjustbox}} && \multicolumn{1}{l}{\begin{adjustbox}{valign=t}$\kern-\nulldelimiterspace\begin{alignedat}[l]{4}&I_{3},\  & &I_{4},\  & &I_{7},\  & &I_{10},\\ &I_{13},\  & &I_{18}\end{alignedat}$\end{adjustbox}} && \multicolumn{1}{l}{\begin{adjustbox}{valign=t}$\kern-\nulldelimiterspace\begin{alignedat}[l]{4}&I_{8},\  & &I_{11},\  & &I_{12},\  & &I_{13},\\ &I_{14},\  & &I_{15},\  & &I_{16}\end{alignedat}$\end{adjustbox}} && \multicolumn{1}{l}{\begin{adjustbox}{valign=t}$\kern-\nulldelimiterspace\begin{alignedat}[l]{5}&I_{1},\  & &I_{17},\  & &I_{23},\  & &I_{24},\  & &I_{26},\\ &I_{27},\  & &I_{28},\  & &I_{29},\  & &I_{31},\  & &I_{32},\\ &I_{34},\  & &I_{35},\  & &I_{36}\end{alignedat}$\end{adjustbox}}\\
\cmidrule{2-11}
                     & \CHSHE &  && \multicolumn{1}{l}{\begin{adjustbox}{valign=t}$\kern-\nulldelimiterspace\begin{alignedat}[l]{4}&I_{6},\  & &I_{7},\  & &I_{9},\  & &I_{10},\\ &I_{15},\  & &I_{17},\  & &I_{18},\  & &I_{21}\end{alignedat}$\end{adjustbox}} &&  && \multicolumn{1}{l}{\begin{adjustbox}{valign=t}$\kern-\nulldelimiterspace\begin{alignedat}[l]{4}&I_{8},\  & &I_{11},\  & &I_{12},\  & &I_{13},\\ &I_{14},\  & &I_{15},\  & &I_{16}\end{alignedat}$\end{adjustbox}} && \multicolumn{1}{l}{\begin{adjustbox}{valign=t}$\kern-\nulldelimiterspace\begin{alignedat}[l]{5}&I_{17},\  & &I_{23},\  & &I_{26},\  & &I_{27},\  & &I_{28},\\ &I_{29},\  & &I_{31},\  & &I_{32},\  & &I_{34},\  & &I_{35},\\ &I_{36}\end{alignedat}$\end{adjustbox}}\\
\cmidrule{2-11}
                     & CHSH &  && \multicolumn{1}{l}{\begin{adjustbox}{valign=t}$\kern-\nulldelimiterspace\begin{alignedat}[l]{4}&I_{17},\  & &I_{18},\  & &I_{21}\end{alignedat}$\end{adjustbox}} &&  && \multicolumn{1}{l}{\begin{adjustbox}{valign=t}$\kern-\nulldelimiterspace\begin{alignedat}[l]{4}&I_{8},\  & &I_{11},\  & &I_{12},\  & &I_{13},\\ &I_{14},\  & &I_{15},\  & &I_{16}\end{alignedat}$\end{adjustbox}} && \multicolumn{1}{l}{\begin{adjustbox}{valign=t}$\kern-\nulldelimiterspace\begin{alignedat}[l]{5}&I_{17},\  & &I_{23},\  & &I_{26},\  & &I_{27},\  & &I_{28},\\ &I_{29},\  & &I_{34},\  & &I_{35},\  & &I_{36}\end{alignedat}$\end{adjustbox}}\\
\cmidrule{2-11}
                     & PPT &  && \multicolumn{1}{l}{\begin{adjustbox}{valign=t}$\kern-\nulldelimiterspace\begin{alignedat}[l]{4}&I_{18}\end{alignedat}$\end{adjustbox}} &&  && \multicolumn{1}{l}{\begin{adjustbox}{valign=t}$\kern-\nulldelimiterspace\begin{alignedat}[l]{4}&I_{8},\  & &I_{11},\  & &I_{12},\  & &I_{13},\\ &I_{14},\  & &I_{15}\end{alignedat}$\end{adjustbox}} && \multicolumn{1}{l}{\begin{adjustbox}{valign=t}$\kern-\nulldelimiterspace\begin{alignedat}[l]{5}&I_{17},\  & &I_{26},\  & &I_{27},\  & &I_{28},\  & &I_{29},\\ &I_{34},\  & &I_{35},\  & &I_{36}\end{alignedat}$\end{adjustbox}}\\
\midrule
\multirow{3}{*}{$\Qunits{3}$} & none & \multicolumn{1}{l}{\begin{adjustbox}{valign=t}$\kern-\nulldelimiterspace\begin{alignedat}[l]{4}&I_{9},\  & &I_{10},\  & &I_{11},\  & &I_{12}\end{alignedat}$\end{adjustbox}} && \multicolumn{1}{l}{\begin{adjustbox}{valign=t}$\kern-\nulldelimiterspace\begin{alignedat}[l]{4}&I_{5},\  & &I_{6},\  & &I_{7},\  & &I_{8},\\ &I_{9},\  & &I_{10},\  & &I_{15},\  & &I_{17},\\ &I_{18},\  & &I_{20},\  & &I_{21}\end{alignedat}$\end{adjustbox}} && \multicolumn{1}{l}{\begin{adjustbox}{valign=t}$\kern-\nulldelimiterspace\begin{alignedat}[l]{4}&I_{3},\  & &I_{4},\  & &I_{7},\  & &I_{10},\\ &I_{13},\  & &I_{18}\end{alignedat}$\end{adjustbox}} && \multicolumn{1}{l}{\begin{adjustbox}{valign=t}$\kern-\nulldelimiterspace\begin{alignedat}[l]{4}&I_{8},\  & &I_{11},\  & &I_{12},\  & &I_{13},\\ &I_{14},\  & &I_{15},\  & &I_{16}\end{alignedat}$\end{adjustbox}} && \multicolumn{1}{l}{\begin{adjustbox}{valign=t}$\kern-\nulldelimiterspace\begin{alignedat}[l]{5}&I_{1},\  & &I_{8},\  & &I_{12},\  & &I_{17},\  & &I_{23},\\ &I_{24},\  & &I_{26},\  & &I_{27},\  & &I_{28},\  & &I_{29},\\ &I_{31},\  & &I_{32},\  & &I_{34},\  & &I_{35},\  & &I_{36}\end{alignedat}$\end{adjustbox}}\\
\cmidrule{2-11}
                     & CGLMP & \multicolumn{1}{l}{\begin{adjustbox}{valign=t}$\kern-\nulldelimiterspace\begin{alignedat}[l]{4}&I_{9},\  & &I_{10},\  & &I_{11},\  & &I_{12}\end{alignedat}$\end{adjustbox}} && \multicolumn{1}{l}{\begin{adjustbox}{valign=t}$\kern-\nulldelimiterspace\begin{alignedat}[l]{4}&I_{5},\  & &I_{6},\  & &I_{7},\  & &I_{8},\\ &I_{9},\  & &I_{10},\  & &I_{15},\  & &I_{17},\\ &I_{18},\  & &I_{20},\  & &I_{21}\end{alignedat}$\end{adjustbox}} && \multicolumn{1}{l}{\begin{adjustbox}{valign=t}$\kern-\nulldelimiterspace\begin{alignedat}[l]{4}&I_{3},\  & &I_{4},\  & &I_{7},\  & &I_{10},\\ &I_{13},\  & &I_{18}\end{alignedat}$\end{adjustbox}} && \multicolumn{1}{l}{\begin{adjustbox}{valign=t}$\kern-\nulldelimiterspace\begin{alignedat}[l]{4}&I_{8},\  & &I_{11},\  & &I_{12},\  & &I_{13},\\ &I_{14},\  & &I_{15},\  & &I_{16}\end{alignedat}$\end{adjustbox}} && \multicolumn{1}{l}{\begin{adjustbox}{valign=t}$\kern-\nulldelimiterspace\begin{alignedat}[l]{5}&I_{1},\  & &I_{8},\  & &I_{12},\  & &I_{17},\  & &I_{23},\\ &I_{24},\  & &I_{26},\  & &I_{27},\  & &I_{28},\  & &I_{29},\\ &I_{31},\  & &I_{32},\  & &I_{34},\  & &I_{35},\  & &I_{36}\end{alignedat}$\end{adjustbox}}\\
\cmidrule{2-11}
                     & \CHSHE &  && \multicolumn{1}{l}{\begin{adjustbox}{valign=t}$\kern-\nulldelimiterspace\begin{alignedat}[l]{4}&I_{5},\  & &I_{6},\  & &I_{7},\  & &I_{8},\\ &I_{9},\  & &I_{10},\  & &I_{15},\  & &I_{17},\\ &I_{18},\  & &I_{21}\end{alignedat}$\end{adjustbox}} && \multicolumn{1}{l}{\begin{adjustbox}{valign=t}$\kern-\nulldelimiterspace\begin{alignedat}[l]{4}&I_{4},\  & &I_{7},\  & &I_{10},\  & &I_{18}\end{alignedat}$\end{adjustbox}} && \multicolumn{1}{l}{\begin{adjustbox}{valign=t}$\kern-\nulldelimiterspace\begin{alignedat}[l]{4}&I_{8},\  & &I_{11},\  & &I_{12},\  & &I_{13},\\ &I_{14},\  & &I_{15},\  & &I_{16}\end{alignedat}$\end{adjustbox}} && \multicolumn{1}{l}{\begin{adjustbox}{valign=t}$\kern-\nulldelimiterspace\begin{alignedat}[l]{5}&I_{8},\  & &I_{12},\  & &I_{17},\  & &I_{23},\  & &I_{24},\\ &I_{26},\  & &I_{27},\  & &I_{28},\  & &I_{29},\  & &I_{31},\\ &I_{32},\  & &I_{34},\  & &I_{35},\  & &I_{36}\end{alignedat}$\end{adjustbox}}\\
\cmidrule{2-11}
                     & PPT &  && \multicolumn{1}{l}{\begin{adjustbox}{valign=t}$\kern-\nulldelimiterspace\begin{alignedat}[l]{4}&I_{17}\end{alignedat}$\end{adjustbox}} &&  && \multicolumn{1}{l}{\begin{adjustbox}{valign=t}$\kern-\nulldelimiterspace\begin{alignedat}[l]{4}&I_{8},\  & &I_{12},\  & &I_{14},\  & &I_{15}\end{alignedat}$\end{adjustbox}} && \multicolumn{1}{l}{\begin{adjustbox}{valign=t}$\kern-\nulldelimiterspace\begin{alignedat}[l]{5}&I_{17},\  & &I_{26},\  & &I_{27},\  & &I_{28},\  & &I_{29},\\ &I_{34},\  & &I_{35},\  & &I_{36}\end{alignedat}$\end{adjustbox}}\\

%% file: ineqs/bell-08D.tex
\begin{dgroup*}
\begin{dmath*}
  I_{0} = H(\A_1\,\B_1\,\C_2)
\end{dmath*}
\begin{dmath*}
  I_{1} = H(\A_1\,\B_2\,\C_1) + H(\A_2\,\B_1\,\C_2) - H(\A_1\,\B_2\,\C_2)
\end{dmath*}
\begin{dmath*}
  I_{2} = H(\A_1\,\B_1\,\C_2) + H(\A_2\,\B_2\,\C_2) - H(\A_2\,\B_1\,\C_2)
\end{dmath*}
\begin{dmath*}
  I_{3} = H(\A_1\,\B_2\,\C_2) + H(\A_2\,\B_1\,\C_2) + H(\A_2\,\B_2\,\C_1) - H(\A_1\,\B_1\,\C_1) - H(\A_2\,\B_2\,\C_2)
\end{dmath*}
\begin{dmath*}
  I_{4} = H(\A_1\,\B_1\,\C_2) + H(\A_1\,\B_2\,\C_1) + H(\A_2\,\B_2\,\C_2) - 2 H(\A_1\,\B_2\,\C_2)
\end{dmath*}
\begin{dmath*}
  I_{5} = H(\A_1\,\B_1\,\C_2) + H(\A_1\,\B_2\,\C_1) + H(\A_2\,\B_2\,\C_2) - H(\A_1\,\B_2\,\C_2) - H(\A_2\,\B_2\,\C_1)
\end{dmath*}
\begin{dmath*}
  I_{6} = 2 H(\A_1\,\B_1\,\C_1) + 2 H(\A_2\,\B_1\,\C_2) + 2 H(\A_2\,\B_2\,\C_1) - H(\A_1\,\B_1\,\C_2) - H(\A_1\,\B_2\,\C_1) - H(\A_2\,\B_1\,\C_1) - H(\A_2\,\B_2\,\C_2)
\end{dmath*}
\end{dgroup*}

%% file: ineqs/bell-12D.tex
\begin{dgroup*}
\begin{dmath*}
  I_{0} = H(\B_1\,\C_2) + H(\B_2\,\C_1) - H(\B_2\,\C_2)
\end{dmath*}
\begin{dmath*}
  I_{1} = H(\A_1\,\B_2) + H(\B_1\,\C_2) - H(\A_1\,\C_2)
\end{dmath*}
\begin{dmath*}
  I_{2} = H(\A_1\,\B_2) + H(\B_1\,\C_2) - H(\B_2\,\C_2)
\end{dmath*}
\begin{dmath*}
  I_{3} = H(\A_1\,\B_1) + H(\B_1\,\C_2) - H(\A_1\,\C_2)
\end{dmath*}
\begin{dmath*}
  I_{4} = H(\A_1\,\B_1) + H(\A_2\,\B_2) + H(\B_2\,\C_1) - H(\A_1\,\B_2) - H(\A_2\,\C_1)
\end{dmath*}
\begin{dmath*}
  I_{5} = H(\A_1\,\B_1) + H(\A_2\,\C_2) + H(\B_1\,\C_2) - H(\A_2\,\B_1) - H(\A_1\,\C_2)
\end{dmath*}
\begin{dmath*}
  I_{6} = H(\A_1\,\B_1) + H(\A_2\,\C_2) + H(\B_2\,\C_2) - H(\A_2\,\B_2) - H(\A_1\,\C_2)
\end{dmath*}
\begin{dmath*}
  I_{7} = H(\A_1\,\B_1) + H(\A_2\,\B_2) + H(\A_2\,\C_2) + H(\B_1\,\C_2) - H(\A_2\,\B_1) - H(\A_1\,\C_2) - H(\B_2\,\C_2)
\end{dmath*}
\begin{dmath*}
  I_{8} = H(\A_1\,\B_1) + H(\A_2\,\C_1) + H(\B_1\,\C_2) + H(\B_2\,\C_1) - H(\A_2\,\B_2) - H(\A_1\,\C_2) - H(\B_1\,\C_1)
\end{dmath*}
\begin{dmath*}
  I_{9} = H(\A_1\,\B_1) + H(\A_2\,\B_1) + H(\A_2\,\B_2) + H(\A_1\,\C_1) + H(\A_1\,\C_2) + H(\B_2\,\C_1) + H(\B_2\,\C_2) - 3 H(\A_1\,\B_2) - H(\A_2\,\C_1) - H(\B_1\,\C_2)
\end{dmath*}
\begin{dmath*}
  I_{10} = 2 H(\A_1\,\B_1) + 2 H(\B_1\,\C_2) + H(\A_1\,\C_1) + H(\A_2\,\C_1) + H(\A_2\,\C_2) - 3 H(\A_1\,\C_2) - H(\A_2\,\B_1) - H(\B_1\,\C_1)
\end{dmath*}
\begin{dmath*}
  I_{11} = 3 H(\A_1\,\B_1) + 2 H(\B_1\,\C_2) + H(\A_1\,\C_1) + H(\A_2\,\C_1) + H(\A_2\,\C_2) + H(\B_2\,\C_1) + H(\B_2\,\C_2) - 4 H(\A_1\,\C_2) - 2 H(\B_1\,\C_1) - H(\A_2\,\B_2)
\end{dmath*}
\begin{dmath*}
  I_{12} = 2 H(\A_2\,\B_2) + 2 H(\A_1\,\C_1) + H(\A_1\,\B_2) + H(\A_2\,\B_1) + H(\A_1\,\C_2) + H(\A_2\,\C_1) + H(\B_1\,\C_1) + H(\B_1\,\C_2) + H(\B_2\,\C_2) - 4 H(\B_2\,\C_1) - 2 H(\A_1\,\B_1) - 2 H(\A_2\,\C_2)
\end{dmath*}
\begin{dmath*}
  I_{13} = 3 H(\B_2\,\C_2) + 2 H(\A_2\,\B_2) + 2 H(\A_1\,\C_2) + H(\A_1\,\B_1) + H(\A_2\,\B_1) + H(\A_1\,\C_1) + H(\A_2\,\C_1) - 4 H(\A_2\,\C_2) - 3 H(\A_1\,\B_2) - H(\B_1\,\C_1)
\end{dmath*}
\end{dgroup*}

%% file: ineqs/bell-14D.tex
\begin{dgroup*}
\begin{dmath*}
  I_{0} = H(\B_1)
\end{dmath*}
\begin{dmath*}
  I_{1} = H(\A_1) + H(\B_1) + H(\C_2) - H(\A_1\,\B_1\,\C_2)
\end{dmath*}
\begin{dmath*}
  I_{2} = H(\A_2\,\B_2\,\C_2) - H(\B_2)
\end{dmath*}
\begin{dmath*}
  I_{3} = H(\A_2\,\B_1\,\C_2) + H(\A_1) - H(\A_1\,\B_1\,\C_2)
\end{dmath*}
\begin{dmath*}
  I_{4} = H(\A_1\,\B_2\,\C_2) + H(\A_2\,\B_2\,\C_1) - H(\A_1\,\B_2\,\C_1) - H(\B_2)
\end{dmath*}
\begin{dmath*}
  I_{5} = H(\A_1\,\B_2\,\C_2) + H(\A_2\,\B_1\,\C_2) + H(\A_2\,\B_2\,\C_2) - H(\A_1\,\B_1\,\C_2) - H(\A_2) - H(\B_2)
\end{dmath*}
\begin{dmath*}
  I_{6} = H(\A_1\,\B_2\,\C_1) + H(\A_2\,\B_1\,\C_1) + H(\A_2\,\B_2\,\C_2) - H(\A_1\,\B_1\,\C_2) - H(\B_2) - H(\C_1)
\end{dmath*}
\begin{dmath*}
  I_{7} = H(\A_1\,\B_1\,\C_1) + H(\A_2\,\B_2\,\C_1) + H(\A_2\,\B_2\,\C_2) - H(\A_1\,\B_1\,\C_2) - H(\B_2) - H(\C_1)
\end{dmath*}
\begin{dmath*}
  I_{8} = H(\A_1\,\B_1\,\C_1) + H(\A_2\,\B_2\,\C_1) + H(\A_2\,\B_2\,\C_2) - H(\A_2\,\B_1\,\C_2) - H(\B_2) - H(\C_1)
\end{dmath*}
\begin{dmath*}
  I_{9} = 2 H(\A_2\,\B_2\,\C_2) + H(\A_1\,\B_2\,\C_1) + H(\A_2\,\B_1\,\C_1) - H(\A_1\,\B_1\,\C_1) - H(\A_2\,\B_2\,\C_1) - H(\A_2) - H(\B_2)
\end{dmath*}
\begin{dmath*}
  I_{10} = 2 H(\A_2\,\B_2\,\C_2) + H(\A_1\,\B_2\,\C_1) + H(\A_2\,\B_1\,\C_1) - H(\A_1\,\B_1\,\C_2) - H(\A_2\,\B_2\,\C_1) - H(\A_2) - H(\B_2)
\end{dmath*}
\begin{dmath*}
  I_{11} = H(\A_1\,\B_1\,\C_1) + H(\A_1\,\B_2\,\C_2) + H(\A_2\,\B_1\,\C_2) + H(\A_2\,\B_2\,\C_2) + H(\C_2) - H(\A_1\,\B_1\,\C_2) - H(\A_1\,\B_2\,\C_1) - H(\A_2) - H(\B_2)
\end{dmath*}
\begin{dmath*}
  I_{12} = H(\A_1\,\B_1\,\C_1) + H(\A_1\,\B_2\,\C_2) + H(\A_2\,\B_1\,\C_1) + H(\A_2\,\B_1\,\C_2) + H(\B_1) - H(\A_1\,\B_1\,\C_2) - H(\A_2\,\B_2\,\C_1) - H(\A_2) - H(\C_1)
\end{dmath*}
\begin{dmath*}
  I_{13} = 2 H(\A_1\,\B_1\,\C_1) + H(\A_1\,\B_2\,\C_2) + H(\A_2\,\B_1\,\C_1) + H(\A_2\,\B_1\,\C_2) - H(\A_1\,\B_1\,\C_2) - H(\A_2\,\B_2\,\C_1) - H(\A_1) - H(\A_2) - H(\C_1)
\end{dmath*}
\begin{dmath*}
  I_{14} = 2 H(\A_1\,\B_1\,\C_1) + H(\A_1\,\B_2\,\C_2) + H(\A_2\,\B_1\,\C_1) + H(\A_2\,\B_1\,\C_2) - H(\A_1\,\B_1\,\C_2) - H(\A_2\,\B_2\,\C_2) - H(\A_1) - H(\A_2) - H(\C_1)
\end{dmath*}
\begin{dmath*}
  I_{15} = 2 H(\A_2\,\B_1\,\C_1) + H(\A_1\,\B_2\,\C_1) + H(\A_1\,\B_2\,\C_2) + H(\A_2\,\B_2\,\C_2) - 2 H(\A_1\,\B_1\,\C_2) - H(\A_2) - H(\B_2) - H(\C_1)
\end{dmath*}
\begin{dmath*}
  I_{16} = H(\A_1\,\B_1\,\C_1) + H(\A_1\,\B_2\,\C_2) + H(\A_2\,\B_1\,\C_1) + H(\A_2\,\B_1\,\C_2) + H(\A_2\,\B_2\,\C_2) - H(\A_1\,\B_1\,\C_2) - H(\A_1\,\B_2\,\C_1) - H(\A_2) - H(\B_2) - H(\C_1)
\end{dmath*}
\begin{dmath*}
  I_{17} = 2 H(\A_1\,\B_1\,\C_1) + 2 H(\A_1\,\B_2\,\C_2) + 2 H(\A_2\,\B_2\,\C_1) - 2 H(\A_2\,\B_1\,\C_2) - H(\A_1\,\B_2\,\C_1) - H(\A_1) - H(\B_2) - H(\C_1)
\end{dmath*}
\begin{dmath*}
  I_{18} = H(\A_1\,\B_1\,\C_1) + H(\A_1\,\B_2\,\C_1) + H(\A_1\,\B_2\,\C_2) + H(\A_2\,\B_1\,\C_1) + H(\A_2\,\B_1\,\C_2) + H(\A_2\,\B_2\,\C_2) - 3 H(\A_1\,\B_1\,\C_2) - H(\A_2) - H(\B_2) - H(\C_1)
\end{dmath*}
\begin{dmath*}
  I_{19} = H(\A_1\,\B_1\,\C_1) + H(\A_1\,\B_2\,\C_1) + H(\A_1\,\B_2\,\C_2) + H(\A_2\,\B_1\,\C_1) + H(\A_2\,\B_1\,\C_2) + H(\A_2\,\B_2\,\C_2) - 2 H(\A_2\,\B_2\,\C_1) - H(\A_1\,\B_1\,\C_2) - H(\A_2) - H(\B_2) - H(\C_1)
\end{dmath*}
\begin{dmath*}
  I_{20} = 2 H(\A_1\,\B_1\,\C_1) + 2 H(\A_1\,\B_2\,\C_2) + 2 H(\A_2\,\B_1\,\C_2) + H(\A_2\,\B_1\,\C_1) - 2 H(\A_2\,\B_2\,\C_1) - H(\A_1\,\B_1\,\C_2) - H(\A_1) - H(\A_2) - H(\C_1) - H(\C_2)
\end{dmath*}
\begin{dmath*}
  I_{21} = 2 H(\A_1\,\B_2\,\C_2) + 2 H(\A_2\,\B_2\,\C_1) + H(\A_1\,\B_1\,\C_1) + H(\A_2\,\B_1\,\C_1) + H(\A_2\,\B_1\,\C_2) - 3 H(\A_1\,\B_1\,\C_2) - 2 H(\B_2) - H(\A_2) - H(\C_1)
\end{dmath*}
\end{dgroup*}

%% file: ineqs/bell-18D.tex
\begin{dgroup*}
\begin{dmath*}
  I_{0} = H(\B_1) + H(\C_1) - H(\B_1\,\C_1)
\end{dmath*}
\begin{dmath*}
  I_{1} = H(\A_1\,\B_2) - H(\B_2)
\end{dmath*}
\begin{dmath*}
  I_{2} = H(\A_1\,\B_2) + H(\A_1\,\C_1) - H(\B_2\,\C_1) - H(\A_1)
\end{dmath*}
\begin{dmath*}
  I_{3} = H(\A_1\,\C_1) + H(\A_2\,\C_1) + H(\A_2\,\C_2) - H(\A_1\,\C_2) - H(\A_2) - H(\C_1)
\end{dmath*}
\begin{dmath*}
  I_{4} = H(\A_1\,\B_2) + H(\A_1\,\C_2) + H(\A_2\,\C_1) + H(\A_2\,\C_2) + H(\B_2\,\C_1) - H(\A_1\,\C_1) - H(\B_2\,\C_2) - H(\A_1) - H(\A_2) - H(\C_1)
\end{dmath*}
\begin{dmath*}
  I_{5} = H(\A_1\,\B_2) + H(\A_1\,\C_1) + H(\A_2\,\C_1) + H(\A_2\,\C_2) + H(\B_2\,\C_2) - H(\A_2\,\B_2) - H(\A_1\,\C_2) - H(\A_2) - H(\B_2) - H(\C_1)
\end{dmath*}
\begin{dmath*}
  I_{6} = H(\A_1\,\B_2) + H(\A_2\,\B_2) + H(\A_1\,\C_1) + H(\A_2\,\C_1) + H(\A_2\,\C_2) - H(\A_1\,\C_2) - H(\B_2\,\C_1) - H(\A_2) - H(\B_2) - H(\C_1)
\end{dmath*}
\begin{dmath*}
  I_{7} = H(\A_1\,\C_1) + H(\A_1\,\C_2) + H(\A_2\,\C_1) + H(\B_1\,\C_1) + H(\B_1\,\C_2) - H(\A_1\,\B_1) - H(\A_2\,\C_2) - 2 H(\C_1) - H(\C_2)
\end{dmath*}
\begin{dmath*}
  I_{8} = H(\A_1\,\B_1) + H(\A_1\,\B_2) + H(\A_2\,\B_1) + H(\A_2\,\B_2) + H(\A_1\,\C_2) + H(\B_1\,\C_1) - H(\A_2\,\C_2) - H(\B_2\,\C_1) - 2 H(\A_1) - H(\A_2) - H(\B_1)
\end{dmath*}
\begin{dmath*}
  I_{9} = H(\A_1\,\B_2) + H(\A_2\,\B_2) + H(\A_1\,\C_2) + H(\A_2\,\C_1) + H(\B_1\,\C_1) + H(\B_1\,\C_2) - H(\A_1\,\C_1) - H(\A_2\,\C_2) - H(\A_1) - H(\B_1) - H(\B_2) - H(\C_1)
\end{dmath*}
\begin{dmath*}
  I_{10} = H(\A_1\,\B_1) + H(\A_1\,\B_2) + H(\A_2\,\B_2) + H(\A_1\,\C_2) + H(\A_2\,\C_1) + H(\A_2\,\C_2) + H(\B_1\,\C_1) - H(\A_2\,\B_1) - H(\A_1\,\C_1) - H(\B_2\,\C_2) - 2 H(\A_1) - H(\A_2) - H(\C_1)
\end{dmath*}
\begin{dmath*}
  I_{11} = H(\A_1\,\B_2) + H(\A_2\,\B_1) + H(\A_2\,\B_2) + H(\A_1\,\C_2) + H(\B_1\,\C_1) + H(\B_1\,\C_2) + H(\B_2\,\C_1) - H(\A_1\,\B_1) - H(\A_2\,\C_1) - H(\B_2\,\C_2) - H(\A_1) - H(\A_2) - H(\B_1) - H(\C_1)
\end{dmath*}
\begin{dmath*}
  I_{12} = H(\A_1\,\B_1) + H(\A_1\,\B_2) + H(\A_2\,\B_2) + H(\A_2\,\C_1) + H(\A_2\,\C_2) + H(\B_1\,\C_1) + H(\B_1\,\C_2) - H(\A_2\,\B_1) - H(\A_1\,\C_1) - H(\B_2\,\C_2) - H(\A_1) - H(\A_2) - H(\B_1) - H(\C_1)
\end{dmath*}
\begin{dmath*}
  I_{13} = H(\A_1\,\B_2) + H(\A_2\,\B_2) + H(\A_1\,\C_2) + H(\A_2\,\C_1) + H(\A_2\,\C_2) + H(\B_1\,\C_1) + H(\B_1\,\C_2) - H(\A_2\,\B_1) - H(\A_1\,\C_1) - H(\B_2\,\C_2) - H(\A_1) - H(\A_2) - H(\C_1) - H(\C_2)
\end{dmath*}
\begin{dmath*}
  I_{14} = H(\A_1\,\B_2) + H(\A_2\,\B_1) + H(\A_2\,\B_2) + H(\A_1\,\C_1) + H(\B_1\,\C_1) + H(\B_1\,\C_2) + H(\B_2\,\C_2) - H(\A_1\,\C_2) - H(\A_2\,\C_2) - H(\B_2\,\C_1) - 2 H(\B_1) - H(\A_1) - H(\B_2)
\end{dmath*}
\begin{dmath*}
  I_{15} = H(\A_1\,\B_2) + H(\A_2\,\B_1) + H(\A_2\,\B_2) + H(\A_1\,\C_1) + H(\A_1\,\C_2) + H(\B_1\,\C_1) + H(\B_1\,\C_2) - H(\A_1\,\B_1) - H(\A_2\,\C_1) - H(\A_2\,\C_2) - H(\A_1) - H(\B_1) - H(\B_2) - H(\C_1)
\end{dmath*}
\begin{dmath*}
  I_{16} = H(\A_1\,\B_2) + H(\A_2\,\B_1) + H(\A_2\,\B_2) + H(\B_1\,\C_1) + H(\B_1\,\C_2) + H(\B_2\,\C_1) + H(\B_2\,\C_2) - H(\A_1\,\B_1) - H(\A_2\,\C_1) - H(\A_2\,\C_2) - 2 H(\B_2) - H(\B_1) - H(\C_1)
\end{dmath*}
\begin{dmath*}
  I_{17} = 2 H(\B_1\,\C_2) + H(\A_1\,\B_2) + H(\A_2\,\B_1) + H(\A_2\,\B_2) + H(\A_1\,\C_2) + H(\B_1\,\C_1) + H(\B_2\,\C_1) - H(\A_1\,\B_1) - H(\A_2\,\C_1) - H(\A_2\,\C_2) - 2 H(\B_1) - H(\A_1) - H(\B_2) - H(\C_1)
\end{dmath*}
\begin{dmath*}
  I_{18} = 2 H(\A_1\,\B_2) + 2 H(\B_1\,\C_1) + H(\A_1\,\B_1) + H(\A_2\,\B_2) + H(\A_1\,\C_2) + H(\A_2\,\C_1) + H(\A_2\,\C_2) + H(\B_1\,\C_2) + H(\B_2\,\C_1) - 2 H(\A_1\,\C_1) - 2 H(\B_2\,\C_2) - H(\A_2\,\B_1) - 2 H(\A_1) - 2 H(\C_1) - H(\A_2) - H(\B_1)
\end{dmath*}
\begin{dmath*}
  I_{19} = 2 H(\A_2\,\B_2) + 2 H(\A_2\,\C_2) + 2 H(\B_1\,\C_1) + H(\A_1\,\B_1) + H(\A_1\,\B_2) + H(\A_1\,\C_2) + H(\A_2\,\C_1) + H(\B_1\,\C_2) + H(\B_2\,\C_1) - 2 H(\A_2\,\B_1) - 2 H(\A_1\,\C_1) - H(\B_2\,\C_2) - 2 H(\B_2) - 2 H(\C_1) - 2 H(\C_2) - H(\A_1)
\end{dmath*}
\begin{dmath*}
  I_{20} = 2 H(\A_2\,\C_1) + 2 H(\A_2\,\C_2) + 2 H(\B_1\,\C_1) + H(\A_1\,\B_1) + H(\A_1\,\B_2) + H(\A_2\,\B_2) + H(\A_1\,\C_2) + H(\B_1\,\C_2) + H(\B_2\,\C_1) - 2 H(\A_2\,\B_1) - 2 H(\A_1\,\C_1) - H(\B_2\,\C_2) - 3 H(\C_1) - 2 H(\C_2) - H(\A_1) - H(\B_2)
\end{dmath*}
\begin{dmath*}
  I_{21} = 3 H(\B_1\,\C_1) + 2 H(\A_1\,\B_1) + 2 H(\A_1\,\B_2) + 2 H(\A_2\,\B_2) + 2 H(\A_2\,\C_1) + 2 H(\A_2\,\C_2) + H(\A_1\,\C_2) + H(\B_1\,\C_2) + H(\B_2\,\C_1) - 3 H(\A_2\,\B_1) - 3 H(\A_1\,\C_1) - H(\B_2\,\C_2) - 3 H(\C_1) - 2 H(\A_1) - 2 H(\B_2) - 2 H(\C_2)
\end{dmath*}
\end{dgroup*}

%% file: ineqs/bell-20D.tex
\begin{dgroup*}
\begin{dmath*}
  I_{0} = H(\A_2\,\B_2) + H(\A_2\,\C_2) + H(\B_2\,\C_2) - 2 H(\A_2\,\B_2\,\C_2)
\end{dmath*}
\begin{dmath*}
  I_{1} = H(\A_2\,\B_2) + H(\A_2\,\C_1) + H(\B_2\,\C_2) - H(\A_2\,\B_2\,\C_1) - H(\A_2\,\B_2\,\C_2)
\end{dmath*}
\begin{dmath*}
  I_{2} = H(\A_2\,\B_1) + H(\A_1\,\C_1) + H(\B_1\,\C_2) - H(\A_1\,\B_1\,\C_1) - H(\A_2\,\B_1\,\C_2)
\end{dmath*}
\begin{dmath*}
  I_{3} = H(\A_1\,\B_1) + H(\A_2\,\B_2) + H(\A_1\,\C_2) + H(\B_2\,\C_1) - H(\A_1\,\B_1\,\C_2) - H(\A_2\,\B_2\,\C_1) - H(\A_1\,\B_2)
\end{dmath*}
\begin{dmath*}
  I_{4} = H(\A_1\,\B_1) + H(\A_1\,\C_1) + H(\B_1\,\C_2) + H(\B_2\,\C_1) - H(\A_1\,\B_1\,\C_2) - H(\A_1\,\B_2\,\C_1) - H(\B_1\,\C_1)
\end{dmath*}
\begin{dmath*}
  I_{5} = H(\A_2\,\B_1\,\C_1) - H(\A_2\,\C_1)
\end{dmath*}
\begin{dmath*}
  I_{6} = H(\A_1\,\B_1\,\C_2) + 2 H(\A_2\,\B_1) + 2 H(\A_2\,\C_2) + H(\B_1\,\C_1) + H(\B_2\,\C_1) + H(\B_2\,\C_2) - 2 H(\A_2\,\B_1\,\C_2) - H(\A_1\,\B_2\,\C_1) - H(\A_2\,\B_2) - H(\A_2\,\C_1) - H(\B_1\,\C_2)
\end{dmath*}
\begin{dmath*}
  I_{7} = H(\A_2\,\B_1\,\C_2) + 2 H(\A_1\,\B_1) + 2 H(\A_1\,\C_2) + H(\B_1\,\C_1) + H(\B_2\,\C_1) + H(\B_2\,\C_2) - 2 H(\A_1\,\B_1\,\C_2) - H(\A_2\,\B_1\,\C_1) - H(\A_1\,\B_2) - H(\A_1\,\C_1) - H(\B_1\,\C_2)
\end{dmath*}
\begin{dmath*}
  I_{8} = H(\A_1\,\B_1\,\C_1) + H(\A_1\,\B_1\,\C_2) + H(\A_2\,\B_1\,\C_2) - H(\A_2\,\B_1\,\C_1) - H(\A_1\,\B_1) - H(\B_1\,\C_2)
\end{dmath*}
\begin{dmath*}
  I_{9} = H(\A_1\,\B_2\,\C_1) + H(\A_1\,\B_2\,\C_2) + H(\A_2\,\B_2\,\C_2) + 2 H(\A_2\,\B_2) + H(\A_2\,\C_1) + H(\B_1\,\C_1) + H(\B_1\,\C_2) + H(\B_2\,\C_1) - 3 H(\A_2\,\B_2\,\C_1) - H(\A_1\,\B_1\,\C_1) - H(\A_1\,\B_2) - H(\A_2\,\B_1) - H(\B_2\,\C_2)
\end{dmath*}
\begin{dmath*}
  I_{10} = H(\A_1\,\B_1\,\C_2) + H(\A_1\,\B_2\,\C_2) + H(\A_2\,\B_1\,\C_2) + 2 H(\A_2\,\B_2) + 2 H(\A_2\,\C_2) + H(\B_1\,\C_1) + H(\B_2\,\C_1) - 3 H(\A_2\,\B_2\,\C_2) - H(\A_1\,\B_2\,\C_1) - H(\A_2\,\B_1) - H(\A_1\,\C_2) - H(\A_2\,\C_1)
\end{dmath*}
\begin{dmath*}
  I_{11} = H(\A_1\,\B_2\,\C_1) + H(\A_1\,\B_2\,\C_2) + H(\A_2\,\B_1\,\C_1) + H(\A_2\,\B_2\,\C_1) - H(\A_2\,\B_1\,\C_2) - H(\A_1\,\B_2) - H(\A_2\,\C_1) - H(\B_2\,\C_1)
\end{dmath*}
\begin{dmath*}
  I_{12} = H(\A_1\,\B_2\,\C_1) + H(\A_1\,\B_2\,\C_2) + H(\A_2\,\B_1\,\C_1) + H(\A_2\,\B_2\,\C_2) - H(\A_1\,\B_1\,\C_2) - H(\A_1\,\B_2) - H(\A_2\,\C_1) - H(\B_2\,\C_2)
\end{dmath*}
\begin{dmath*}
  I_{13} = H(\A_1\,\B_1\,\C_2) + H(\A_1\,\B_2\,\C_1) + H(\A_2\,\B_1\,\C_2) + H(\A_2\,\B_2\,\C_2) - H(\A_2\,\B_2\,\C_1) - H(\A_1\,\B_2) - H(\A_2\,\C_2) - H(\B_1\,\C_2)
\end{dmath*}
\begin{dmath*}
  I_{14} = H(\A_1\,\B_2\,\C_1) + H(\A_2\,\B_1\,\C_1) + H(\A_2\,\B_2\,\C_1) + H(\A_2\,\B_2\,\C_2) - H(\A_1\,\B_1\,\C_2) - H(\A_2\,\B_2) - H(\A_2\,\C_1) - H(\B_2\,\C_1)
\end{dmath*}
\begin{dmath*}
  I_{15} = 2 H(\A_1\,\B_2\,\C_1) + H(\A_1\,\B_1\,\C_2) + H(\A_1\,\B_2\,\C_2) + H(\A_2\,\B_1\,\C_1) + H(\A_2\,\B_2\,\C_1) - H(\A_1\,\B_1\,\C_1) - H(\A_2\,\B_2\,\C_2) - H(\A_1\,\B_2) - H(\A_1\,\C_2) - H(\A_2\,\C_1) - H(\B_2\,\C_1)
\end{dmath*}
\begin{dmath*}
  I_{16} = 2 H(\A_2\,\B_2\,\C_2) + H(\A_1\,\B_1\,\C_2) + H(\A_1\,\B_2\,\C_1) + H(\A_2\,\B_1\,\C_2) + H(\A_2\,\B_2\,\C_1) - H(\A_1\,\B_1\,\C_1) - H(\A_1\,\B_2\,\C_2) - H(\A_2\,\B_2) - H(\A_2\,\C_2) - H(\B_1\,\C_2) - H(\B_2\,\C_1)
\end{dmath*}
\end{dgroup*}

%% file: ineqs/bell-26D.tex
\begin{dgroup*}
\begin{dmath*}
  I_{0} = H(\B_1) + H(\C_1) - H(\B_1\,\C_1)
\end{dmath*}
\begin{dmath*}
  I_{1} = H(\A_1\,\B_1) + H(\A_1\,\B_2) + H(\A_2\,\B_1) - H(\A_2\,\B_2) - H(\A_1) - H(\B_1)
\end{dmath*}
\begin{dmath*}
  I_{2} = H(\A_1\,\C_2) + H(\B_2\,\C_2) - H(\A_1\,\B_2\,\C_2) - H(\C_2)
\end{dmath*}
\begin{dmath*}
  I_{3} = H(\A_1\,\B_1) + H(\A_2\,\B_1) + H(\A_1\,\C_1) + H(\A_1\,\C_2) + H(\A_2\,\C_1) - H(\A_2\,\B_1\,\C_1) - H(\A_2\,\C_2) - H(\A_1) - H(\B_1) - H(\C_1)
\end{dmath*}
\begin{dmath*}
  I_{4} = H(\A_1\,\B_2) + H(\A_2\,\B_1) + H(\A_2\,\B_2) + H(\A_1\,\C_2) + H(\A_2\,\C_2) - H(\A_2\,\B_2\,\C_2) - H(\A_1\,\B_1) - H(\A_2) - H(\B_2) - H(\C_2)
\end{dmath*}
\begin{dmath*}
  I_{5} = H(\A_1\,\B_2\,\C_2) - H(\B_2\,\C_2)
\end{dmath*}
\begin{dmath*}
  I_{6} = H(\A_2\,\B_1\,\C_1) + H(\A_1\,\B_1) + H(\A_1\,\B_2) + H(\A_2\,\B_2) - H(\A_2\,\B_2\,\C_1) - H(\A_2\,\B_1) - H(\A_1) - H(\B_2)
\end{dmath*}
\begin{dmath*}
  I_{7} = H(\A_1\,\B_2\,\C_2) + H(\A_1\,\C_1) + H(\A_2\,\C_1) + H(\A_2\,\C_2) - H(\A_2\,\B_2\,\C_1) - H(\A_1\,\C_2) - H(\A_2) - H(\C_1)
\end{dmath*}
\begin{dmath*}
  I_{8} = H(\A_1\,\B_2\,\C_2) + H(\A_1\,\B_1) + H(\A_2\,\B_1) + H(\A_1\,\C_1) + H(\A_2\,\C_1) - H(\A_1\,\B_1\,\C_1) - H(\A_2\,\B_2\,\C_2) - H(\A_1) - H(\B_1) - H(\C_1)
\end{dmath*}
\begin{dmath*}
  I_{9} = H(\A_2\,\B_1\,\C_1) + H(\A_1\,\B_1) + H(\A_1\,\B_2) + H(\B_1\,\C_2) + H(\B_2\,\C_2) - H(\A_1\,\B_2\,\C_2) - H(\A_2\,\B_2\,\C_1) - H(\A_1) - H(\B_1) - H(\C_2)
\end{dmath*}
\begin{dmath*}
  I_{10} = H(\A_2\,\B_1\,\C_1) + H(\A_1\,\B_1) + H(\A_1\,\B_2) + H(\A_2\,\B_2) + H(\B_1\,\C_2) + H(\B_2\,\C_2) - H(\A_1\,\B_2\,\C_2) - H(\A_2\,\B_2\,\C_1) - H(\A_2\,\B_1) - H(\A_1) - H(\B_2) - H(\C_2)
\end{dmath*}
\begin{dmath*}
  I_{11} = H(\A_2\,\B_1\,\C_2) + H(\A_2\,\B_1) + H(\A_2\,\B_2) + H(\B_1\,\C_1) + H(\B_2\,\C_1) + H(\B_2\,\C_2) - H(\A_2\,\B_1\,\C_1) - H(\A_2\,\B_2\,\C_2) - H(\B_1\,\C_2) - H(\A_2) - H(\B_2) - H(\C_1)
\end{dmath*}
\begin{dmath*}
  I_{12} = H(\A_1\,\B_1\,\C_1) + H(\A_1\,\B_2) + H(\A_2\,\B_1) + H(\A_2\,\B_2) + H(\B_1\,\C_2) + H(\B_2\,\C_2) - H(\A_1\,\B_2\,\C_1) - H(\A_2\,\B_1\,\C_2) - H(\A_1\,\B_1) - H(\A_2) - H(\B_2) - H(\C_2)
\end{dmath*}
\begin{dmath*}
  I_{13} = H(\A_1\,\B_1\,\C_2) + H(\A_1\,\B_2) + H(\A_2\,\B_1) + H(\A_2\,\B_2) + H(\A_1\,\C_2) + H(\A_2\,\C_2) - H(\A_2\,\B_1\,\C_2) - H(\A_2\,\B_2\,\C_2) - H(\A_1\,\B_1) - H(\A_2) - H(\B_2) - H(\C_2)
\end{dmath*}
\begin{dmath*}
  I_{14} = H(\A_1\,\B_1\,\C_2) + H(\A_2\,\B_1\,\C_2) + H(\A_1\,\B_2) + H(\A_2\,\B_2) + H(\A_1\,\C_1) - H(\A_2\,\B_2\,\C_2) - H(\A_2\,\C_1) - H(\B_1\,\C_2) - H(\A_1) - H(\B_2)
\end{dmath*}
\begin{dmath*}
  I_{15} = H(\A_1\,\B_2\,\C_1) + H(\A_2\,\B_2\,\C_2) + H(\A_1\,\B_1) + H(\A_2\,\B_1) + H(\A_1\,\C_1) + H(\A_1\,\C_2) + H(\A_2\,\C_1) - H(\A_2\,\B_1\,\C_1) - H(\A_2\,\B_2\,\C_1) - H(\A_1\,\B_2) - H(\A_2\,\C_2) - H(\A_1) - H(\B_1) - H(\C_1)
\end{dmath*}
\begin{dmath*}
  I_{16} = H(\A_1\,\B_1\,\C_2) + H(\A_1\,\B_2\,\C_2) + H(\A_1\,\B_1) + H(\A_2\,\B_1) + H(\A_2\,\B_2) + H(\A_1\,\C_1) + H(\A_2\,\C_1) - H(\A_2\,\B_1\,\C_1) - H(\A_2\,\B_1\,\C_2) - H(\A_1\,\B_2) - H(\A_1\,\C_2) - H(\A_2) - H(\B_1) - H(\C_1)
\end{dmath*}
\begin{dmath*}
  I_{17} = H(\A_1\,\B_1\,\C_1) + H(\A_1\,\B_1\,\C_2) + H(\A_2\,\B_1\,\C_1) - H(\A_2\,\B_1\,\C_2) - H(\A_1\,\B_1) - H(\B_1\,\C_1)
\end{dmath*}
\begin{dmath*}
  I_{18} = H(\A_1\,\B_1\,\C_1) + H(\A_1\,\B_2\,\C_1) + H(\A_1\,\B_2\,\C_2) + H(\A_2\,\C_1) + H(\A_2\,\C_2) - H(\A_1\,\B_1\,\C_2) - H(\A_2\,\B_2\,\C_1) - H(\A_1\,\B_2) - H(\A_2) - H(\C_1)
\end{dmath*}
\begin{dmath*}
  I_{19} = H(\A_1\,\B_1\,\C_1) + H(\A_1\,\B_2\,\C_1) + H(\A_1\,\B_2\,\C_2) + H(\A_2\,\C_1) + H(\A_2\,\C_2) - H(\A_1\,\B_1\,\C_2) - H(\A_2\,\B_2\,\C_2) - H(\A_1\,\B_2) - H(\A_2) - H(\C_1)
\end{dmath*}
\begin{dmath*}
  I_{20} = H(\A_1\,\B_1\,\C_1) + H(\A_1\,\B_2\,\C_1) + H(\A_1\,\B_2\,\C_2) + H(\A_1\,\B_1) + H(\A_2\,\B_1) + H(\A_2\,\B_2) - H(\A_1\,\B_1\,\C_2) - H(\A_2\,\B_1\,\C_1) - H(\A_1\,\B_2) - H(\A_1\,\C_1) - H(\A_2) - H(\B_1)
\end{dmath*}
\begin{dmath*}
  I_{21} = H(\A_1\,\B_1\,\C_1) + H(\A_1\,\B_2\,\C_1) + H(\A_1\,\B_2\,\C_2) + H(\A_1\,\B_1) + H(\A_2\,\B_1) + H(\A_2\,\B_2) - H(\A_1\,\B_1\,\C_2) - H(\A_2\,\B_2\,\C_1) - H(\A_1\,\B_2) - H(\A_1\,\C_1) - H(\A_2) - H(\B_1)
\end{dmath*}
\begin{dmath*}
  I_{22} = H(\A_1\,\B_1\,\C_1) + H(\A_1\,\B_2\,\C_1) + H(\A_2\,\B_1\,\C_2) + H(\A_2\,\B_1) + H(\A_2\,\B_2) + H(\B_2\,\C_2) - H(\A_2\,\B_1\,\C_1) - H(\A_2\,\B_2\,\C_2) - H(\A_1\,\C_1) - H(\B_1\,\C_2) - H(\A_2) - H(\B_2)
\end{dmath*}
\begin{dmath*}
  I_{23} = H(\A_1\,\B_1\,\C_2) + H(\A_1\,\B_2\,\C_2) + H(\A_2\,\B_1\,\C_1) + H(\A_2\,\B_1) + H(\A_2\,\B_2) + H(\B_2\,\C_1) - H(\A_2\,\B_2\,\C_1) - H(\A_2\,\B_2\,\C_2) - H(\A_1\,\C_2) - H(\B_1\,\C_1) - H(\A_2) - H(\B_2)
\end{dmath*}
\begin{dmath*}
  I_{24} = H(\A_1\,\B_1\,\C_2) + H(\A_1\,\B_2\,\C_2) + H(\A_2\,\B_1\,\C_1) + H(\A_1\,\B_2) + H(\A_2\,\B_2) + H(\A_1\,\C_1) + H(\A_2\,\C_1) + H(\A_2\,\C_2) - 2 H(\A_2\,\B_2\,\C_1) - 2 H(\A_1\,\C_2) - H(\A_2\,\B_1) - H(\A_2) - H(\B_2) - H(\C_1)
\end{dmath*}
\begin{dmath*}
  I_{25} = H(\A_1\,\B_1\,\C_2) + H(\A_1\,\B_2\,\C_1) + H(\A_2\,\B_1\,\C_1) + 2 H(\A_1\,\B_2) + 2 H(\A_2\,\B_2) + 2 H(\A_1\,\C_2) + 2 H(\A_2\,\C_2) + 2 H(\B_2\,\C_2) + H(\A_2\,\B_1) + H(\A_2\,\C_1) + H(\B_1\,\C_2) + H(\B_2\,\C_1) - 4 H(\A_2\,\B_2\,\C_2) - H(\A_1\,\B_1\,\C_1) - H(\A_1\,\B_2\,\C_2) - H(\A_1\,\B_1) - H(\A_1\,\C_1) - 3 H(\A_2) - 3 H(\B_2) - 3 H(\C_2)
\end{dmath*}
\begin{dmath*}
  I_{26} = H(\A_1\,\B_1\,\C_1) + H(\A_1\,\B_1\,\C_2) + H(\A_1\,\B_2\,\C_1) + H(\A_2\,\B_1\,\C_1) - H(\A_2\,\B_2\,\C_2) - H(\A_1\,\B_1) - H(\A_1\,\C_1) - H(\B_1\,\C_1)
\end{dmath*}
\begin{dmath*}
  I_{27} = H(\A_1\,\B_1\,\C_1) + H(\A_1\,\B_2\,\C_1) + H(\A_2\,\B_1\,\C_1) + H(\A_2\,\B_2\,\C_2) - H(\A_1\,\B_1\,\C_2) - H(\A_2\,\B_2) - H(\A_1\,\C_1) - H(\B_1\,\C_1)
\end{dmath*}
\begin{dmath*}
  I_{28} = H(\A_1\,\B_1\,\C_1) + H(\A_1\,\B_2\,\C_1) + H(\A_2\,\B_1\,\C_1) + H(\A_2\,\B_2\,\C_2) - H(\A_1\,\B_2\,\C_2) - H(\A_2\,\B_2) - H(\A_1\,\C_1) - H(\B_1\,\C_1)
\end{dmath*}
\begin{dmath*}
  I_{29} = H(\A_1\,\B_1\,\C_1) + H(\A_1\,\B_2\,\C_1) + H(\A_2\,\B_2\,\C_1) + H(\A_2\,\B_2\,\C_2) - H(\A_2\,\B_1\,\C_2) - H(\A_2\,\B_2) - H(\A_1\,\C_1) - H(\B_2\,\C_1)
\end{dmath*}
\begin{dmath*}
  I_{30} = H(\A_1\,\B_1\,\C_1) + H(\A_1\,\B_2\,\C_1) + H(\A_1\,\B_2\,\C_2) + H(\A_2\,\B_1\,\C_2) + H(\A_1\,\B_1) + H(\A_2\,\B_1) + H(\A_2\,\B_2) - H(\A_2\,\B_1\,\C_1) - H(\A_2\,\B_2\,\C_2) - H(\A_1\,\B_2) - H(\A_1\,\C_1) - H(\B_1\,\C_2) - H(\A_2) - H(\B_1)
\end{dmath*}
\begin{dmath*}
  I_{31} = H(\A_1\,\B_1\,\C_2) + H(\A_1\,\B_2\,\C_1) + H(\A_1\,\B_2\,\C_2) + H(\A_2\,\B_1\,\C_1) + H(\A_1\,\B_1) + H(\A_2\,\B_1) + H(\A_2\,\B_2) - H(\A_2\,\B_2\,\C_1) - H(\A_2\,\B_2\,\C_2) - H(\A_1\,\B_2) - H(\A_1\,\C_2) - H(\B_1\,\C_1) - H(\A_2) - H(\B_1)
\end{dmath*}
\begin{dmath*}
  I_{32} = H(\A_1\,\B_1\,\C_1) + H(\A_1\,\B_1\,\C_2) + H(\A_1\,\B_2\,\C_2) + H(\A_2\,\B_1\,\C_1) + H(\A_1\,\B_2) + H(\A_2\,\B_1) + H(\A_2\,\B_2) - H(\A_2\,\B_2\,\C_1) - H(\A_2\,\B_2\,\C_2) - H(\A_1\,\B_1) - H(\A_1\,\C_2) - H(\B_1\,\C_1) - H(\A_2) - H(\B_2)
\end{dmath*}
\begin{dmath*}
  I_{33} = 2 H(\A_1\,\B_1\,\C_1) + 2 H(\A_1\,\B_2\,\C_1) + H(\A_1\,\B_2\,\C_2) + H(\A_1\,\B_1) + H(\A_2\,\B_1) + H(\A_2\,\B_2) + H(\A_1\,\C_2) + H(\A_2\,\C_2) - 2 H(\A_2\,\B_1\,\C_1) - H(\A_1\,\B_1\,\C_2) - 2 H(\A_1\,\B_2) - 2 H(\A_1\,\C_1) - H(\A_2) - H(\B_1) - H(\C_2)
\end{dmath*}
\begin{dmath*}
  I_{34} = 2 H(\A_1\,\B_1\,\C_1) + H(\A_1\,\B_1\,\C_2) + H(\A_1\,\B_2\,\C_2) + H(\A_2\,\B_1\,\C_1) + H(\A_2\,\B_2\,\C_1) - H(\A_1\,\B_2\,\C_1) - H(\A_2\,\B_1\,\C_2) - H(\A_1\,\B_1) - H(\A_1\,\C_2) - H(\A_2\,\C_1) - H(\B_1\,\C_1)
\end{dmath*}
\begin{dmath*}
  I_{35} = 2 H(\A_1\,\B_2\,\C_2) + H(\A_1\,\B_1\,\C_1) + H(\A_1\,\B_2\,\C_1) + H(\A_2\,\B_1\,\C_2) + H(\A_2\,\B_2\,\C_2) - H(\A_1\,\B_1\,\C_2) - H(\A_2\,\B_1\,\C_1) - H(\A_1\,\B_2) - H(\A_1\,\C_1) - H(\A_2\,\C_2) - H(\B_2\,\C_2)
\end{dmath*}
\begin{dmath*}
  I_{36} = 2 H(\A_1\,\B_1\,\C_1) + 2 H(\A_2\,\B_1\,\C_2) + 2 H(\A_2\,\B_2\,\C_1) + H(\A_1\,\B_1\,\C_2) + H(\A_1\,\B_2\,\C_1) + H(\A_2\,\B_2\,\C_2) - 2 H(\A_1\,\B_2\,\C_2) - H(\A_2\,\B_1\,\C_1) - H(\A_1\,\B_1) - H(\A_2\,\B_2) - H(\A_1\,\C_1) - H(\A_2\,\C_2) - H(\B_1\,\C_2) - H(\B_2\,\C_1)
\end{dmath*}
\end{dgroup*}

%% file: ineqs/cca-4.tex
\begin{dgroup*}
\begin{dmath*}
  I_{0} = H_{24} - H_{2} - H_{4}
\end{dmath*}
\begin{dmath*}
  I_{1} = H_{123} + H_{124} + H_{234} - H_{12} - H_{23} - H_{14} - H_{34}
\end{dmath*}
\begin{dmath*}
  I_{2} = H_{123} + H_{124} + H_{134} + 2 H_{13} - H_{12} - H_{23} - H_{14} - H_{34} - 2 H_{1} - 2 H_{3}
\end{dmath*}
\begin{dmath*}
  I_{3} = 2 H_{124} + 2 H_{134} + 2 H_{234} + H_{13} - 2 H_{12} - 2 H_{23} - 2 H_{14} - 2 H_{34} - H_{1} - H_{3}
\end{dmath*}
\begin{dmath*}
  I_{4} = 3 H_{124} + 3 H_{234} + 2 H_{123} + 2 H_{134} - H_{1234} - 3 H_{12} - 3 H_{23} - 3 H_{14} - 3 H_{34}
\end{dmath*}
\begin{dmath*}
  I_{5} = 3 H_{123} + 3 H_{234} + 2 H_{124} + 2 H_{134} - H_{1234} - 3 H_{12} - 3 H_{23} - 3 H_{14} - 3 H_{34}
\end{dmath*}
\begin{dmath*}
  I_{6} = 3 H_{123} + 3 H_{134} + 2 H_{124} + 2 H_{234} + H_{13} - H_{1234} - 3 H_{12} - 3 H_{23} - 3 H_{14} - 3 H_{34} - H_{1} - H_{3}
\end{dmath*}
\begin{dmath*}
  I_{7} = 3 H_{123} + 3 H_{124} + 2 H_{134} + 2 H_{234} + H_{13} - H_{1234} - 3 H_{12} - 3 H_{23} - 3 H_{14} - 3 H_{34} - H_{1} - H_{3}
\end{dmath*}
\begin{dmath*}
  I_{8} = 3 H_{124} + 3 H_{134} + 2 H_{123} + 2 H_{234} + 2 H_{13} - H_{1234} - 3 H_{12} - 3 H_{23} - 3 H_{14} - 3 H_{34} - 2 H_{1} - 2 H_{3}
\end{dmath*}
\begin{dmath*}
  I_{9} = 5 H_{234} + 4 H_{123} + 4 H_{124} + 4 H_{134} - 2 H_{1234} - 5 H_{12} - 5 H_{23} - 5 H_{14} - 5 H_{34}
\end{dmath*}
\begin{dmath*}
  I_{10} = 5 H_{124} + 4 H_{123} + 4 H_{134} + 4 H_{234} + H_{13} - 2 H_{1234} - 5 H_{12} - 5 H_{23} - 5 H_{14} - 5 H_{34} - H_{1} - H_{3}
\end{dmath*}
\begin{dmath*}
  I_{11} = 5 H_{123} + 4 H_{124} + 4 H_{134} + 4 H_{234} + H_{13} - 2 H_{1234} - 5 H_{12} - 5 H_{23} - 5 H_{14} - 5 H_{34} - H_{1} - H_{3}
\end{dmath*}
\begin{dmath*}
  I_{12} = 6 H_{123} + 6 H_{124} + 6 H_{134} + 6 H_{234} + H_{13} - 3 H_{1234} - 7 H_{12} - 7 H_{23} - 7 H_{14} - 7 H_{34} - H_{1} - H_{3}
\end{dmath*}
\end{dgroup*}

%% file: ineqs/cca-5.tex
\begin{dgroup*}
\begin{dmath*}
  I_{0} = H_{35} - H_{3} - H_{5}
\end{dmath*}
\begin{dmath*}
  I_{1} = H_{134} - H_{34} - H_{1}
\end{dmath*}
\begin{dmath*}
  I_{2} = H_{235} + H_{3} - H_{23} - H_{35}
\end{dmath*}
\begin{dmath*}
  I_{3} = H_{124} + H_{245} - H_{12} - H_{24} - H_{45}
\end{dmath*}
\begin{dmath*}
  I_{4} = H_{1234} + H_{1245} + H_{1345} + H_{2345} + H_{124} + H_{245} + H_{35} - H_{123} - H_{234} - H_{125} - H_{145} - H_{345} - H_{12} - H_{24} - H_{45} - H_{3} - H_{5}
\end{dmath*}
\begin{dmath*}
  I_{5} = H_{1234} + H_{1235} + H_{1345} + H_{2345} + 3 H_{124} + 2 H_{35} + 3 H_{2} - H_{123} - H_{234} - H_{125} - H_{145} - H_{345} - 3 H_{12} - 3 H_{24} - 2 H_{3} - 2 H_{5}
\end{dmath*}
\begin{dmath*}
  I_{6} = H_{1234} + H_{1235} + H_{1245} + H_{1345} + 4 H_{124} + H_{134} + H_{135} + H_{235} + 4 H_{2} + H_{3} + H_{5} - H_{123} - H_{234} - H_{125} - H_{145} - H_{345} - 4 H_{12} - 4 H_{24} - 2 H_{35} - H_{23} - H_{34} - H_{15} - H_{1}
\end{dmath*}
\begin{dmath*}
  I_{7} = H_{1234} + H_{1235} + H_{1245} + H_{2345} + 7 H_{124} + H_{35} + 7 H_{2} - H_{123} - H_{234} - H_{125} - H_{145} - H_{345} - 7 H_{12} - 7 H_{24} - H_{3} - H_{5}
\end{dmath*}
\begin{dmath*}
  I_{8} = 2 H_{1235} + 2 H_{1245} + 2 H_{1345} + 2 H_{2345} + 5 H_{124} + 2 H_{245} + H_{135} + 3 H_{2} + H_{5} - 2 H_{123} - 2 H_{234} - 2 H_{125} - 2 H_{145} - 2 H_{345} - 5 H_{12} - 5 H_{24} - 2 H_{45} - H_{15} - H_{35}
\end{dmath*}
\begin{dmath*}
  I_{9} = 3 H_{1245} + 3 H_{1345} + 3 H_{2345} + 2 H_{1234} + 2 H_{1235} + 3 H_{124} + H_{245} + 2 H_{2} - H_{12345} - 3 H_{123} - 3 H_{234} - 3 H_{125} - 3 H_{145} - 3 H_{345} - 3 H_{12} - 3 H_{24} - H_{45}
\end{dmath*}
\begin{dmath*}
  I_{10} = 3 H_{1235} + 3 H_{1345} + 3 H_{2345} + 2 H_{1234} + 2 H_{1245} + 5 H_{124} + H_{35} + 5 H_{2} - H_{12345} - 3 H_{123} - 3 H_{234} - 3 H_{125} - 3 H_{145} - 3 H_{345} - 5 H_{12} - 5 H_{24} - H_{3} - H_{5}
\end{dmath*}
\begin{dmath*}
  I_{11} = 3 H_{1234} + 3 H_{1345} + 3 H_{2345} + 2 H_{1235} + 2 H_{1245} + 4 H_{124} + H_{245} + 2 H_{35} + 3 H_{2} - H_{12345} - 3 H_{123} - 3 H_{234} - 3 H_{125} - 3 H_{145} - 3 H_{345} - 4 H_{12} - 4 H_{24} - H_{45} - 2 H_{3} - 2 H_{5}
\end{dmath*}
\begin{dmath*}
  I_{12} = 3 H_{1234} + 3 H_{1235} + 3 H_{1345} + 2 H_{1245} + 2 H_{2345} + 5 H_{124} + H_{235} + H_{35} + 5 H_{2} - H_{12345} - 3 H_{123} - 3 H_{234} - 3 H_{125} - 3 H_{145} - 3 H_{345} - 5 H_{12} - 5 H_{24} - H_{23} - 2 H_{5} - H_{3}
\end{dmath*}
\begin{dmath*}
  I_{13} = 3 H_{1234} + 3 H_{1245} + 3 H_{1345} + 2 H_{1235} + 2 H_{2345} + 5 H_{124} + H_{134} + H_{235} + H_{245} + 4 H_{2} - H_{12345} - 3 H_{123} - 3 H_{234} - 3 H_{125} - 3 H_{145} - 3 H_{345} - 5 H_{12} - 5 H_{24} - H_{23} - H_{34} - H_{45} - H_{1} - H_{5}
\end{dmath*}
\begin{dmath*}
  I_{14} = 3 H_{1235} + 3 H_{1245} + 3 H_{1345} + 2 H_{1234} + 2 H_{2345} + 6 H_{124} + H_{134} + H_{235} + 6 H_{2} + H_{3} - H_{12345} - 3 H_{123} - 3 H_{234} - 3 H_{125} - 3 H_{145} - 3 H_{345} - 6 H_{12} - 6 H_{24} - H_{23} - H_{34} - H_{35} - H_{1}
\end{dmath*}
\begin{dmath*}
  I_{15} = 3 H_{1235} + 3 H_{1245} + 3 H_{2345} + 2 H_{1234} + 2 H_{1345} + 8 H_{124} + 8 H_{2} - H_{12345} - 3 H_{123} - 3 H_{234} - 3 H_{125} - 3 H_{145} - 3 H_{345} - 8 H_{12} - 8 H_{24}
\end{dmath*}
\begin{dmath*}
  I_{16} = 3 H_{1234} + 3 H_{1245} + 3 H_{2345} + 2 H_{1235} + 2 H_{1345} + 7 H_{124} + H_{245} + H_{35} + 6 H_{2} - H_{12345} - 3 H_{123} - 3 H_{234} - 3 H_{125} - 3 H_{145} - 3 H_{345} - 7 H_{12} - 7 H_{24} - H_{45} - H_{3} - H_{5}
\end{dmath*}
\begin{dmath*}
  I_{17} = 3 H_{1234} + 3 H_{1235} + 3 H_{1245} + 2 H_{1345} + 2 H_{2345} + 8 H_{124} + H_{235} + 8 H_{2} + H_{3} - H_{12345} - 3 H_{123} - 3 H_{234} - 3 H_{125} - 3 H_{145} - 3 H_{345} - 8 H_{12} - 8 H_{24} - H_{23} - H_{35}
\end{dmath*}
\begin{dmath*}
  I_{18} = 3 H_{1234} + 3 H_{1235} + 3 H_{2345} + 2 H_{1245} + 2 H_{1345} + 9 H_{124} + H_{35} + 9 H_{2} - H_{12345} - 3 H_{123} - 3 H_{234} - 3 H_{125} - 3 H_{145} - 3 H_{345} - 9 H_{12} - 9 H_{24} - H_{3} - H_{5}
\end{dmath*}
\begin{dmath*}
  I_{19} = 5 H_{1345} + 5 H_{2345} + 4 H_{1234} + 4 H_{1235} + 4 H_{1245} + 6 H_{124} + H_{245} + H_{35} + 5 H_{2} - 2 H_{12345} - 5 H_{123} - 5 H_{234} - 5 H_{125} - 5 H_{145} - 5 H_{345} - 6 H_{12} - 6 H_{24} - H_{45} - H_{3} - H_{5}
\end{dmath*}
\begin{dmath*}
  I_{20} = 5 H_{1235} + 5 H_{1345} + 4 H_{1234} + 4 H_{1245} + 4 H_{2345} + 7 H_{124} + H_{235} + 7 H_{2} - 2 H_{12345} - 5 H_{123} - 5 H_{234} - 5 H_{125} - 5 H_{145} - 5 H_{345} - 7 H_{12} - 7 H_{24} - H_{23} - H_{5}
\end{dmath*}
\begin{dmath*}
  I_{21} = 5 H_{1245} + 5 H_{1345} + 4 H_{1234} + 4 H_{1235} + 4 H_{2345} + 7 H_{124} + H_{134} + H_{235} + H_{245} + 6 H_{2} + H_{3} - 2 H_{12345} - 5 H_{123} - 5 H_{234} - 5 H_{125} - 5 H_{145} - 5 H_{345} - 7 H_{12} - 7 H_{24} - H_{23} - H_{34} - H_{35} - H_{45} - H_{1}
\end{dmath*}
\begin{dmath*}
  I_{22} = 5 H_{1245} + 5 H_{2345} + 4 H_{1234} + 4 H_{1235} + 4 H_{1345} + 9 H_{124} + H_{245} + 8 H_{2} - 2 H_{12345} - 5 H_{123} - 5 H_{234} - 5 H_{125} - 5 H_{145} - 5 H_{345} - 9 H_{12} - 9 H_{24} - H_{45}
\end{dmath*}
\begin{dmath*}
  I_{23} = 5 H_{1234} + 5 H_{1245} + 4 H_{1235} + 4 H_{1345} + 4 H_{2345} + 9 H_{124} + H_{235} + H_{245} + 8 H_{2} - 2 H_{12345} - 5 H_{123} - 5 H_{234} - 5 H_{125} - 5 H_{145} - 5 H_{345} - 9 H_{12} - 9 H_{24} - H_{23} - H_{45} - H_{5}
\end{dmath*}
\begin{dmath*}
  I_{24} = 5 H_{1235} + 5 H_{1245} + 4 H_{1234} + 4 H_{1345} + 4 H_{2345} + 10 H_{124} + H_{235} + 10 H_{2} + H_{3} - 2 H_{12345} - 5 H_{123} - 5 H_{234} - 5 H_{125} - 5 H_{145} - 5 H_{345} - 10 H_{12} - 10 H_{24} - H_{23} - H_{35}
\end{dmath*}
\begin{dmath*}
  I_{25} = 5 H_{1235} + 5 H_{2345} + 4 H_{1234} + 4 H_{1245} + 4 H_{1345} + 11 H_{124} + 11 H_{2} - 2 H_{12345} - 5 H_{123} - 5 H_{234} - 5 H_{125} - 5 H_{145} - 5 H_{345} - 11 H_{12} - 11 H_{24}
\end{dmath*}
\begin{dmath*}
  I_{26} = 5 H_{1234} + 5 H_{1235} + 4 H_{1245} + 4 H_{1345} + 4 H_{2345} + 11 H_{124} + H_{235} + 11 H_{2} - 2 H_{12345} - 5 H_{123} - 5 H_{234} - 5 H_{125} - 5 H_{145} - 5 H_{345} - 11 H_{12} - 11 H_{24} - H_{23} - H_{5}
\end{dmath*}
\begin{dmath*}
  I_{27} = 7 H_{1245} + 6 H_{1234} + 6 H_{1235} + 6 H_{1345} + 6 H_{2345} + 11 H_{124} + H_{235} + H_{245} + 10 H_{2} + H_{3} - 3 H_{12345} - 7 H_{123} - 7 H_{234} - 7 H_{125} - 7 H_{145} - 7 H_{345} - 11 H_{12} - 11 H_{24} - H_{23} - H_{35} - H_{45}
\end{dmath*}
\begin{dmath*}
  I_{28} = 7 H_{1235} + 6 H_{1234} + 6 H_{1245} + 6 H_{1345} + 6 H_{2345} + 13 H_{124} + H_{235} + 13 H_{2} + H_{3} - 3 H_{12345} - 7 H_{123} - 7 H_{234} - 7 H_{125} - 7 H_{145} - 7 H_{345} - 13 H_{12} - 13 H_{24} - H_{23} - H_{35}
\end{dmath*}
\begin{dmath*}
  I_{29} = 10 H_{1234} + 10 H_{1345} + 8 H_{1235} + 8 H_{1245} + 8 H_{2345} + 15 H_{124} + 2 H_{134} + 2 H_{235} + 2 H_{245} + H_{35} + 13 H_{2} - 4 H_{12345} - 10 H_{123} - 10 H_{234} - 10 H_{125} - 10 H_{145} - 10 H_{345} - 15 H_{12} - 15 H_{24} - 2 H_{23} - 2 H_{34} - 2 H_{45} - 3 H_{5} - 2 H_{1} - H_{3}
\end{dmath*}
\begin{dmath*}
  I_{30} = 10 H_{1234} + 10 H_{2345} + 8 H_{1235} + 8 H_{1245} + 8 H_{1345} + 19 H_{124} + 2 H_{245} + 3 H_{35} + 17 H_{2} - 4 H_{12345} - 10 H_{123} - 10 H_{234} - 10 H_{125} - 10 H_{145} - 10 H_{345} - 19 H_{12} - 19 H_{24} - 2 H_{45} - 3 H_{3} - 3 H_{5}
\end{dmath*}
\begin{dmath*}
  I_{31} = 14 H_{1345} + 12 H_{1234} + 12 H_{1235} + 12 H_{1245} + 12 H_{2345} + 19 H_{124} + 2 H_{134} + 2 H_{235} + 2 H_{245} + 17 H_{2} + H_{3} - 6 H_{12345} - 14 H_{123} - 14 H_{234} - 14 H_{125} - 14 H_{145} - 14 H_{345} - 19 H_{12} - 19 H_{24} - 2 H_{23} - 2 H_{34} - 2 H_{45} - H_{35} - 2 H_{1} - H_{5}
\end{dmath*}
\begin{dmath*}
  I_{32} = 14 H_{2345} + 12 H_{1234} + 12 H_{1235} + 12 H_{1245} + 12 H_{1345} + 23 H_{124} + 2 H_{245} + H_{35} + 21 H_{2} - 6 H_{12345} - 14 H_{123} - 14 H_{234} - 14 H_{125} - 14 H_{145} - 14 H_{345} - 23 H_{12} - 23 H_{24} - 2 H_{45} - H_{3} - H_{5}
\end{dmath*}
\begin{dmath*}
  I_{33} = 14 H_{1234} + 12 H_{1235} + 12 H_{1245} + 12 H_{1345} + 12 H_{2345} + 23 H_{124} + 2 H_{235} + 2 H_{245} + H_{35} + 21 H_{2} - 6 H_{12345} - 14 H_{123} - 14 H_{234} - 14 H_{125} - 14 H_{145} - 14 H_{345} - 23 H_{12} - 23 H_{24} - 2 H_{23} - 2 H_{45} - 3 H_{5} - H_{3}
\end{dmath*}
\begin{dmath*}
  I_{34} = 16 H_{1234} + 16 H_{1235} + 16 H_{1245} + 16 H_{1345} + 16 H_{2345} + 27 H_{124} + 2 H_{235} + 2 H_{245} + 25 H_{2} + H_{3} - 8 H_{12345} - 18 H_{123} - 18 H_{234} - 18 H_{125} - 18 H_{145} - 18 H_{345} - 27 H_{12} - 27 H_{24} - 2 H_{23} - 2 H_{45} - H_{35} - H_{5}
\end{dmath*}
\begin{dmath*}
  I_{35} = H_{12345} + H_{124} + H_{235} + H_{245} - H_{1235} - H_{12} - H_{23} - H_{24} - H_{45} - H_{5}
\end{dmath*}
\begin{dmath*}
  I_{36} = H_{12345} + 3 H_{124} + 2 H_{35} + 3 H_{2} - H_{1245} - 3 H_{12} - 3 H_{24} - 2 H_{3} - 2 H_{5}
\end{dmath*}
\begin{dmath*}
  I_{37} = H_{12345} + 2 H_{124} + H_{135} + H_{245} + H_{2} + H_{5} - H_{1234} - 2 H_{12} - 2 H_{24} - H_{15} - H_{35} - H_{45}
\end{dmath*}
\begin{dmath*}
  I_{38} = H_{12345} + 5 H_{124} + 5 H_{2} - H_{1345} - 5 H_{12} - 5 H_{24}
\end{dmath*}
\begin{dmath*}
  I_{39} = H_{12345} + 3 H_{124} + H_{134} + H_{135} + H_{235} + 3 H_{2} + H_{3} + H_{5} - H_{2345} - 3 H_{12} - 3 H_{24} - 2 H_{35} - H_{23} - H_{34} - H_{15} - H_{1}
\end{dmath*}
\end{dgroup*}